\documentclass[trackchanges]{aastex701}
\usepackage{amsmath}
\usepackage{physics}
\usepackage{multirow}

\begin{document}

\title{A Stratification in Magnetic Field Structures: The Radio Outflow in NGC\,4151}

\author[0009-0000-1447-5419,sname='Salmoli Ghosh']{Salmoli Ghosh}
\affiliation{National Centre for Radio Astrophysics (NCRA) - Tata Institute of Fundamental Research (TIFR), 
S. P. Pune University Campus, Ganeshkhind, Pune 411007, Maharashtra, India}
\email[show]{salmoli@ncra.tifr.res.in}  

\author[0000-0003-3203-1613]{Preeti Kharb}
\affiliation{National Centre for Radio Astrophysics (NCRA) - Tata Institute of Fundamental Research (TIFR), 
S. P. Pune University Campus, Ganeshkhind, Pune 411007, Maharashtra, India}
\email{kharb@ncra.tifr.res.in}

\author[0000-0001-8470-749X]{Elisa Costantini}
\affiliation{SRON Space Research Organisation Netherlands, Niels Bohrweg 4, 2333CA, Leiden, The Netherlands}
\affiliation{Anton Pannekoek Institute for Astronomy, Postbus 94249, 1090 GE Amsterdam, The Netherlands}
\email{E.Costantini@sron.nl}

\author[0000-0002-6972-2760]{Jack F. Gallimore}
\affil{Department of Physics and Astronomy, Bucknell University, Lewisburg, PA 17837, USA}
\email[]{jgallimo@bucknell.edu}

\author[0000-0001-7361-0246]{David Williams-Baldwin}
\affiliation{Jodrell Bank Centre for Astrophysics, School of Physics and Astronomy,
The University of Manchester Alan Turing Building, Oxford Road, Manchester M13 9PY
Lancashire, United Kingdom}
\email{david.williams-7@manchester.ac.uk}

\author[0000-0002-4992-4664]{Missagh Mehdipour}
\affiliation{Department of Astronomy, University of Michigan, 1085 South University Avenue, Ann Arbor, MI, 48109, USA}
\email{missagh.mehdipour@gmail.com}


\begin{abstract}
The nature of radio outflows in radio-quiet AGN remains poorly understood. In this study, we present kpc-scale polarization observations of the Seyfert galaxy NGC\,4151 using the Karl G. Jansky Very Large Array (VLA) in B-array at 3 and 10 GHz. We find that the inferred magnetic (B-) field structures show a stratification: a `spine-sheath'-like structure, with fields perpendicular to the jet direction in the `spine' and parallel in the `sheath', is observed in the higher resolution ($0.5\arcsec$) image at 10~GHz. In addition, a `wind'-like component with B-fields perpendicular to the radio outflow is observed in the 3~GHz image (resolution $2\arcsec$); this feature is prominent along the `receding' (eastern) jet direction. Rotation measure (RM) ranges from $-230$ to 250~rad~m$^{-2}$ over the polarized regions, indicating a low-electron-density ($10^{-2}-10^{-3}$ cm$^{-3}$) tenuous medium surrounding the source causing Faraday rotation. A {tentative} RM gradient of $+75$ to $-25$ rad~m$^{-2}$ is observed 
transverse to the northern `wind' component, while a similar gradient with opposite sign is seen across the southern `wind' component, suggestive of a helical magnetic field threading the outflow. Based on an analysis of the available radio and X-ray data, we conclude that the stratified radio outflow in NGC~4151 is magnetically-driven. The bi-conical radio `wind' is found to be massive ($1050-3200~M_\odot$) with a high mass outflow rate ($0.01-0.03$~M$_\odot$ yr$^{-1}$) but low in kinetic power ($<0.01$\% of L$_{\rm{bol}}$), making it less impactful for galactic-scale feedback. Our study suggests that radio-quiet AGN may also host magnetically dominant jets and winds, even while their jets are smaller and weaker compared to radio-loud AGN.
\end{abstract}



\section{Introduction}
While radio-quiet (RQ) Seyfert galaxies constitute the majority of nearby active galactic nuclei (AGN), the nature of their radio emission remains poorly understood \citep{Ishibashi2011,Panessa2019,Radcliffe2021}. Unlike their radio-loud {(RL)  counterparts \citep[$R=S_{5~GHz}/S_{B-band}>10$;][]{Kellermann1989}, Seyfert galaxies} typically host small and relatively weak radio jets that remain confined to their host galaxies. The origin of their radio emission on kpc scales has been suggested to be either AGN-jet or wind-related \citep{Colbert1996b, Sebastian2020} or starburst-wind-related \citep{Baum1993, Cecil2001} or a combination of both \citep{Veilleux2005, Silpa2021}. Very long baseline interferometry (VLBI) observations of Seyfert galaxies support the AGN-jet scenario in several Seyfert galaxies \citep{Giroletti2009, Kharb2021}, highlighting the greater complexity of radio outflows in RQ AGN compared to RL AGN. 

NGC\,4151 is a well-studied nearby ($z=0.0033$) Seyfert type 1.5 galaxy \citep{Williams2017}. This barred spiral galaxy has also recently been identified as a changing-look AGN \citep[CLAGN;][]{Feng2024}. A wedge-shaped torus, a stratified broad-line region (BLR), an ultra-fast outflow from near the {
supermassive black hole \citep[SMBH; $\mathrm{M_{BH}} = 
1.66^{+0.48}_{-0.34}\times10^7\,M_\odot$
in NGC\,4151;][]{Bentz2022},} and a powerful accretion-disk-driven wind have been suggested to be present in this galaxy \citep{Tombesi2011, Gianolli2023, Xiang2025}. 
Mildly relativistic gas outflows ($v\sim0.03c-0.2c$) are suggested to be interacting with the ambient media and producing strong shocks, which are accelerating particles to high energies. This has been suggested to cause X-ray absorption in NGC\,4151 as well as the emission of gamma rays and neutrinos \citep{Crenshaw2007, Tombesi2011, Peretti2025}. {It is interesting to note that the `radio-quiet' NGC\,4151 exhibits properties similar to radio-loud sources that possess highly relativistic jets.} 

NGC\,4151 has a relatively low star formation rate (SFR) of $0.25-0.95$~M$_\sun$~yr$^{-1}$ \citep{Peretti2025}. There are suggestions of an over-density of gas in the bar fueling the AGN, and `AGN feedback' quenching the star formation \citep{Pedlar1992, Mundell1999b}. NGC\,4151 shows a multi-layered jet structure at radio frequencies extending over 10~$\arcsec$ ($\sim$900 pc) at wavelengths of $6-20$~cm, and aligned with the polarized optical continuum emission \citep{Johnston1982}. Higher resolution images at 1.7~GHz show an elongated structure of size $3.5\arcsec\times0.5\arcsec$, with a double source at the centre separated by 0.45~$\arcsec$ \citep{Booler1982,Carral1990}. The images by \citet{Ulvestad2005} reveal knotty jet components aligned in east-west direction at a position angle (PA) of $\sim77\degr$. 

A 1.4~GHz study of NGC\,4151 by \citet{Mundell1995, Mundell2001} shows absorption against the eastern jet components, supporting the interpretation that the eastern jet is receding while the western jet is approaching us. Observations with the enhanced Multi Element Remotely Linked Interferometer Network (\textit{e}-MERLIN) array reveal a factor of 2 increase in the radio peak flux density of the central-most component (first jet component) compared to what was observed 2 decades ago, indicating its variable nature. Furthermore, the flat spectral index core suggests that the jets are currently being powered by the AGN \citep{Williams2017,Williams2020}.  

The emission-line region gas is extended with both aligned and non-aligned components with the radio outflow, suggesting the effects of both photoionization from the AGN and shock ionization by the jets \citep{Mundell2003,Wang2011a,Williams2017}. Filamentary structures are observed in different emission lines as well, consistent with episodic activity and precession \citep[e.g.,][]{Williams2017}. The jet speed inferred from VLBI observations is low: an upper limit of 0.05c is inferred at a distance of 0.16 pc from the core \citep{Ulvestad2005, Williams2017}. The low jet speed has been suggested to arise from strong jet medium interaction \citep{Ulvestad2005}. X-ray observations of NGC\,4151 have revealed quasi-periodically oscillating signals, most likely due to the accretion disk instabilities \citep{Yongkang2025}. A long-term optical emission-line and continuum variability has also been studied for this galaxy, consistent with changes in accretion rate or an accelerating outflow from near the SMBH \citep[e.g.,][]{Shapovalova2010}.

In this paper, we present a radio polarimetric study of NGC\,4151 at 3 and 10 GHz with the  Karl G. Jansky Very Large array (VLA). We present rotation measure (RM), spectral index and depolarization images of the radio outflow. The paper consists of the following sections. Section \ref{sec:obs} includes the observational details and the data reduction procedure. Section \ref{sec:results} presents our findings with Sections \ref{sec:radiomorph}, \ref{sec:depolRM}, and \ref{sec:wind} mentioning details about the radio structure and inferred B-fields, depolarization models, and properties of the observed `wind' component. Section \ref{sec:discussion} discusses the implications of our findings. Section \ref{sec:conclusion} presents the conclusions from this work.
Throughout this paper, we adopt the $\lambda$ cold dark matter cosmology with H$_0=$67.8~km~s$^{-1}$~Mpc$^{-1}$, $\Omega_{mat}=0.308$, $\Omega_{vac}=0.692$. Spectral index, $\alpha$, is defined such that flux density at a frequency $\nu$ is $S_\nu\propto\nu^\alpha$.

\begin{figure*}
\includegraphics[width=6.5cm,trim=80 220 120 100]{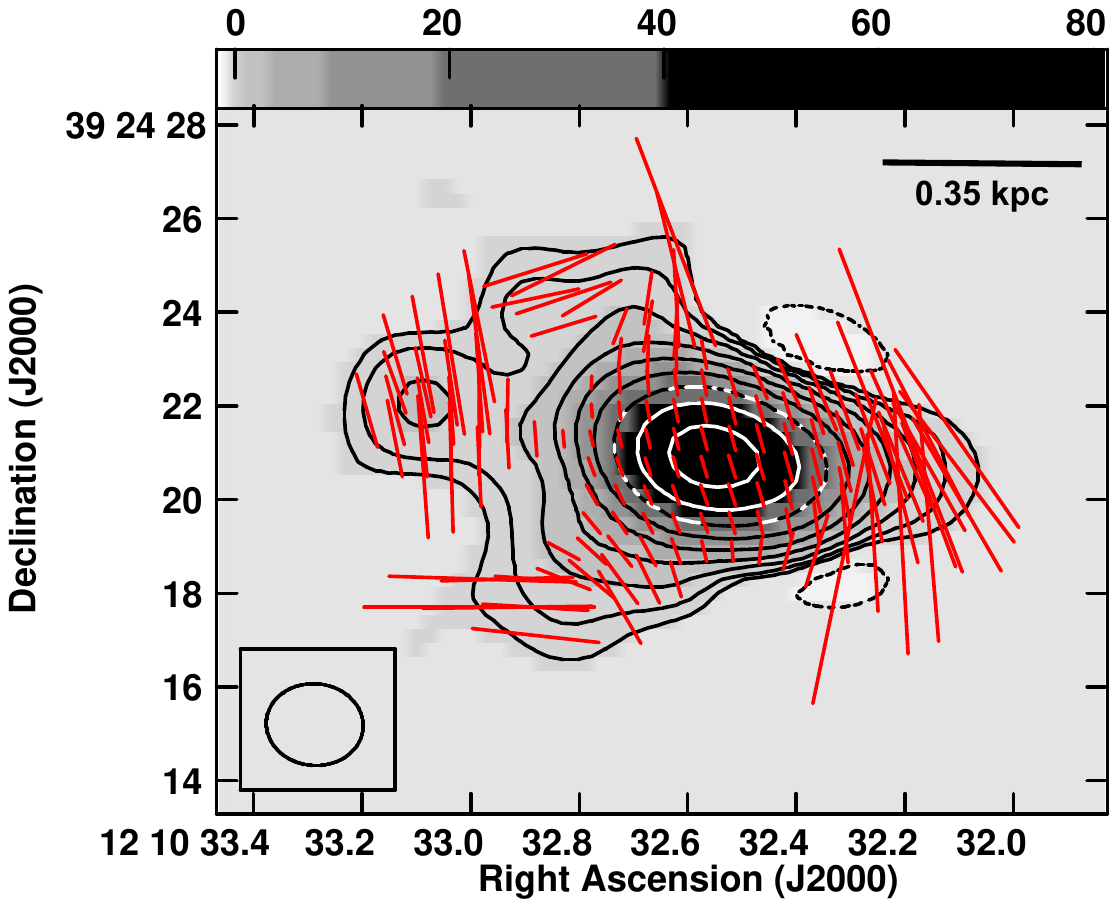}
\includegraphics[width=8.5cm,trim=100 240 40 200]{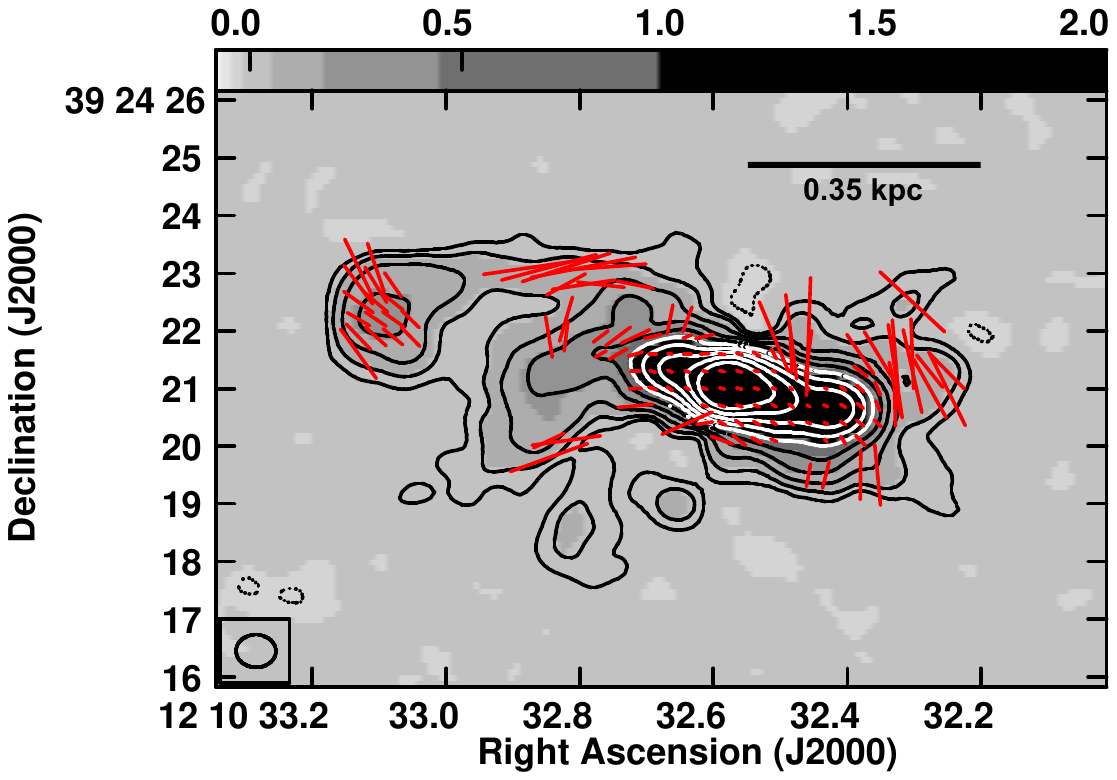}
\includegraphics[width=8.2cm,trim=160 30 120 50]{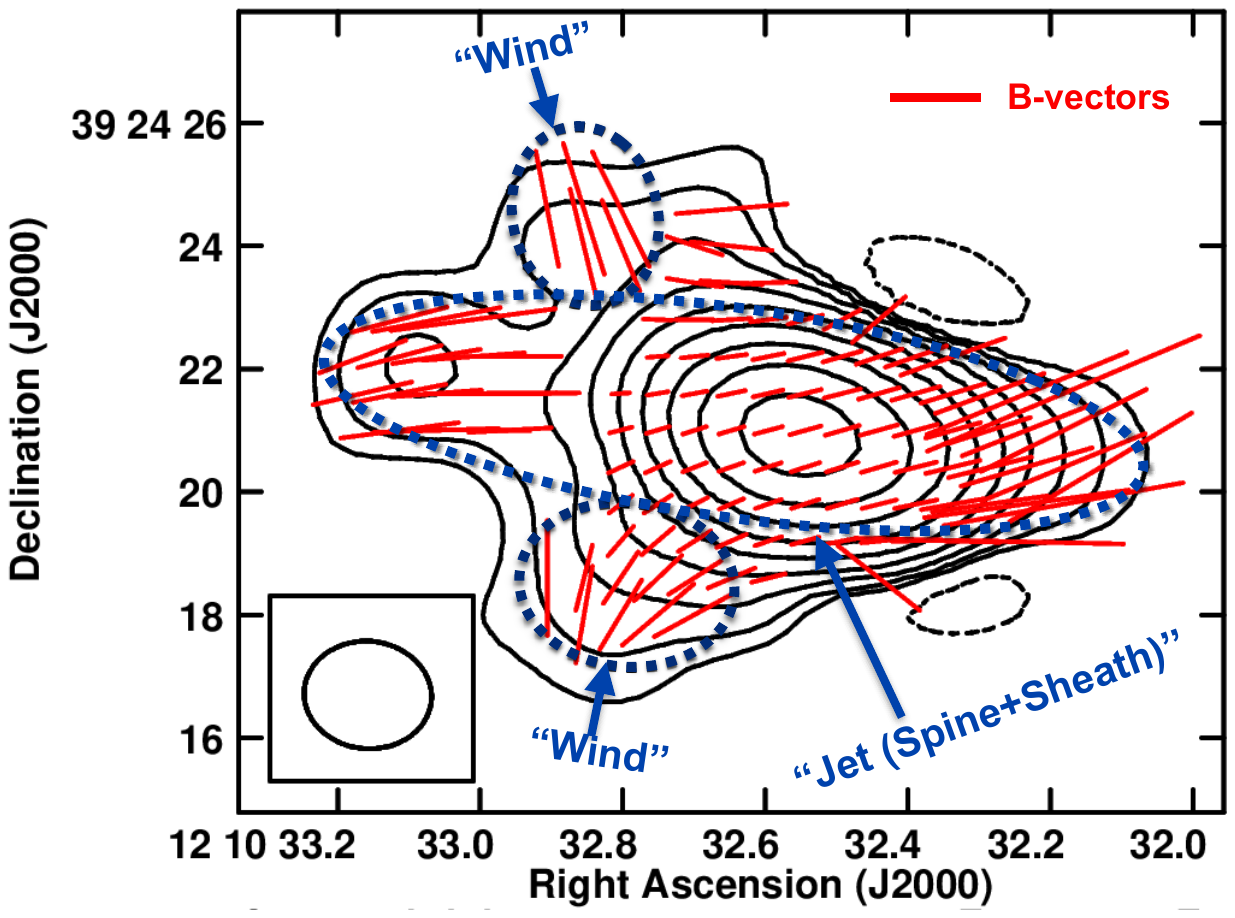}
\includegraphics[width=9.9cm,trim=30 30 20 80]{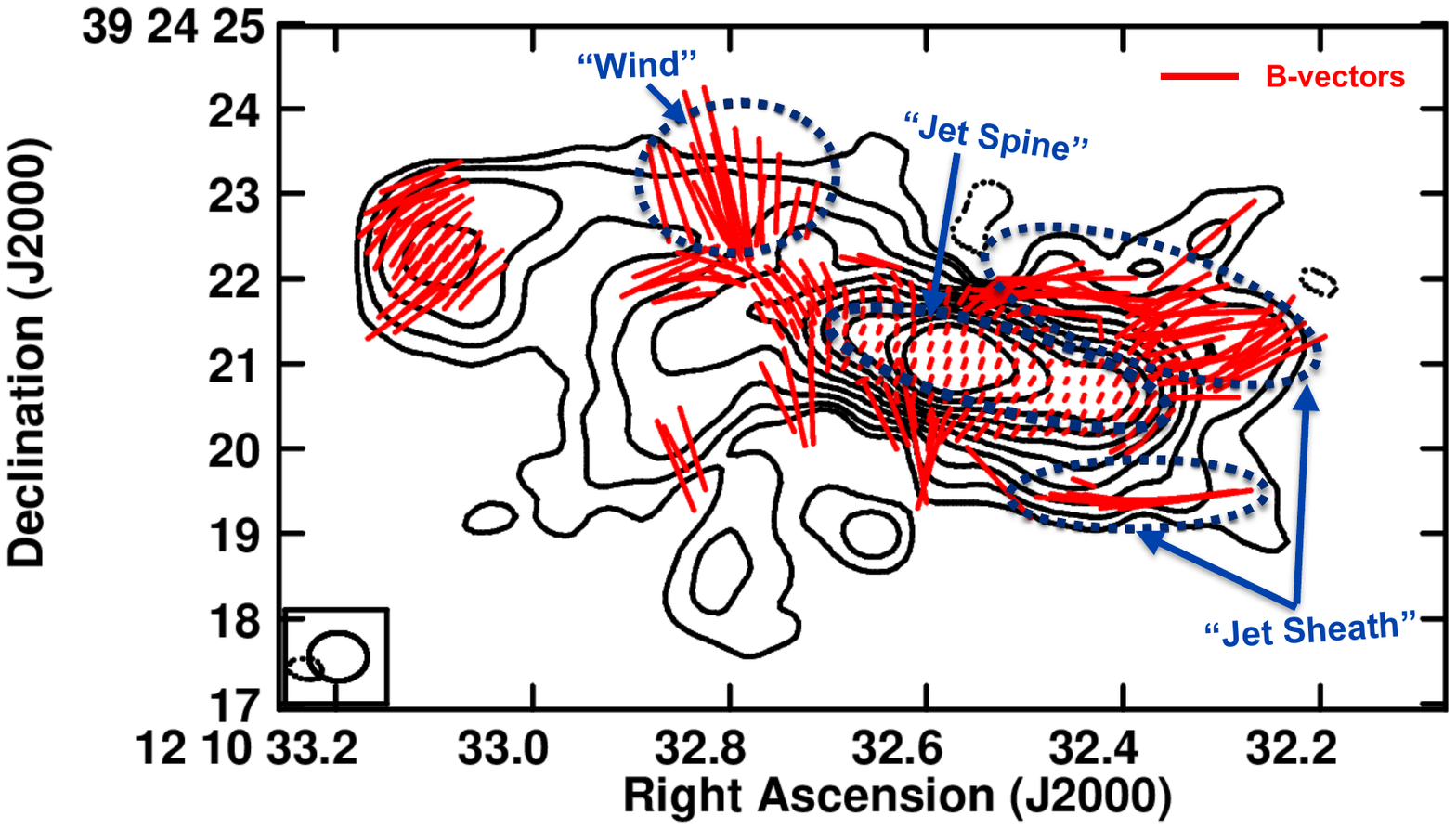}
\caption{\small Top Left: 3 GHz image of NGC\,4151 at a resolution of $\sim$ 2 arcsec, obtained using the VLA B-array.  The polarization vectors are shown in red ticks with 6 $\arcsec$ of length proportional to a fractional polarization of 10\%. The contour levels shown in black are at 5$\sigma\times$ ($\pm1$, 2, 4, 8, 16, 32, 64, 128, 256) with $\sigma=80~\mu$Jy~beam$^{-1}$. The gray-scale varies from $-0.8$ to 80 mJy beam$^{-1}$ in logarithmic scale. The dark horizontal line marks the distance scale of $\sim$ 350 pc (4 arcsec). The synthesized beam shown at the bottom left corner of the image is of size 2.1 arcsec $\times$ 1.7 arcsec at a PA of 85$\degr$. Top Right: 10 GHz image of NGC\,4151 at a resolution of $\sim$ 0.5 arcsec obtained using the VLA B-array.  The polarization vectors are shown in red ticks with 2~arcsec proportional to a fractional polarization of 50\%. The contour levels shown in black are at 3$\sigma\times$ ($\pm1$, 2, 4, 8, 16, 32, 64, 128, 256, 512) with $\sigma=8~\mu$Jy~beam$^{-1}$. The gray-scale varies from $-0.07$ to 2 mJy beam$^{-1}$ in logarithmic scale. The synthesized beam is of size 0.7 arcsec $\times$ 0.6 arcsec at a PA of 89$\degr$. Bottom Left \& Right: Inferred magnetic-field vectors in the NGC\,4151 outflow at 3 (left) and 10~GHz (right) by rotating EVPAs by 90$\degr$. The B-field shows a stratification and is found to be similar to what is predicted for a jet `spine', jet `sheath' and a magnetically driven `wind' \citep[e.g.,][]{Laing2014, Mehdipour2019}. {North is to the top and East to the left in this and all subsequent figures.}}
\label{fig:3GHzimage}
\end{figure*}

\begin{figure*}
\centering
\includegraphics[width=8cm]{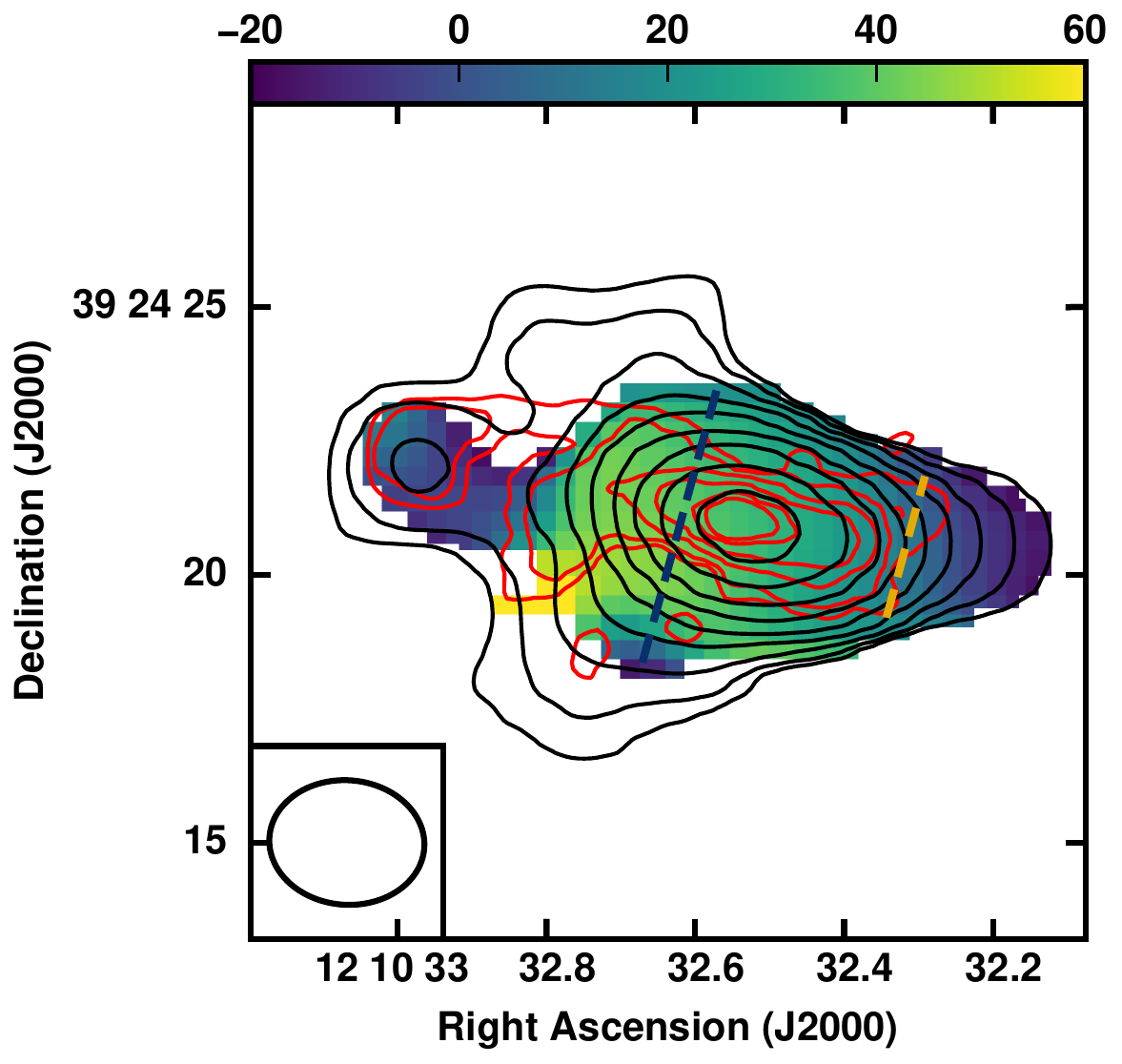}
\includegraphics[width=8cm]{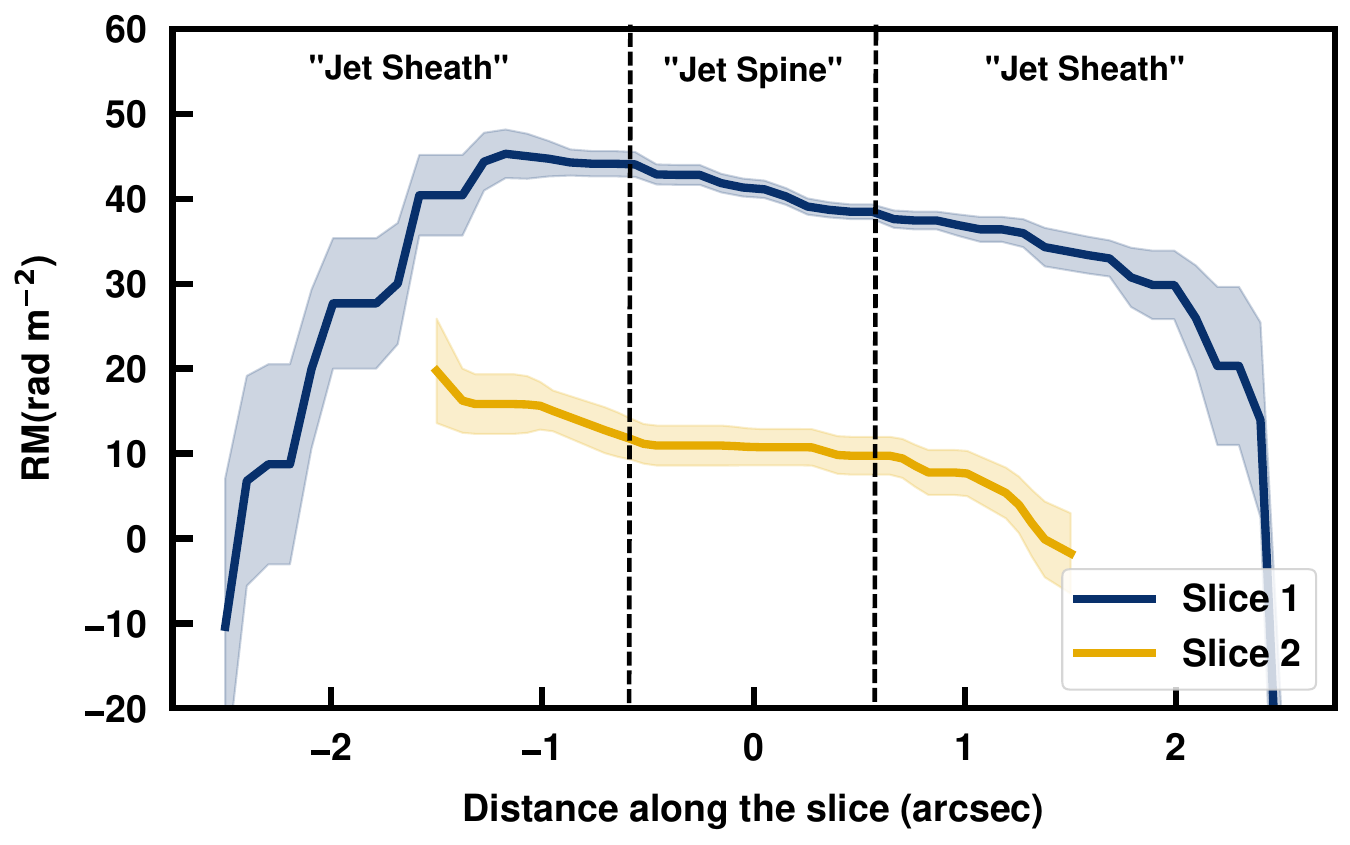}
\caption{\small {Left: The in-band 3~GHz RM image, blanked by an RM error of 20 rad m$^{-2}$. The 3 GHz contours are shown in black, the 10 GHz contours in red. The blue and the yellow dashed lines indicate the transverse slices taken across the jet. Right: The RM and RM error values are shown with a solid line and a shaded region, respectively, corresponding to the slices drawn on the left image. The slices taken from bottom to top on the left figure are represented from left to right on the right figure.}}
\label{fig:3GHzRM}
\end{figure*}

\begin{figure*}
\centering
\includegraphics[width=8cm]{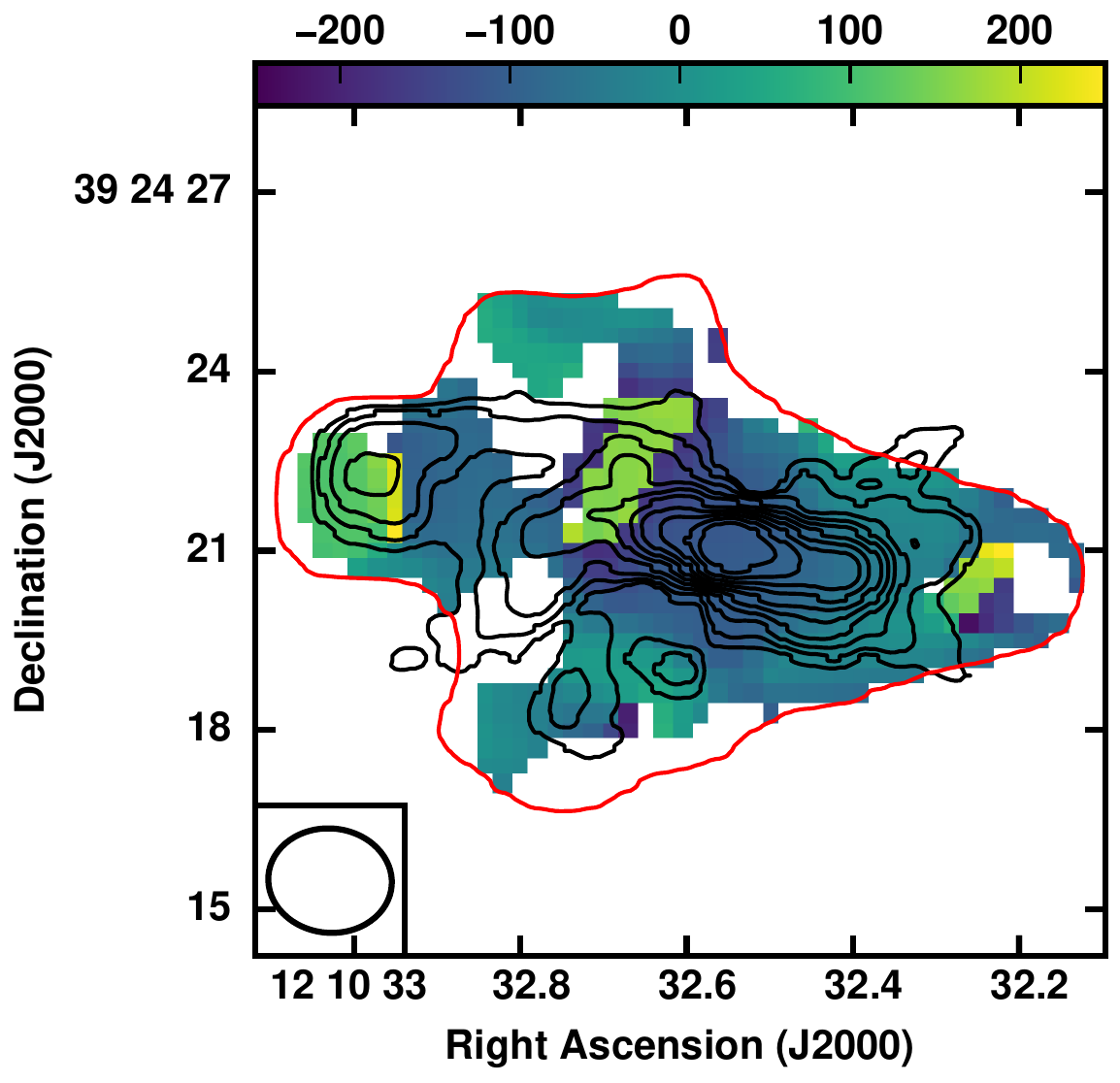}
\includegraphics[width=8cm]{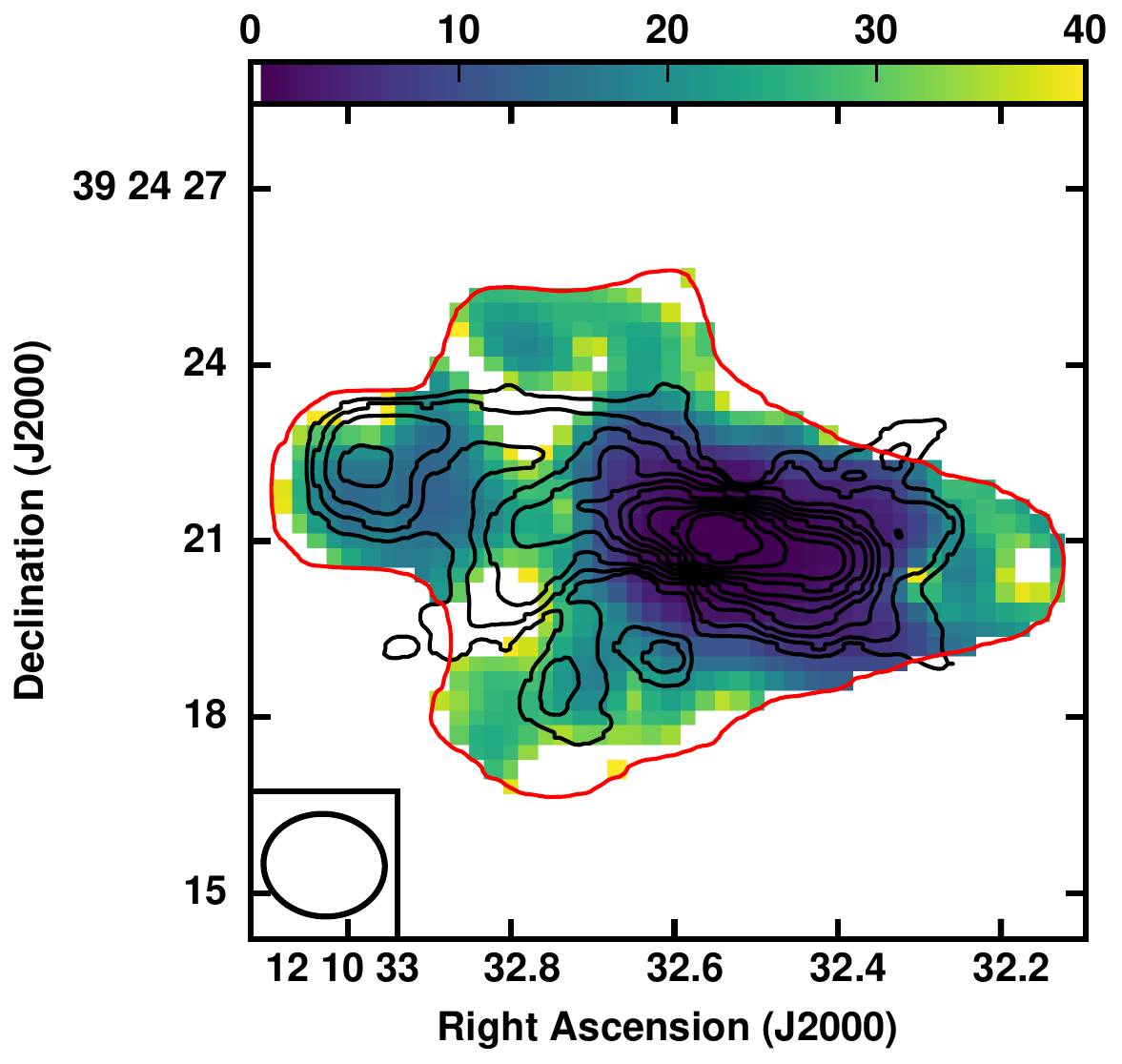}
\includegraphics[width=8cm]{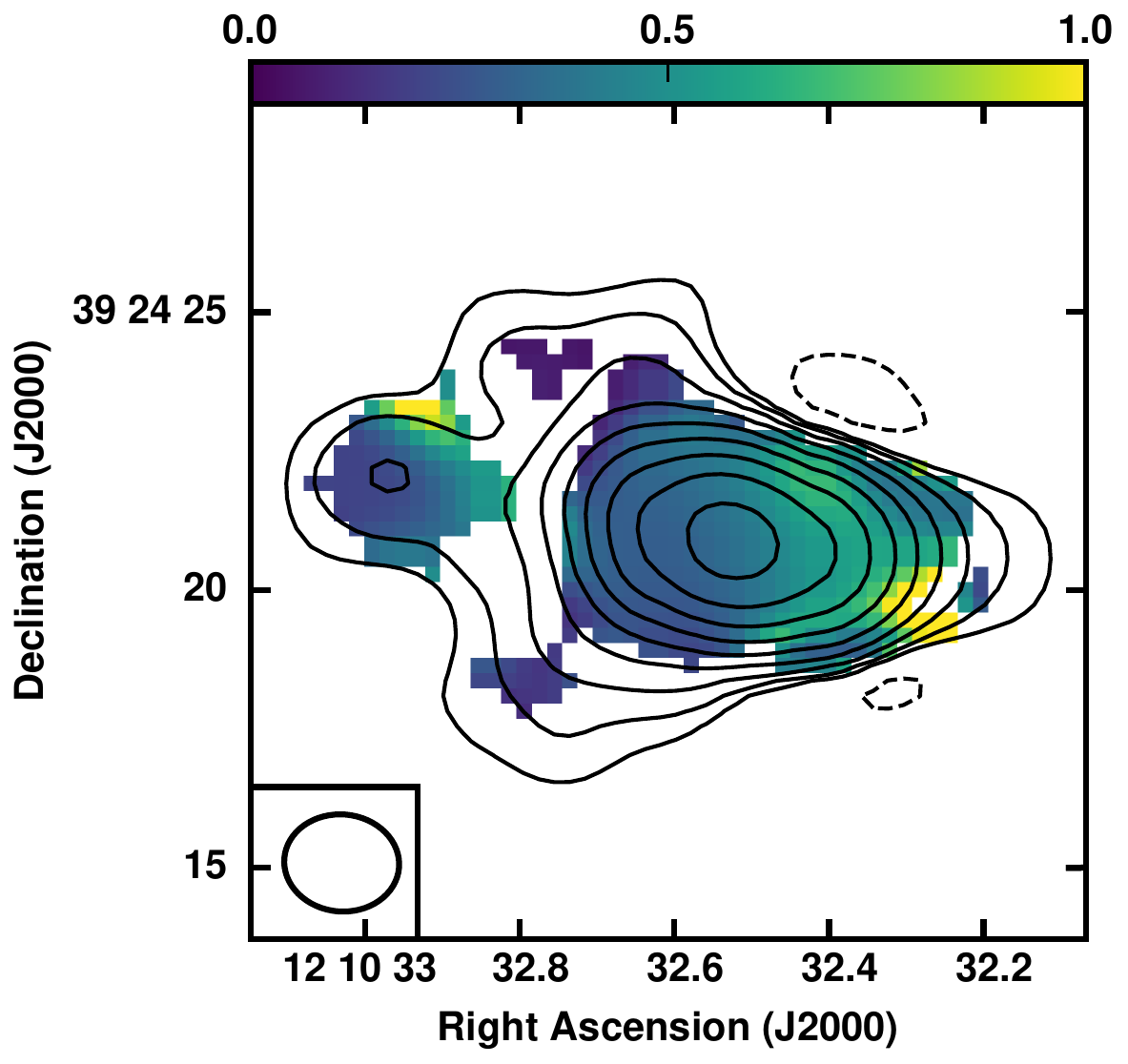}
\includegraphics[width=8cm]{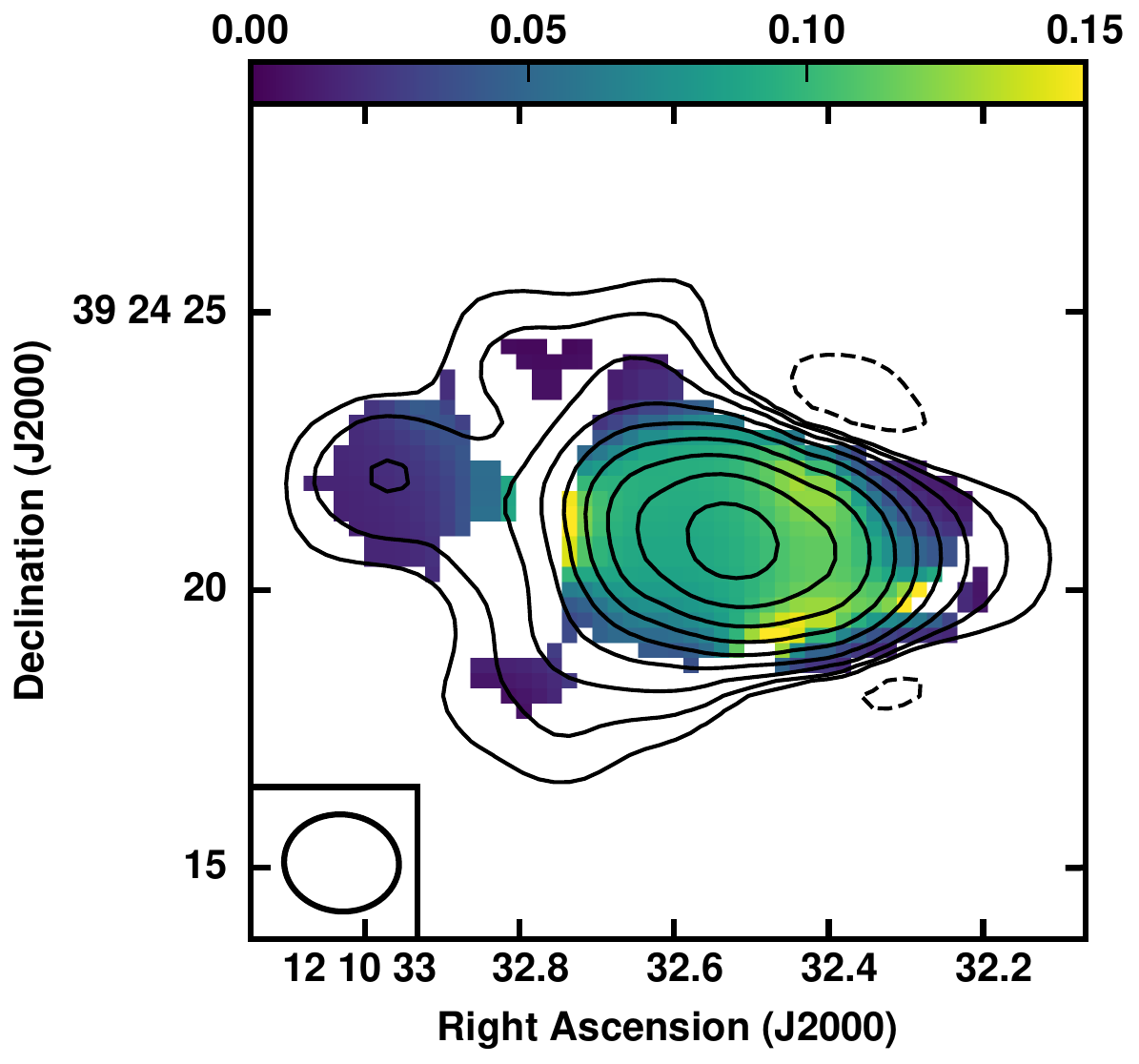}
\caption{\small Top left: RM image of NGC\,4151, using polarization data at 3 GHz, 9 GHz and 11 GHz, shown in color. The contours from 10 GHz data (Figure \ref{fig:3GHzimage}, right) are overplotted in black. The lowest contour from 3 GHz data has been shown in red. The RM values are shown from $-200$ to $+200$ rad m$^{-2}$. {Top right: The RM error image shown in colour with values ranging from 0 to 40 rad m$^{-2}$.} Bottom left: The depolarization image of NGC\,4151 with values ranging from 0 to 1. The contours from 3 GHz data (Figure \ref{fig:3GHzimage}, left) are overplotted in black. {Bottom right: The depolarization error image shown in colour with values ranging from 0 to 0.15.}}
\label{fig:RM}
\end{figure*}

\begin{figure*}
\centering
\includegraphics[width=8cm]{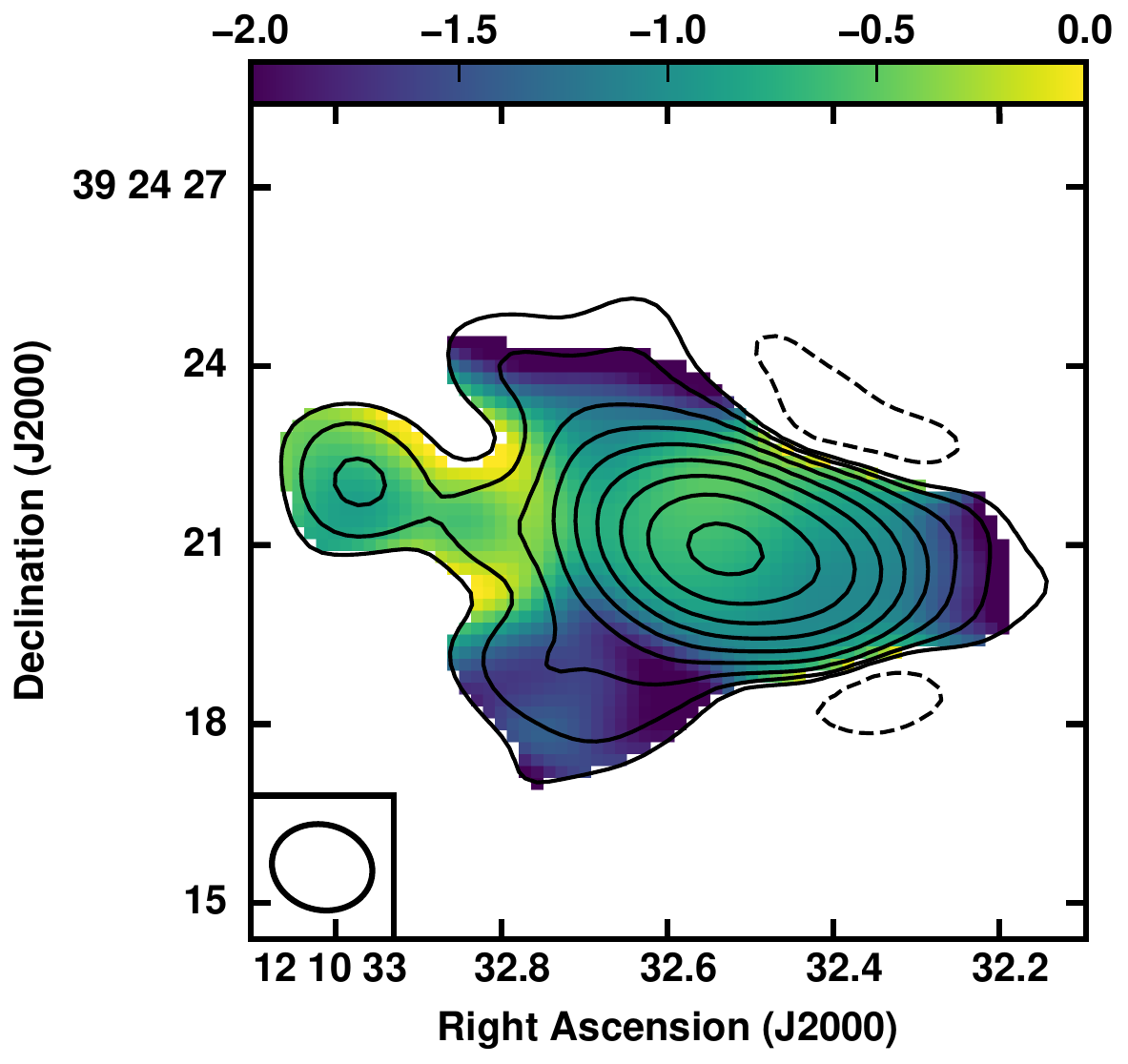}
\includegraphics[width=8cm]{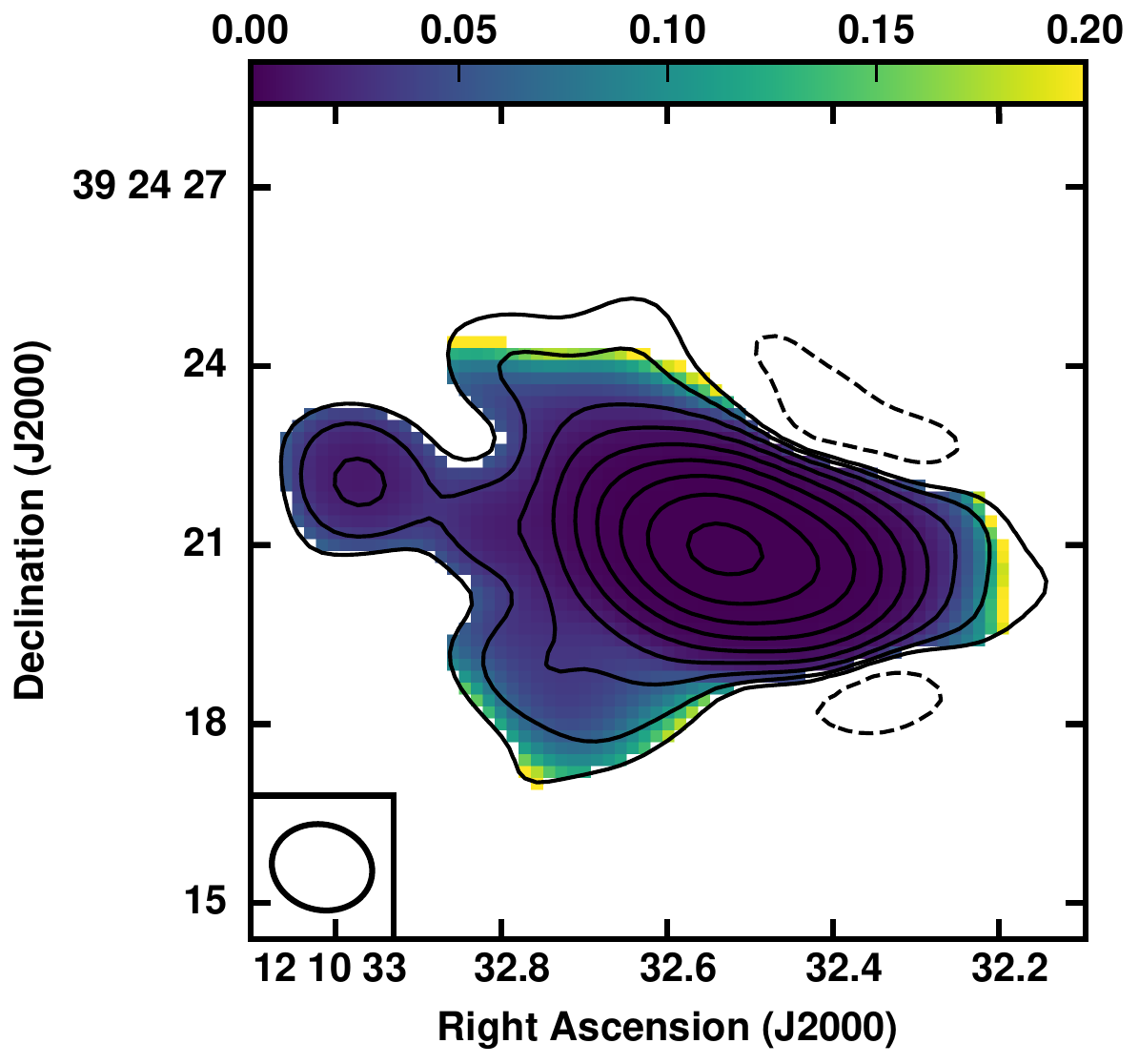}
\caption{\small Left: Spectral index image made from 3 GHz and 10 GHz total intensity images. The spectral index values are shown in colour with values ranging from $-2$ to 0. The uniformly weighted 3 GHz Stokes I image has been overlaid with black contours with levels similar to Figure \ref{fig:3GHzimage}, left panel. The synthesized beam is of size 1.7 arcsec $\times$ 1.4 arcsec at a PA of 75$\degr$. {Right: The spectral index error image in colour with values ranging from $0$ to $0.2$.}}
\label{fig:spix}
\end{figure*}

\section{Observations and Data Analysis}\label{sec:obs}
We observed NGC\,4151 with the VLA at 3 GHz {(with a bandwidth of 2~GHz)} and 10 GHz {(with a bandwidth of 4~GHz)} in B-array configuration on 2 August 2024 and 17 August 2024 {(Project ID: 24A-122; PI: Salmoli Ghosh)}, yielding resolutions of $\sim2\arcsec$ (corresponding to $\sim0.2$ kpc) and $\sim0.5\arcsec$ (corresponding to $\sim$0.4 kpc), respectively. We used the {\tt Common Astronomical Software Application} \citep[CASA, version 6.6;][]{CASA2022} to carry out the basic calibration for radio data, comprising initial data flagging, instrument-specific corrections, and solving for delays, bandpass, and complex gain terms. The solutions were refined with intermediate flagging until convergence was achieved. After applying the basic calibration, we attempted the polarization calibration on the data following the CASA-based VLA radio-polarimetric data reduction script\footnote{\href{https://github.com/astrouser-salm/radio-imaging/blob/main/VLA_polarization_pipeline.py}{https://github.com/astrouser-salm/radio-imaging/blob/main/VLA\_polarization\_pipeline.py}}. The standard calibrator 3C286 with known polarization model was used to determine both leakage and the absolute polarization angle. The leakages were also derived using an unpolarized calibrator OQ208. Although both 3C286 and OQ208 produced similar values for leakages, solutions using OQ208 have been used, as they were better behaved and showed fewer outliers.

After applying the polarization calibration solutions, we imaged NGC\,4151 using the {\tt TCLEAN} task in {\tt CASA} with the {\tt Multi-term Multi-Frequency Synthesis} {\citep[MT-MFS;][]{Bhatnagar2013}} deconvolver, a {\tt Robust} weighting of $+0.5$ and manual masking. We then carried out three rounds of phase-only self-calibration and two to three rounds of combined amplitude and phase self-calibration. The final output includes Stokes I, Q, U, and V images with a synthesized beam size of $2.1\arcsec\times1.7\arcsec$ and $0.7\arcsec\times0.6\arcsec$ and beam position angles of $85\degr$ and $89\degr$ with an r.m.s. noise of 80 $\mu$Jy beam$^{-1}$ and 8 $\mu$Jy beam$^{-1}$ at 3 and 10 GHz, respectively. The polarization quantities are estimated as: linear polarization fraction ($f_p$) = $\sqrt{Q^2+U^2}/I$, and linear polarization angle ($\chi$) = 0.5~tan$^{-1}$(U/Q). The final polarization images have been made by plotting the electric vector position angles (EVPAs) with the vector lengths proportional to $f_p$ (see Figure \ref{fig:3GHzimage}; top panel). Figure~\ref{fig:3GHzimage} bottom panel presents the inferred B-field vectors with lengths proportional to $f_p$, obtained by rotating the EVPAs by 90$\degr$. These images have been corrected for Galactic RM. The Galactic plus source RM corrected images are presented in Figure~\ref{fig:RM-correctedB-vectors} in the Appendix and discussed ahead. We note that these images do not capture some of the diffuse polarized emission with higher RM uncertainties, in contrast to Figure~\ref{fig:3GHzimage}.

{We generated the in-band 3~GHz RM image by splitting the 3~GHz data into three sub-bands and running the task {\tt RMFIT} in {\tt CASA} to fit the resultant QU cubes. The in-band RM image has been blanked by RM error $>20$ rad m$^{-2}$ (see Figure~\ref{fig:3GHzRM}, left panel). The right panel in Figure \ref{fig:3GHzRM} shows the in-band 3~GHz RM values with the corresponding error values taken across the jet.} The (Figure \ref{fig:RM}, top left panel) is generated by splitting the 10~GHz data into two sub-bands of 9 and 11~GHz, and the 3~GHz data. All images were made with the same image-size, cell-size, and beam. {\tt RMFIT} is used to fit the three QU cubes made at the three frequencies for each pixel to obtain an RM image and an RM error image. The final RM image was made at a resolution of $2.1\arcsec\times1.7\arcsec$ and a beam position angle of $89\degr$, and blanked above an error value of $>30$ rad m$^{-2}$. It was also corrected for a Galactic RM value of $0.1\pm1.2$ rad~m$^{-2}$ \citep{Xu2014}. {The RM error image has been shown with error values from 0 to 40 rad m$^{-2}$ in the top right panel of Figure \ref{fig:RM}.} The depolarization image (Figure \ref{fig:RM}, bottom left panel) was obtained by dividing the fractional polarization image at the lower frequency (3 GHz) by that at the higher frequency (10 GHz). The depolarization image shows values extending from 0 to 1, with 1 indicating the lowest and 0 indicating the highest depolarization regions. {The error on the estimated depolarization has been shown in the bottom right panel of Figure~\ref{fig:RM}.}

The RM image was constructed using only pixels with intensities above 3$\sigma$ in each image, so the EVPAs and inferred B‐field vectors in Figure \ref{fig:3GHzimage} are shown without correction for the derived RM because most pixels are blanked when combining the RM, the $f_p$ and the $\chi$ images. The 10 GHz EVPAs (and inferred B-field vectors) remain unchanged within their uncertainties (as expected at shorter wavelengths; see Equation~\ref{eq:1}). At 3~GHz, the EVPA orientations are broadly similar except in the core and hotspot regions, where RM values of $\sim100$ rad m$^{-2}$ cause $\sim50\degr$ rotations, aligning the core vectors to the vectors found in the 10~GHz jet `spine'. 
Source RM-corrected B-field images for both 3 and 10~GHz, as shown in Figure~\ref{fig:RM-correctedB-vectors} in the Appendix, have been created by plotting the EVPA vectors with magnitude $f_p$ at individual frequencies and with angles from the pa0 (polarization angle at 0 wavelength, i.e., source RM-corrected) image produced by the {\tt rmfit} task in {\tt CASA} blanked by its respective error image (with errors $>10\degr$), and finally rotated by 90$\degr$ to obtain the B-vectors. 

The two-frequency spectral index image has been created using the 3 GHz and 10 GHz total intensity images made with the same {\tt uvrange} and {\tt uvtaper} in terms of wavelength, and a robust parameter of $-1.5$ (for 3~GHz) and $+1.5$ (for 10~GHz). Finally convolving them using an identical {restoring beam} in {\tt tclean}, the spectral index image was created at a synthesized beam size of 1.7 arcsec $\times$ 1.4 arcsec, at a PA of 75$\degr$. The spectral index image was then blanked by the spectral index noise image with spectral index errors $>$ 0.3, which has been further blanked by a total intensity value less than $3\sigma$, to obtain the final image (Figure~\ref{fig:spix}, left panel). {The corresponding error image is shown in the right panel of Figure~\ref{fig:spix}.}

\section{Results}\label{sec:results}
A multi-component radio outflow is revealed in our VLA observations. The inferred magnetic field structures imply the presence of three distinct components, which we suggest to be a jet `spine', jet `sheath', and a possible `wind' component. Using `minimum energy' estimates and RM, we derive constraints on the depolarizing medium in NGC\,4151. 

\subsection{Radio Morphology \& B-field Structures} \label{sec:radiomorph}
The lower-resolution VLA image at 3 GHz reveals two primary components: a jet (PA $\approx80\degr$) and a transverse extension on both sides of the jet, which we identify to be the `wind'. The jet shows an inferred B-field that is almost aligned with the outflow direction, with a fractional polarization of $f_p\approx0.9\%$. 
In contrast, the `wind' component (PA $\approx90\degr$) exhibits higher polarization ($f_p\approx3\%$) with EVPAs $\chi=\pm70\degr$, negative on the northern side and positive on the southern side. This implies a B-field orientation of $\approx\pm20\degr$, respectively, which is roughly perpendicular to the outflow. The 10~GHz image further resolves the jet into two components: an inner `spine' with inferred B-fields perpendicular to the outflow and an outer `sheath' layer with inferred B-field aligned with the outflow. We observe a maximum lateral extent of $\sim9\arcsec$ ($\sim0.8$ kpc) on the east side of the core, and $\sim4 \arcsec$ ($\sim$0.3 kpc) on the western side. The eastern outflow extends longer as well compared to the western side by about 2.5 $\arcsec$ ($\sim0.2$ kpc). 

{We note that the lower resolution 3~GHz image shows polarized structures that are consistent with those observed in the higher resolution 10~GHz image. The regions with the highest fractional polarization at 10~GHz dominate the polarized emission in the 3~GHz image, as seen especially clearly in the regions at the jet extremities. In the other regions, a vector averaged B-field structure is observed (over and above the global Faraday rotation). This is consistent with the 10 GHz image re-created using the 3~GHz  synthesized beam. Moreover, the in-band RM image centered around 3~GHz (see Section~\ref{sec:depolRM}) shows the signatures of two regions with different RM values, consistent with the spine-sheath structure observed at 10~GHz, as discussed in greater detail ahead.}

The jet `spine' on the western side is brighter than the eastern one, suggesting that the western jet is the approaching one and is {mildly Doppler-boosted as suggested by the VLBI observations of \citet{Williams2017}}. We have estimated the jet-to-counterjet brightness ratio ($R_J$) by comparing the flux densities of the western versus eastern jet-spine. Considering an inclination angle ($\theta$) of 40$\degr$ \citep{Ruiz2003}, the velocity ($\beta$c) obtained using $\beta = \frac{{R_J}^{1/(2-\alpha)} - 1}{\cos{\theta}({R_J}^{1/(2-\alpha)} + 1)}$ \citep{UrryPadovani1995} is $\sim$ 0.16c at 0.1 kpc from the nucleus. The `wind' component can only be observed on the receding side of the jet. Similar structures have been observed in other astrophysical sources as well; for instance, in the protostar Hops 315 \citep{Vleugels2025}. \citet{Laing2006} note that if the jets are relativistic and faster on-axis rather than at their edges (i.e. a velocity stratification), then the approaching jet appears more centre-brightened than the receding one. 

The observed two-frequency spectral index image reveals a core spectral index of $\alpha\approx-0.66$, and a jet spectral index of $\alpha\approx-0.8$, consistent with synchrotron emission in a collimated outflow. The shocked region on the eastern side appears to have flatter $\alpha$ of $\approx-0.64$, indicating a re-acceleration of charged particles in that region. By contrast, the southeastern wind-feature is markedly steeper with $\alpha\approx-1.55$, suggesting a rapidly ageing electron population undergoing significant radiative, inverse Compton, and adiabatic losses as it expands. The jet kinetic power, estimated from the core flux density at 5 GHz (extrapolated from 45.5 mJy at 10 GHz assuming $\alpha = -0.66$ at the core) using the empirical relation of \citet{Merloni2007}, 
is found to be $\sim10^{43}$ erg s$^{-1}$. This places NGC 4151 toward the upper end of kinetic powers observed in typical radio-quiet AGN \citep[e.g.,][]{Kharb2021}. The depolarization image {(Figure~\ref{fig:RM}, right panel)} indicates that the winds are predominantly depolarized. The receding jet is more depolarized compared to the approaching one, likely due to the Laing-Garrington effect \citep{Laing1988, Garrington1988}; that is, due to the presence of excess depolarizing material \citep[see][which describes the presence of an HI disk]{Mundell1995} towards the receding jet.

\begin{table*}
{\tiny
\caption{Observed radio properties of NGC\,4151} \label{tab:radioprop}
\begin{flushleft}
\begin{tabular}
{cccccc|ccccc} \hline 
\multicolumn{6}{c|}{3 GHz}&\multicolumn{5}{c}{10 GHz}\\
\hline
Region&$S_{tot}$&$f_p$&$\chi$&$\alpha$&RM&$S_{tot}$&$f_p$&$\chi$&$\alpha$&RM\\
 &(mJy)&(\%)&($\degr$)& &(rad m$^{-2}$)&(mJy)&(\%)&($\degr$)& &(rad m$^{-2}$)\\
\hline
Total&$221\pm11$&$3.25\pm0.6$&\nodata&\nodata&\nodata&$88\pm4$&$12\pm3$&\nodata&\nodata&\nodata\\
Core&$148\pm25$&$0.87\pm0.02$&$16\pm5$&$-0.66\pm0.03$&$-104\pm2$&$45.5\pm1$&$2.2\pm0.03$&$73\pm1$&$-0.63\pm0.03$&$-104\pm2$\\

Eastern&\multirow{2}{*}{$1.7\pm0.3$}&\multirow{2}{*}{$2.8\pm0.6$}&\multirow{2}{*}{$10\pm6$}&\multirow{2}{*}{$-0.64\pm0.02$}&\multirow{2}{*}{$70\pm20$}&\multirow{2}{*}{$0.18\pm0.01$}&\multirow{2}{*}{$12.5\pm2$}&\multirow{2}{*}{$50\pm2$}&\multirow{2}{*}{$-0.62\pm0.04$}&\multirow{2}{*}{$50\pm30$}\\
Hotspot& & & & & & & & & &\\
Western&\multirow{2}{*}{$1.0\pm0.2$}&\multirow{2}{*}{$6\pm1$}&\multirow{2}{*}{$31\pm6$}&\multirow{2}{*}{$-1.9\pm0.1$}&\multirow{2}{*}{\nodata}&\multirow{2}{*}{$0.10\pm\ 0.01$}&\multirow{2}{*}{$28\pm8$}&\multirow{2}{*}{$20\pm4$}&\multirow{2}{*}{$-1.1\pm0.1$}&\multirow{2}{*}{$-24\pm10$}\\
Jet-edge& & & & & & & & & &\\
Western&\multirow{2}{*}{$25\pm8$}&\multirow{2}{*}{$1.2\pm0.1$}&\multirow{2}{*}{$14\pm1$}&\multirow{2}{*}{$-0.9\pm0.1$}&\multirow{2}{*}{$-40\pm4$}&\multirow{2}{*}{$4\pm1$}&\multirow{2}{*}{$1.8\pm0.4$}&\multirow{2}{*}{$60\pm2$}&\multirow{2}{*}{$-0.95\pm0.03$}&\multirow{2}{*}{$70\pm10$}\\
Jet& & & & & & & & & &\\
Eastern&\multirow{2}{*}{$25\pm2$}&\multirow{2}{*}{$0.7\pm0.2$}&\multirow{2}{*}{$17\pm2$}&\multirow{2}{*}{$-0.70\pm0.02$}&\multirow{2}{*}{$-60\pm25$}&\multirow{2}{*}{$3\pm1$}&\multirow{2}{*}{$2.9\pm0.5$}&\multirow{2}{*}{$66\pm3$}&\multirow{2}{*}{{$-0.66\pm 0.04$}}&\multirow{2}{*}{$-131\pm5$}\\
Jet& & & & & & & & & &\\
Southern&\multirow{2}{*}{$0.9\pm0.2$}&\multirow{2}{*}{$2.4\pm0.6$}&\multirow{2}{*}{$63\pm8$}&\multirow{2}{*}{${-}1.55\pm0.05$}&\multirow{2}{*}{$-28\pm8$}&\multirow{2}{*}{$0.12\pm0.01$}&\multirow{2}{*}{\nodata}&\multirow{2}{*}{\nodata}&\multirow{2}{*}{$-1.5\pm0.4$}&\multirow{2}{*}{$-29\pm3$}\\
Wind& & & & & & & & & &\\
Northern&\multirow{2}{*}{$0.6\pm0.2$}&\multirow{2}{*}{$3.3\pm0.3$}&\multirow{2}{*}{$-70\pm8$}&\multirow{2}{*}{$-2.2\pm0.2$}&\multirow{2}{*}{$28\pm7$}&\multirow{2}{*}{$0.18\pm0.05$}&\multirow{2}{*}{$33\pm1$}&\multirow{2}{*}{$-68\pm4$}&\multirow{2}{*}{$-0.8\pm0.2$}&\nodata\\
Wind& & & & & & & & & & \\
\hline
\end{tabular}
\end{flushleft}}
{\small Note: Column 1: Region of interest. {The regions are chosen as square boxes of width $\sim$1.5 $\arcsec$ and $\sim$1 $\arcsec$ for 3 and 10~GHz images, respectively} and are defined as follows. `Core' is 
unresolved base of the jet in Figure \ref{fig:3GHzimage}. `Eastern Hotspot' is the region of shocked emission $\sim$ 6.3 arcsec away from the core. `Western Jet-edge' is the region of high polarization fraction (the jet 'sheath' observed at 10~GHz) $\sim$ 4.6 arcsec to the west of the core at 3 GHz ($\sim$3.5 arcsec west of the core at 10 GHz). `Western Jet' is the western part of the jet as shown in Figure \ref{fig:3GHzimage}, left panel, $\sim$ 2.2 arcsec from the core towards the west. The `Eastern Jet' denotes the region $\sim$ 2 arcsec to the east of the core. The `Western Jet' and `Eastern Jet' at 10 GHz are the regions along the jet `spine' as shown in Figure \ref{fig:3GHzimage}, right panel, at $\sim$ 1.7 and $\sim$ 1.3 arcsec from the core in the 10~GHz image. The `Southern Wind' and the `Northern Wind' at 3 GHz include the `wind' regions shown in Figure \ref{fig:3GHzimage} $\sim$ 3 and $\sim$ 4 arcsec to the south and the north of the core, respectively. The `Northern Wind' and `Southern Wind' at 10 GHz denote regions about $\sim$ 2 arcsec north, $\sim$ 2.5 arcsec east and $\sim$ 2.6 arcsec south, $\sim$ 2.7 arcsec east of the core, respectively. Column 2: Total flux density {of the chosen region} in mJy. For compact features such as the `core' and the `Eastern Hotspot', the peak intensity has been noted. Column 3: {Mean} fractional polarization in the region of interest. Column 4: {Mean} polarization angle in the considered region, only if the vectors are aligned in a similar direction. No data implies that vectors are oriented in multiple directions within that region, or polarization is not detected. Column 5: The {mean} 3~GHz $-$ 10~GHz spectral index in the region. Column 6: The mean rotation measure estimate in the region of choice. No data indicates that the RM has a rapid change in values and a high error ($>$60\% of the absolute value) due to the low SNR of polarized emission in one of the three frequencies, or is undetected. Column $7-11$: Same as in Column $1-6$, estimated from the 10~GHz data. The $\alpha$ values here are the {mean} 3~GHz $-$ 10~GHz spectral index values for the regions of choice.}
\end{table*}

\subsection{Deriving Constraints on the Depolarizing Media from RM}\label{sec:depolRM}
When polarized light traverses a magneto-ionized medium, it undergoes Faraday rotation {\citep[e.g.,][]{Beck2013}}. The difference between the observed angle of polarization ($\chi_{obs}$) and the intrinsic angle of polarization ($\chi_{int}$) is given by 
\begin{equation}
    \chi_{obs}-\chi_{int}=\mathrm{RM}\times \lambda^2
    \label{eq:1}
\end{equation}
 where $\lambda$ is the wavelength. RM is dependent on the electron density ($n_e$), line of sight B-field ($B_\parallel$) and the length of the Faraday rotating medium (L) as 
\begin{equation}
\mathrm{RM}~(\mathrm{rad\, m^{-2}})=812\,n_e (\mathrm{cm^{-3}})\,B_\parallel\ (\mathrm{\mu G}) L\,(\mathrm{kpc})
\end{equation}

We created an in-band RM image using the 3~GHz polarization data as well as a broadband RM image using both the 3~GHz and 10~GHz polarization data. The in-band RM image is affected by differential Faraday rotation within the 3~GHz emitting regions (see Figure \ref{fig:3GHzRM}), even though these `layers' are not distinctly resolved in the 3~GHz polarization image (Figure \ref{fig:3GHzimage}, left panel). 
{We find that RM values are higher in the inner jet region and show a decrease towards the edges. We note that the RM uncertainty as reported by the CASA task {\tt rmfit}, and plotted in Figure~\ref{fig:3GHzRM}, corresponds to the 1$\sigma$ error from a weighted linear fit of polarization angle versus wavelength squared, with weights derived from the Q and U noise. While the wider transverse slice (Slice 1 in Figure~\ref{fig:3GHzRM}) shows an RM difference in the spine–sheath region at or greater than the $\sim2\sigma$ level, the RM values become similar in the spine-sheath region in the narrower slice (Slice 2 in Figure~\ref{fig:3GHzRM}) at the $\sim2\sigma$ level. While marginal, the in-band RM structure supports the picture of jet stratification.}
The lower RM values in the `sheath' may indicate a reduction in either the magnetic field strength, the electron density, or the effective path length of the Faraday-rotating medium. The enhanced depolarization observed in the `sheath' compared to the `spine' (see Figure \ref{fig:RM}, bottom left panel) may arise from gas entrainment and `mass loading', leading to enhanced electron density. Assuming comparable path lengths for the Faraday-rotating medium in both the jet `spine' and `sheath', the observed decrease in RM going from the `spine' to the `sheath' could also point to greater magnetic field strength along the jet axis and lesser towards the edges \citep[e.g.,][]{Laing1980, Laing2002}. A combination of `magnetic flux' and `mass loading' have indeed been suggested to influence the creation of collimated outflows in protostellar as well as AGN jets in simulations \citep[e.g.,][]{Fendt2006,Pudritz2006}.

The RM in NGC\,4151 is found to change its value and sign {rapidly} in the kpc-scale region, with values ranging from $-230$ to 250~rad~m$^{-2}$. {A transverse RM gradient is observed across the `wind' components (see Section \ref{sec:wind}). Since each `wind' region is comparable to the size of the synthesized beam, the RM is likely to be influenced by systematic uncertainties. However, the northern and southern `wind' regions display RM of comparable magnitude but opposite signs. Taken together, their combined RM structure would be consistent with a transverse gradient assuming that these regions belonged to a single physical component (e.g., a wide-angled bi-conical wind).} RM gradients transverse to parsec-scale jets have been observed in several RL AGN \citep{Asada2002,Hovatta2012}, on the $\sim100$ parsec scale jet in M87 \citep{Pasetto2021}, as well as some RQ AGN \citep{Ghosh2025b}. These gradients have been suggested to arise due to the presence of helical B-fields \citep{O'Sullivan2009,Gabuzda2004,Kharb2009}. The mean RM observed at the 0.5 $\arcsec$ and 2 $\arcsec$ cores are $-109.5\pm0.5$ and $-103.1\pm0.7$~rad~m$^{-2}$.

We have estimated the magnetic field strength ($B_\mathrm{min}$), total energy ($E_\mathrm{total}$) and the electron lifetimes ($\tau$) in the core of NGC~4151, at two different resolutions, assuming the `minimum energy' condition and using the relations from \citet{vanderlaan1969, Pacholczyk1976,OdeaOwen1987}. At the core, the volume-filling factor ($\phi$) of the radio plasma is assumed to be 1. The proton-to-electron energy ratio (k) is varied from 1 to 100 \citep[e.g.,][]{Rosa2006, Williams2017, Ghosh2025b}. The core peak flux density was estimated by fitting a Gaussian component to the central region of width equal to the FWHM of the synthesized beam. The $\alpha$ values are obtained as the average spectral index in the core region. The `minimum energy' B-field strength for the $0.5\arcsec$ and $2\arcsec$ cores at 10 and 3 GHz was estimated as $100-300\, \mu$G and $50-150\, \mu$G for k = $1-100$, respectively. Similarly, the total energies are found to be $(3-30)~\times10^{52}$ erg and $(20-200)~\times10^{52}$ erg for k = $1-100$, in the cases of $0.5\arcsec$ and $2\arcsec$ cores, respectively. The electron lifetimes for synchrotron and CMB losses are obtained as $(2.5-0.5)~\times10^5$ yrs and $(1.5-0.3)~\times10^6$ yrs, respectively.

Assuming $B_{min}\equiv B_\parallel$ \citep[see][]{Feain2009, Kharb2009}, and L to be the extent of the core, we estimate the electron densities to be in the range of $0.006-0.02$ cm$^{-3}$ and $0.004-0.01$ cm$^{-3}$, respectively. The inferred electron density of $10^{-2}-10^{-3}$ cm$^{-3}$ for a Faraday rotating medium $0.06-0.2$ kpc in extent, with B-fields of strength $10^2-10^3\,\mu$G, is low compared to typical $n_e$ values observed in the warm ionized medium \citep[$n_e\sim$0.1 cm$^{-3}$;][]{Cordes2002, Gaensler2008}. Such low electron densities suggest that the Faraday screen may consist of a tenuous, magnetized sheath or cocoon around the jet \citep[e.g.,][]{Laing2008, Lisakov2021}. 

\subsubsection{Media Characteristics for External and Internal Depolarization}\label{sec:extintdepol}
The polarized emission from the sources is modified by the intervening medium through Faraday rotation, which alters the observed polarization angle and can also reduce the degree of polarization. When the Faraday-rotating medium lies between the source and the observer, the effect is referred to as external depolarization. Conversely, when the rotating medium is mixed with the source plasma itself, the effect is referred to as internal depolarization. 

In the case of external depolarization, the fractional polarization ($f_p$) at a particular wavelength ($\lambda$) can be expressed following \citet{Burn1966,vanBreugel1984} as
\begin{equation}
f_p(\lambda^2)=f_{pi}\,\exp{-2\,(812\,n_eB_\parallel)^2\,d\,R\,\lambda^4}
\label{eq:externaldepolarization}
\end{equation}
For the 2 $\arcsec$ core, we considered an $n_e = 0.5 \times 10^{-2}$ cm$^{-3}$ (see Section \ref{sec:depolRM}), and for a length ($R$) of 0.2 kpc (length of the core), and the line of sight magnetic field ($B_\parallel$) of $100-300\,\mu$G, and estimated a polarization fraction of $0.87\%\pm0.02\%$ and $2.2\%\pm0.03\%$ at 0.1 m and 0.03 m of wavelength ($\lambda$), respectively. This yields an intrinsic polarization fraction ($f_{pi}$) 
of 2.22\% and a fluctuation scale ($d$) of $0.01-0.1$ kpc. 

In the case of internal depolarization, the mixing of the plasma introduces a randomized (but isotropic) component ($B_{\rm{rand}}$) of the B-field apart from the uniform component ($B_{\rm{uni}}$). The total B-field can be represented as $B_{\rm{tot}}^2=B_{\rm{uni}}^2+B_{\rm{rand}}^2$. The complex fractional polarization at a particular wavelength can now be given as \citep[see, ][]{Burn1966, Pacholczyk1970, Sullivan2013a}
\begin{equation}
P(\lambda^2)=f_{pi}\frac{1-e^{-S}}{S}
\label{eqn3}
\end{equation}
where 
\begin{equation}
S=2(812\,n_e\,B_{\rm{rand}})^2\,d\,L\,\lambda^4-1624i\,n_e\,B_{\rm{uni}}\,L\,\lambda^2
\label{eqn4}
\end{equation}
and
\begin{equation}
f_{pi}=f_{pi}^\prime(B_{\rm{uni}}^2/B_{\rm{tot}}^2).
\end{equation}
The theoretical polarization fraction for $B_{\rm{rand}}=0$ for synchrotron radiation in optically thin regions ($f_{pi}^\prime$) is given as $\frac{3-3\alpha}{5-3\alpha}$. For an $\alpha$ of $\approx-0.66$ for the core, we obtained $f_{pi}^\prime=70\%\pm1\%$ and $B_{\rm{rand}}/B_{\rm{uni}} = 5.55\pm0.04$. Assuming, $B_{\rm{uni}} = \sqrt{3}B_\parallel$ \citep[see][]{Burn1966,Sullivan2013a}, and $B_\parallel$ to be $100-300$ $\mu$G, we obtain $B_{\rm{uni}}$ to be $170-520$ $\mu$G and $B_{\rm{rand}}$ to be $0.96 - 2.88$~mG and therefore $B_{\rm{tot}}$ to be $0.97 - 2.93$~mG. We use Equation~\ref{eqn4} to find the electron density, $n_e \approx 6\times 10^{-3}\,\mathrm{cm}^{-3}$. If thermal gas is mixed with the synchrotron plasma and causes depolarization, the total mass of it in the `core' region can be estimated as $M_{tg} \sim n_e\phi m_H V$, where the volume filling factor $\phi$ is assumed to be unity, total volume of the lobes $V\approx 4.4\times10^{62}$ cm$^3$ (assuming a cylindrical volume), and mass of ionised hydrogen is $m_H = 1.67\times10^{-24}$ g. We find that a thermal gas mass of $M_{tg} \approx (2.21\pm0.18)\times 10^3$ M$_\sun$ could be mixed in the `core' region causing `core' depolarization.

\subsection{`Wind' Kinematics in NGC\,4151}\label{sec:wind}
We now derive the kinetic power and mass outflow rate of the `wind' in NGC\,4151, assuming it to be like a hollow cone around the jet. When the jets interact with the surrounding media, they inflate overpressured cocoons of relativistic plasma {as proposed for Cygnus\,A by \citet{Begelman1989} and the hydrodynamical simulations of \citet{Cioffi1992}.} The cocoon then drives shocks into the ambient gas, accelerating the neutral and ionized gas into high-speed outflows (up to thousands of km~s$^{-1}$). {Alternatively, cocoons have been suggested to originate from precessing jets, as observed in the Galactic microquasar, SS\,433 \citep{Kochanek1990, Blundell2005}. Interestingly, jet precession has been invoked in NGC\,4151 by \citet{Ulvestad1998}. However, using the formalism of \citet{Hjellming1981}, we find that the entire radio structure cannot be fitted by a single continuously precessing jet. Rather, at least two separate episodes of jet precession are required to fully model the inner jet and the outer S-shaped lobes. We therefore suggest that the wide-angled `wind' component is anchored to the accretion disk, similar to the radio jet. A conical morphology can naturally arise from initially weakly collimated outflows, shaped by the outward magnetic pressure of the accelerating plasma and/or by radial gradients in the magnetic field strength anchored in the accretion disk. Such conditions can give rise to hollow, conical winds with half-opening angles of $\sim 30$–$40^\circ$ \citep[e.g.,][]{Romanova2009}.}

The kinetic energy, momentum and the mass outflow rate are given by
\begin{equation}
    E_{\mathrm{kin}} = \frac{1}{2} M_{\mathrm{out}} v_{\mathrm{out}}^2
\end{equation}    
\begin{equation}
    p_{\mathrm{out}} = M_{\mathrm{out}} \cdot v_{\mathrm{out}}
\end{equation}
\begin{equation}
    \dot{M}_{\mathrm{out}} = \frac{M_{\mathrm{out}}\, v_{\mathrm{out}}}{R_{\mathrm{out}}}
\end{equation}
Some of the gas may be entrained within the turbulent boundary layer of the jet, accelerating to significant speeds without being fully ionized. The thermal pressure inside the cocoon can also push it outward to form the outflow, which is effective within radii of 1 to 10 kpc, where cooling and mixing can occur. Comparing the flux densities of the inferred jet (`spine+sheath'; $\approx$ 90 mJy) and the `wind' component ($\approx$ 9.9 mJy), the `wind' power is found to be $10-12$\% of that of the jet. In the `wind' region $\sim$ 400 pc from the centre, the absolute value of $|\mathrm{RM}|$ is found to be around 28~rad~m$^{-2}$. Assuming the `minimum energy' condition and k = $1-100$, $\phi=0.5$ \citep{Blustin2009} for a clumpy wind medium, and a hollow cone of outer, inner radius and a height of $\sim$ 400 pc, $\sim$ 130 pc, and $\sim$ 280 pc, respectively, the B-field in the wind region is found to be $27-82\,\mu$G with $n_e\sim0.002-0.006$ cm$^{-3}$. From \citet{Blustin2009,Faucher2012}, $M_{wind} \approx n_p f m_p V$, where $f$ is the filling factor. We assume $f$ to be 0.5 for homogeneous emission and $n_p\approx n_e$. Therefore, the mass of the `wind' component is found to be $1050-3200~M_\sun$. Assuming the velocity of the `wind' to be 10\% of that of the jet, i.e., 0.008c \citep[also see][]{Fiore2017, Perucho2017}, we estimate the wind mass outflow rate to be $0.01-0.03~M_\sun$ yr$^{-1}$, and the wind kinetic energy to be $(12-36) \times 10^{52}$ erg. {The spread in the derived estimates emerges from the range of k values adopted in the calculations.} 

A jet-driven secondary shock outflow, as may be present in NGC\,4151, has also been suggested for other Seyfert galaxies by \citet{May2017, Hopkins2010}. The typical starburst-driven galactic wind outflows are found to be much more powerful, with mass outflow rates up to 10 $M_\odot$ yr$^{-1}$ and velocities of $100-500$~km~s$^{-1}$ \citep{Strickland2000,Veilleux2005}. 
Seyfert galaxy winds are often highly ionized, with lower mass outflow rates of $\sim0.1-1~M_\sun$ yr$^{-1}$ but higher velocities (up to thousands of km~s$^{-1}$), and greater kinetic energy per unit mass \citep{Veilleux2005}. 
Therefore, for NGC\,4151, the observed wind appears to be weak. The kinetic power associated with the outflow is given by $\dot{E}_{\mathrm{kin}} = \frac{1}{2} \dot{M}_{\mathrm{out}} v_{\mathrm{out}}^2$, is $(3-10)\times10^{40}$ erg s$^{-1}$. 

{Following \citet{Zakamska2014}, we independently estimated the kinetic power of an AGN-driven wind from the radio emission, adopting a conversion efficiency of $3.6\times10^{-5}$. We computed the radio luminosity of the wind region at 1.4 GHz as $\nu L_\nu$, where $L_\nu=4\pi D_L^2 S\nu (1+z)^{-(\alpha+1)}$ with {$D_L$, $S_\nu$, $z$, and $\alpha$ being the luminosity distance, flux density at frequency $\nu$, redshift, and spectral index, respectively}. We adopted an $\alpha$ of $-1.8$, corresponding to the mean value measured for the northern and southern wind components (Table \ref{tab:radioprop}), and extrapolated the 1.4~GHz flux density from the 3~GHz data by summing the flux densities of the two (north and south) wind regions {(the wind flux density estimated at 3 GHz is $1.5\pm0.4$ mJy)}. Using this approach, we obtained a monochromatic radio luminosity of $L_{\rm{radio}}$ of {$\sim 2.2 \times 10^{36}$ erg s$^{-1}$}, and a wind kinetic power ($\dot{E}_{\mathrm{kin}}$) of {$\sim 5.9\pm1.6 \times 10^{40}$ erg s$^{-1}$. The uncertainty here is propagated from the uncertainty in the flux density}. This value is consistent with our independent estimate of the wind kinetic power derived above from the outflow mass and velocity {within the derived uncertainties for both the estimates}.} 

Hence, the coupling efficiency $\epsilon_f=\dot{E}_{\mathrm{kin}}/L_{bol}$, using a bolometric luminosity of 10$^{44}$ erg s$^{-1}$ \citep{Kraemer2021}, is found to be {$10^{-4}-0.005$}. That is, the wind kinetic power is only {$0.01-0.5\%$} of $L_{bol}$. {This coupling efficiency lies at the lower end of the 0.1–10\% range derived for radiatively-driven AGN winds \citep{Fiore2017, Zakamska2014}, supporting the possibility of ionised winds creating the radio-emitting cocoon.} However, this is too small for galaxy-scale AGN feedback, which has typically been observed in systems with kinetic power that is at least 0.5\%~$L_{bol}$ \citep[e.g.,][]{Merloni2007}. Rather, it is consistent with a modest AGN-driven magnetohydrodynamic (MHD) wind, which may influence the immediate ISM but not provide galaxy-wide feedback. We note that the properties of the radio `wind' are consistent with the warm absorbers detected in X-rays as described in Section~\ref{sec:warmabsorbers}.

\section{Discussion}\label{sec:discussion}
\subsection{A stratification in the B-field structure} The inferred B-field vectors show sharp changes in orientation with respect to the jet direction. We identify three layers as a jet `spine', a jet `sheath', and a `wind' component. The inferred B-field is perpendicular in the jet spine. This is consistent with a jet dominated by shocks, as has been inferred in other studies of NGC~4151 \citep[e.g.,][]{Williams2020}. The B-field is aligned with the jet direction in the `sheath', consistent with shear due to jet-medium interaction. If true, the magnetic field stratification could be consistent with a velocity stratification in the radio outflow of NGC~4151, similar to what is observed in FRI radio galaxies \citep[e.g., 3C~31;][]{Laing2002}. We get to observe a combined effect of jet `spine $+$ sheath' in the lower resolution 3~GHz image of NGC~4151. Along the eastern receding outflow, a transverse component with B-fields perpendicular to the jet is observed. We identify this to be a `wind' component, threaded by toroidal or helical B-fields similar to that suggested in \citet{Mehdipour2019}. The suggestion of a helical field is consistent with the RM gradient that is observed in this feature.

The `jet spine' can be traced up to $\sim$ 3~$\arcsec$ ($\sim$ 0.03 kpc), while the sheath extends further, up to $\sim$ 4~$\arcsec$ ($\sim$0.04 kpc) from the centre. The persistence of the sheath to larger distances likely arises from jet–medium interaction at the boundaries, which re-energizes particles and produces synchrotron emission. There is a lack of adequate resolution in the spectral index and polarization images from the current data to see these boundary layers and test this hypothesis. High-resolution polarization observations with the e-MERLIN telescope are being planned for this purpose. The absence of a clear `wind' component on the approaching jet side may be attributed to either stronger interactions in inhomogeneous media or the prominence of the jet, which outshines the relatively faint `wind' emission. 

\subsection{A Magnetically driven outflow in NGC~4151?}
We now check whether the outflow in NGC~4151 is magnetically-driven based on the radio and X-ray data. Winds in AGN can be powered by radiation pressure, magnetic pressure, or thermal pressure, as denoted by the equation of motion, $\rho \frac{D v}{D t} + \rho \nabla \Phi = -\nabla P + \frac{1}{4\pi} (\nabla \times B) \times B + \rho F^{\text{rad}}$. Terms on the left-hand side represent the force per unit volume to accelerate the fluid, and the gravitational force per unit volume restricting the fluid from flowing out. The terms on the right side represent the force due to gas pressure, the Lorentz force, and the radiation pressure force, respectively \citep{Crenshaw2003, Proga2007}. For an outflow to originate, any of the forces on the right-hand side must exceed the gravitational force. Radiative winds are typically observed in massive AGN with high optical/UV luminosity, and are less important in the case of low-luminosity AGN {\citep{Veilleux2005,Proga2007,Farcy2025} like NGC~4151 \citep{Williams2020}}. To understand whether these winds are magnetically or thermally powered, it is necessary to estimate both the magnetic and thermal gas pressure. The magnetic pressure is defined as $B^2/8\pi$ \citep[see][]{Longair2011}. From the minimum energy condition, the B-field in the `wind' region is found to be $28-86\,\mu$G, and at the `core' it is $100-300\,\mu$G. Since this provides the B-field strength in one  direction, we should multiply this by $\sqrt3$ to obtain the total B-field strength, using the condition of isotropy. Hence, the magnetic field pressure in the `wind' region is $(1-9)\times10^{-10}$ dyne cm$^{-2}$, and in the `core' is $(12-107)\times10^{-10}$ dyne cm$^{-2}$. 

The `wind' component observed in NGC\,4151 extends out to $\sim400$~parsec. This structure exhibits a toroidal B-field with a relatively high degree of polarization (fractional polarization in the `wind' region is found to be $\sim33\%$ at 10 GHz). This supports the suggestion that the wind could be an AGN accretion disk wind rather than a galactic wind from starburst activity \citep{Colbert1996a, Beck1996, Veilleux2005}, where the fractional polarization is typically $\leq$5\%, both at lower \citep[e.g., at $\sim$1.5 GHz in NGC~1134, NGC~253, UGC~903;][]{Sebastian2020} and at higher frequencies \citep[e.g., at $\sim$8 GHz in M~82;][]{Reuter1994}, due to the lack of large-scale organised B-fields in the outflowing gas. Moreover, we find that the RM values are positive in the northern `wind' component but negative in the southern `wind' component (see Figure~\ref{fig:helicalfield}). {Closer to the nucleus, the RM values on the northern and southern sides become comparable, before exhibiting a reversal in sign.} This strongly supports the picture of a helical B-field threading a wide-angle bi-conical `wind' component \citep[also see Figure~1 in][]{Ghosh2025IAU}.

Diffuse soft X-ray gas has been observed in NGC\,4151 extending up to $\sim2$~kpc, suggested as a thermal plasma with a temperature of about $0.25-0.58$~keV. The hot gas pressure is estimated as $\sim 2 n_e kT$ and is found as $6.8\times10^{-10}$~dyne~cm$^{-2}$ from close to the NLR. It can be as low as $5\times10^{-12}$ dyne cm$^{-2}$ when reaching equilibrium with the neutral hydrogen gas in the galaxy \citep{Wang2010, Wang2011a}. Hence, the magnetic pressure exceeds the gas pressure in NGC\,4151. The outflow, therefore, appears to be magnetically driven \citep[also see][]{Kraemer2020}. Magnetocentrifugal winds from the accretion disk usually corotate unless the poloidal magnetic field is extremely powerful. As the rotation in the disk happens differentially, the toroidal component of the magnetic field is easily built (and is much stronger than the poloidal one), the magnetic pressure of which drives these slow-moving outflowing winds \citep{Blandford1982, 
Uchida1985, Pudritz1986, Emmering1992, Contopoulos1995}. These results are also consistent with the suggestion of toroidal or helical B-fields from the RM gradient in NGC\,4151. \citet{Gallimore2024} have been able to find direct evidence for such a hydromagnetic wind in the Seyfert galaxy, NGC\,1068, by studying the ${H_2}O$ megamasers with the High Sensitivity Array (HSA). 

\begin{figure*}
\centering
    \includegraphics[width=9.2cm, angle=90,trim=0 40 0 0]{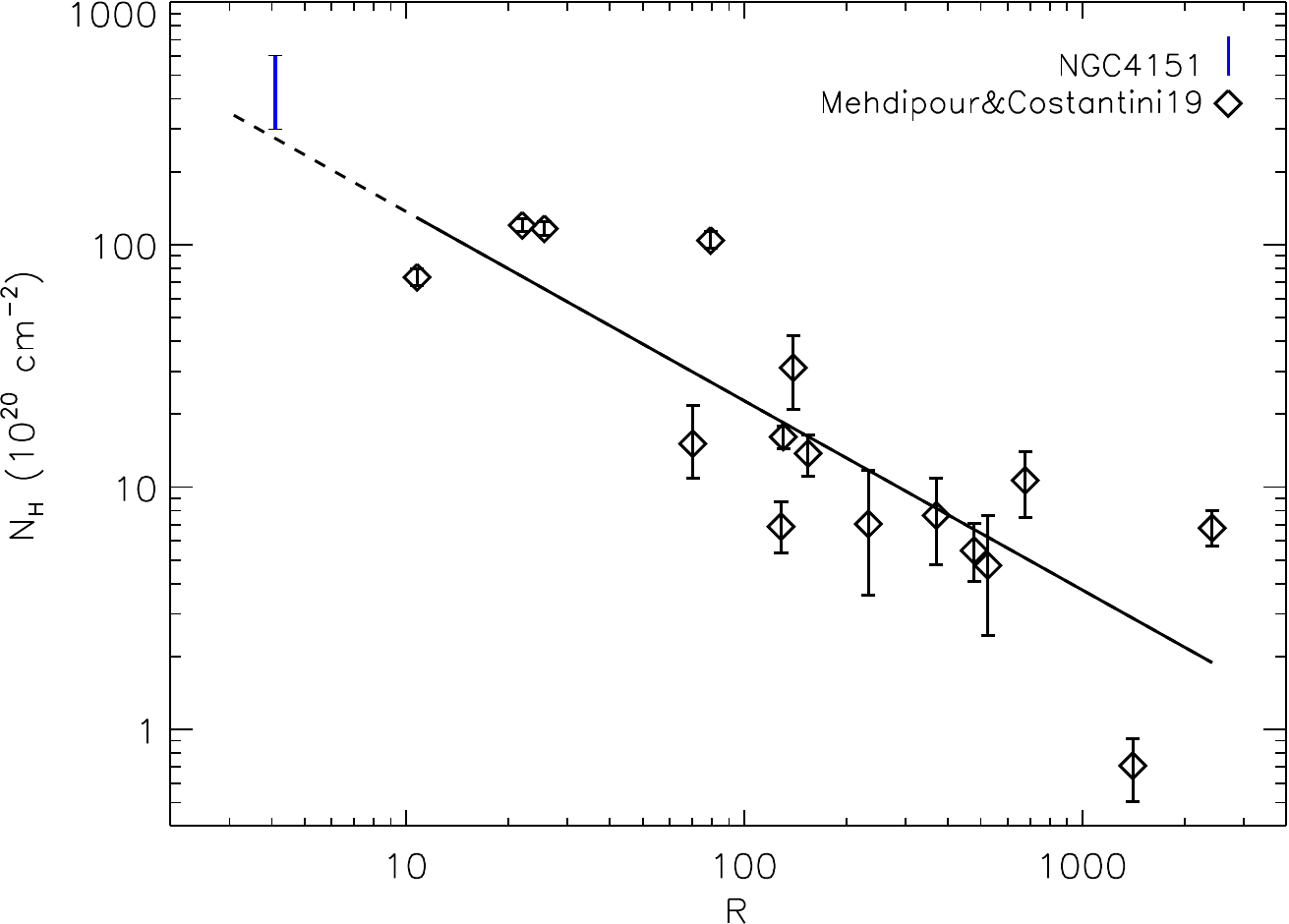}
\caption{\small Values of $N_{\rm H}$ and the radio loudness parameter, $R$, estimated for NGC\,4151 (blue bar) superimposed to the anti-correlation (black line) found for radio-loud sources \citep[black diamonds,][]{Mehdipour2019}. The dashed line simply extends the correlation to our estimate for NGC\,4151.}
\label{fig:nh_r}
\end{figure*}

\subsection{Constraints on the Warm Absorber}\label{sec:warmabsorbers}
The soft X-ray spectrum of NGC\,4151 is dominated by emission lines \citep[e.g.,][]{Schurch2004, Kraemer2006}, making it less straightforward to identify all the absorption components. However, two absorption components have been consistently reported by the {\it Chandra} observatory \citep[e.g.][]{Couto2016}. In particular, one of them has been directly put in relation with the absorption components detected in the UV \citep{Kraemer2006}, indicating a low-ionization component. In \citet{Mehdipour2019}, the column density of the low-ionization components of the warm absorber have been put in relation with the radio loudness parameter, $R$, for a sample of radio-loud AGN, finding a robust anti-correlation between the amount of material available to form a warm absorber in the disk and the radio power. We attempted to locate NGC\,4151 \citep[$R\sim3.09$;][]{Ho2001} in this anti-correlation (Figure~\ref{fig:nh_r}), adopting the column density of the low-ionization component reported in \citet{Couto2016} as a lower limit for the warm absorber that might have been detected in the RGS band ($N_{\rm H}\sim 3-6\times10^{22}$\,cm$^{-2}$, taking into account the moderate variability of this component). Other, even lower ionization components might still be present. Importantly, Figure~\ref{fig:nh_r} shows that the disk-wind connection with the jet emission found for radio-loud sources \citep{Mehdipour2019} may be extended to radio-quiet objects.

The highly ionized X-ray absorber with a column density of $N_{\mathrm{H}} \sim 3.2\times10^{22}~\mathrm{cm^{-2}}$ \citep{Crenshaw2012, Couto2016} represents the innermost component of the warm absorber, likely originating from MHD disk winds. The outflow, with a velocity of $v_r \approx -500~\mathrm{km~s^{-1}}$, is located at a radial distance of $r \approx (2.5 - 30)\times10^{16}~\mathrm{cm}$ from the central engine, corresponding to gas densities of $n_{\mathrm{H}}\sim10^{6}~\mathrm{cm^{-3}}$ \citep{Crenshaw2012, Couto2016}. Assuming a spherically symmetric outflow \citep[consistent with][]{Crenshaw2003, Crenshaw2012}, the total gas mass can be expressed as
\begin{equation}
M_{\mathrm{gas}} = 4\pi r^{2} C_{\mathrm{f}}\, \mu\, m_{\mathrm{p}}\, N_{\mathrm{H}},
\end{equation}
where $\mu = 1.4$ (for solar composition) is the mean atomic weight, $m_{\mathrm{p}}$ is the proton mass, and $C_{\mathrm{f}}$ is the global covering fraction. For $C_{\mathrm{f}}\sim0.5$, the estimated gas mass lies in the range $M_{\mathrm{gas}}\sim(0.15 - 21)\,M_{\odot}$. The corresponding mass outflow rate,
\begin{equation}
\dot{M}_{\mathrm{gas}} = 4\pi r C_{\mathrm{f}}\, \mu\, m_{\mathrm{p}}\, N_{\mathrm{H}} v_r,
\end{equation}
is $\dot{M}_{\mathrm{gas}} \approx 0.01 - 0.1\,M_{\odot}\,\mathrm{yr^{-1}}$, implying a kinetic luminosity
\begin{equation}
L_{\mathrm{KE}} = \tfrac{1}{2}\dot{M}_{\mathrm{gas}} v_r^2 = 2\pi r C_{\mathrm{f}}\, \mu\, m_{\mathrm{p}}\, N_{\mathrm{H}} v_r^3,
\end{equation}
of $L_{\mathrm{KE}} \approx (0.7 - 8.2)\times10^{39}$~erg~s$^{-1}$. 

Although the mass outflow rate is comparable to or exceeds the inferred accretion rate by a factor of $1-10$, the mechanical power constitutes less than $0.01\%$ of the bolometric luminosity, as was also found for the radio `wind' component (see Section~\ref{sec:wind}). The force multiplier (= ratio of the total photo-absorption cross-section to the Thomson cross-section) for this source, obtained by {\tt{CLOUDY}} modeling, is found to be near unity, making radiative driving inefficient in sub-Eddington AGN, and hence implying that mechanisms such as MHD-driven outflow must be responsible for their acceleration \citep{Crenshaw2012, Kraemer2018}. This indicates that, while the outflow is mass-loaded, its kinetic coupling efficiency is low, consistent with a slow, magnetically driven wind rather than a radiatively accelerated one. The relatively small gas mass ($M_{\mathrm{gas}}$) compared to that required for the mixed thermal plasma inferred from internal depolarization modeling (see Section~\ref{sec:extintdepol}{)} further suggests that at least one external Faraday-rotating screen contributes to the observed depolarization. Thus, the physical conditions point toward a magnetically dominated, low-efficiency outflow in the nuclear region, characteristic of radio-quiet Seyfert nuclei.

\section{Conclusions}\label{sec:conclusion}
We have observed the Seyfert galaxy NGC\,4151 with the VLA B-array configuration at 3 and 10 GHz and detected a stratified outflow based on the inferred magnetic field structures. We summarize the primary findings of our study below.

\begin{enumerate}
\item The 3 GHz VLA image of NGC\,4151 reveals the well-known east–west radio outflow, with a hotspot-like region in the eastern lobe. Additional diffuse emission is detected perpendicular to the east-west outflow. The inferred B-fields, after source RM correction, remain aligned with the outflow at the jet edge and perpendicular towards the inner region. Besides, the B-fields are found perpendicular to the transverse outflow component surrounding the jet. These structures and inferred B-fields are consistent with the jet+wind model of \citet{Mehdipour2019}.

\item The 10 GHz VLA image resolves the outflow into a jet `spine' with perpendicular B-fields and a jet `sheath' with aligned B-fields. The hotspot-like region is clearly seen with B-fields resembling oblique shocks. The base of the wind-like feature seen in the 3 GHz image is also visible at 10 GHz. The perpendicular B-fields in the jet `spine' are consistent with shocks as have been suggested in the case of NGC\,4151 from various studies in the literature \citep{Wang2010, Williams2020}. The aligned B-fields in the `sheath' are consistent with shearing due to jet-medium interaction. 

\item The jet velocity $\sim$0.1 kpc away from the nucleus is estimated to be $\sim0.16$c from the jet-to-counterjet intensity ratio assuming Doppler boosting effects and a jet inclination of 40 degrees. The jet kinetic power, derived from the core radio luminosity, is $\sim10^{43}$~erg~s$^{-1}$, which is at the higher end of jet powers typically observed in radio-quiet AGN. 

\item We derive an electron density of $0.004-0.02$~cm$^{-3}$ and a characteristic extent of $0.06-0.2$ kpc for the Faraday rotating medium around the jet in NGC\,4151. These estimates are consistent with either a sparse warm ionized medium in the host galaxy or a magnetized sheath or cocoon surrounding the jet and acting as the Faraday screen. Assuming internal depolarization in the wider `wind' outflow of NGC\,4151 as thermal gas is mixed with synchrotron-emitting plasma (consistent with strong depolarization in the `wind' region), we estimate the thermal gas mass to be $\sim10^{3}\,M_\odot$. From the warm absorber mass estimates in NGC\,4151, it is evident that the mixing of this gas cannot be solely responsible for the observed depolarization. It is essential to have an additional external screen to fully explain the observed depolarization.

\item The identified radio `wind' component is found to be heavy with a mass of $1050 - 3200~M_\odot$, with an outflow rate ($=0.01 - 0.03~M_\odot$ yr$^{-1}$) higher than the black hole accretion rate, although the {kinetic power ($\sim0.01-0.5$\% of $L_{\rm{bol}}$)} is small, suggesting the presence of massive but kinematically weak winds, ineffective in providing galactic-scale feedback. We also observe a transverse RM gradient ranging from $+75$ to $-25$ rad~m$^{-2}$ across the (northern side of) `wind' component, suggestive of a helical magnetic field threading it. {Although limited spatial resolution implies that systematic effects exist, similar RM values with opposite signs in the northern and southern `wind' components point to an underlying coherent magnetic field structure. Taken together these components support the presence of a large-scale transverse RM gradient and thereby the presence of a helical magnetic field across the wide-angled bi-conical wind.}

\item The `minimum energy' B-field strength ranges from $100-300~\mu$G in the `core' region to $28-86~\mu$G in the `wind' region, $\sim$0.5 kpc from the core. The corresponding magnetic pressure decreases from $(12-107)\times10^{-10}$ dyne cm$^{-2}$ in the core to $(1-9)\times10^{-10}$~dyne~cm$^{-2}$ in the `wind'. On the other hand, the thermal gas pressure decreases from $6.8\times10^{-10}$~dyne~cm$^{-2}$ from close to the NLR to $5\times10^{-12}$~dyne~cm$^{-2}$ in the galactic outskirts. Across different spatial scales, the magnetic pressure exceeds the thermal gas pressure, indicating that the outflow in NGC\,4151 is magnetically-driven. The presence of MHD winds in this source has been previously discussed by \citet{Crenshaw2012, Kraemer2018}. 

\item We find that the amount of material available to launch an UV/X-ray wind, given the radio-to-optical flux density ratio ($R$) for this source, follows the same trend as found for radio-loud objects. This suggests that a disk-wind radio connection might also be at play for relatively radio-weak sources. Moreover, the kinetic power of the high-ionization gas present at ($2.5 - 30) \times 10^{16}$ cm is found to be low i.e., $L_{\mathrm{KE}} \approx (0.7 - 8.2)\times10^{39}~erg~s^{-1} <0.01\% L_{\rm{bol}}$, but similar to the radio-detected `wind' in NGC\,4151.
\end{enumerate}

\section{Acknowledgement}
{We thank the referee for their suggestions which have significantly improved our manuscript.} SG and PK acknowledge the support of the Department of Atomic Energy, Government of India, under the project 12-R\&D-TFR-5.02-0700. This research has utilised data from the National Radio Astronomy Observatory (NRAO) facility. The NRAO and Green Bank Observatory are facilities of the U.S. National Science Foundation operated under cooperative agreement by Associated Universities, Inc. 

\section{Appendix}
\begin{figure*}
\centering
\includegraphics[width=8cm]{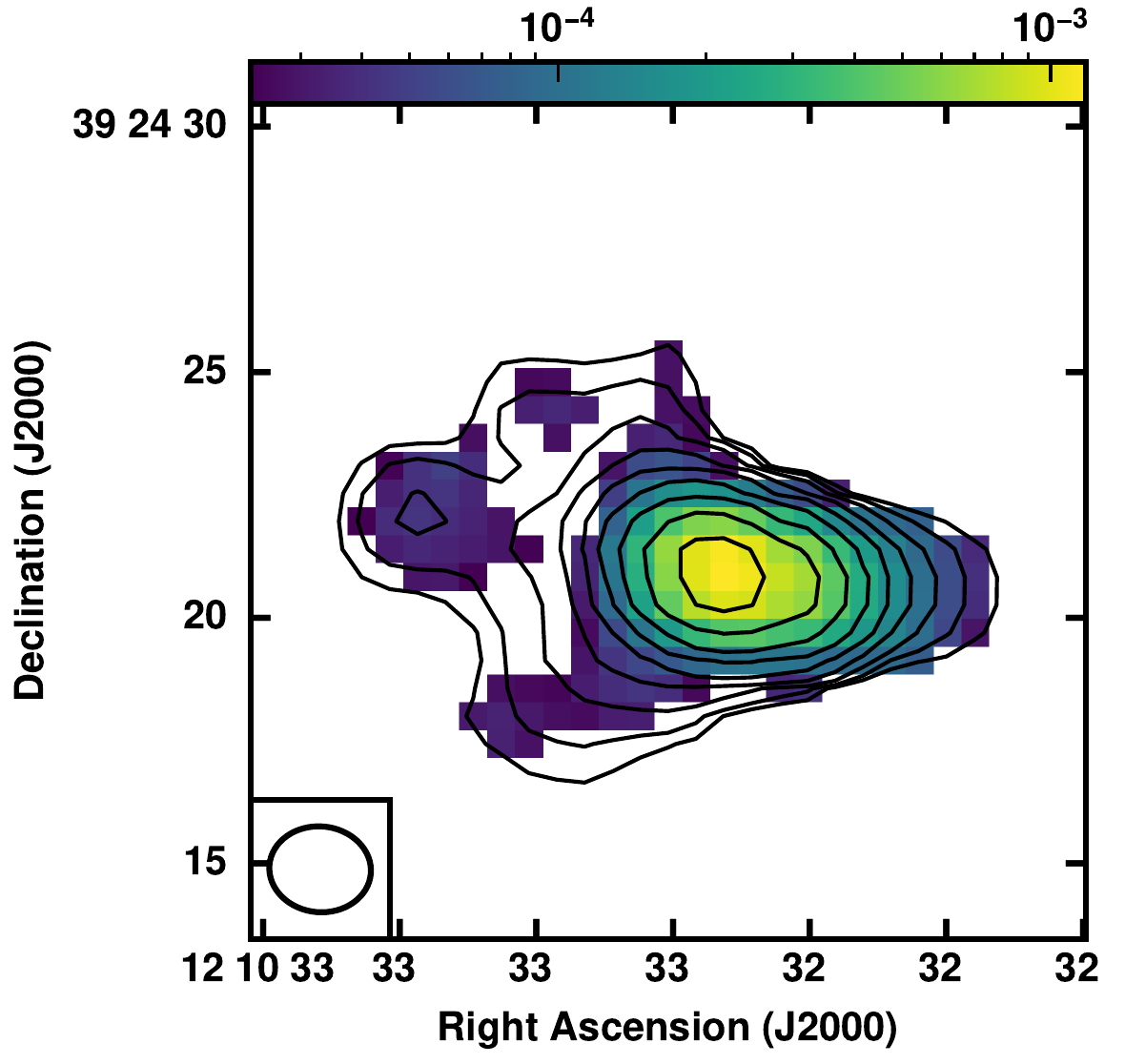}
\includegraphics[width=8cm]{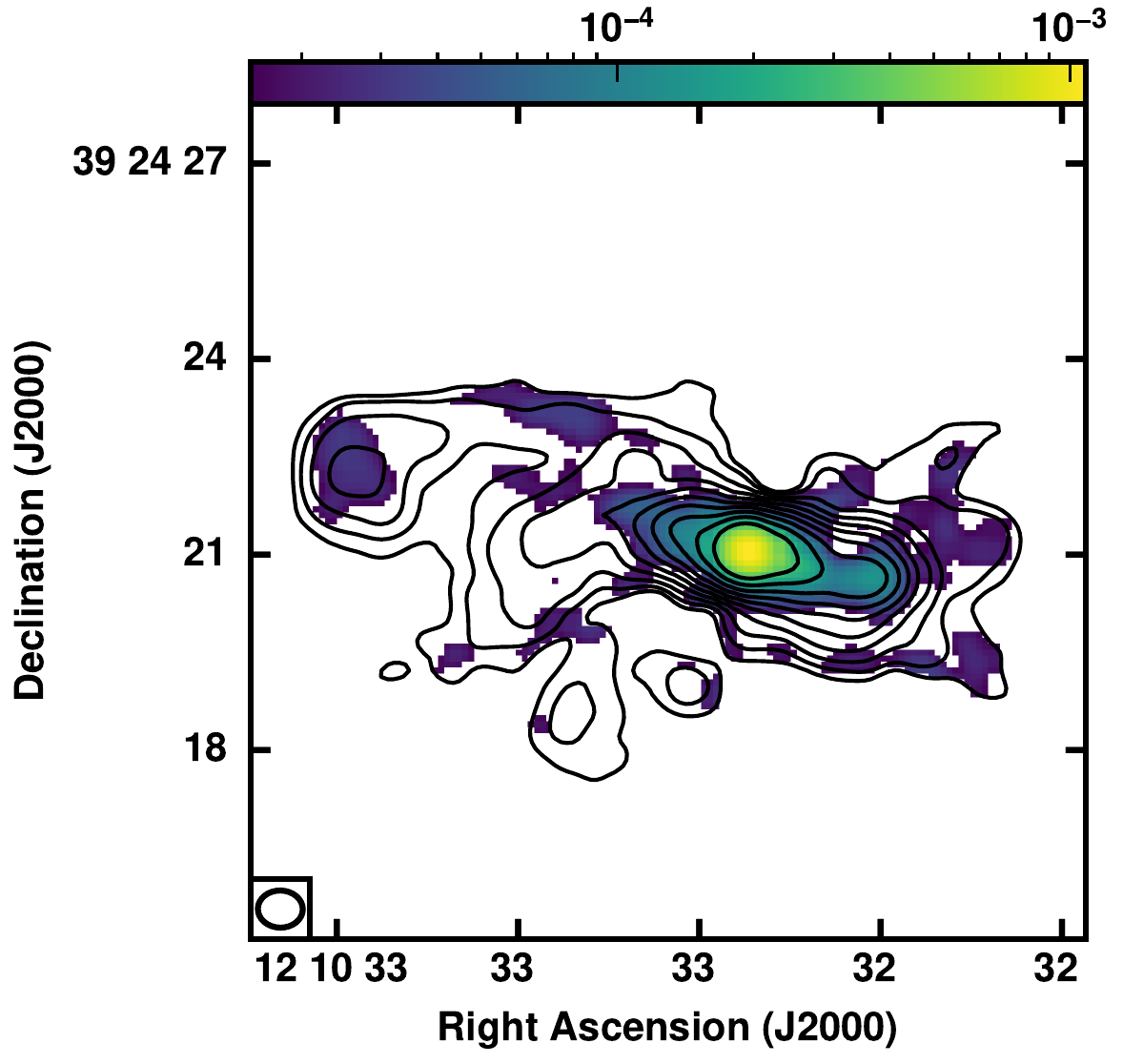}
\caption{\small Left: 3 GHz polarisation intensity in colour in logarithmic scale, ranging from 24 $\mu$Jy beam$^{-1}$ to 1.2 mJy beam$^{-1}$. The total intensity contours at 3 GHz are shown in black as above. Right:10 GHz polarization intensity in colour in logarithmic scale, ranging from 15 $\mu$Jy beam$^{-1}$ to 1.1  mJy beam$^{-1}$. The total intensity contours at 10 GHz are shown in black as above.}
\label{fig:Ipolmap}
\end{figure*}

\begin{figure*}
\centering
\includegraphics[width=8cm]{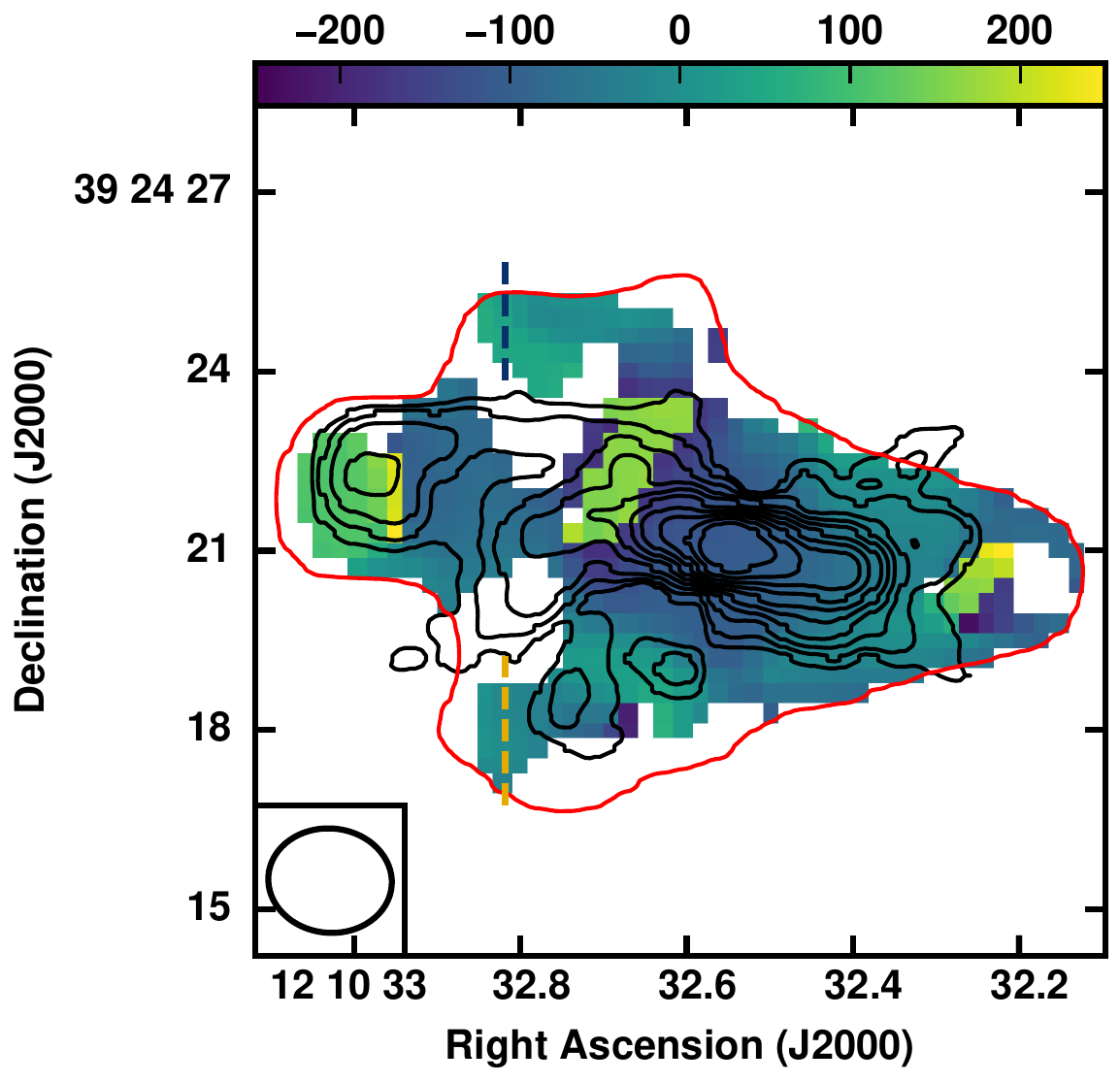}
\includegraphics[width=8cm]{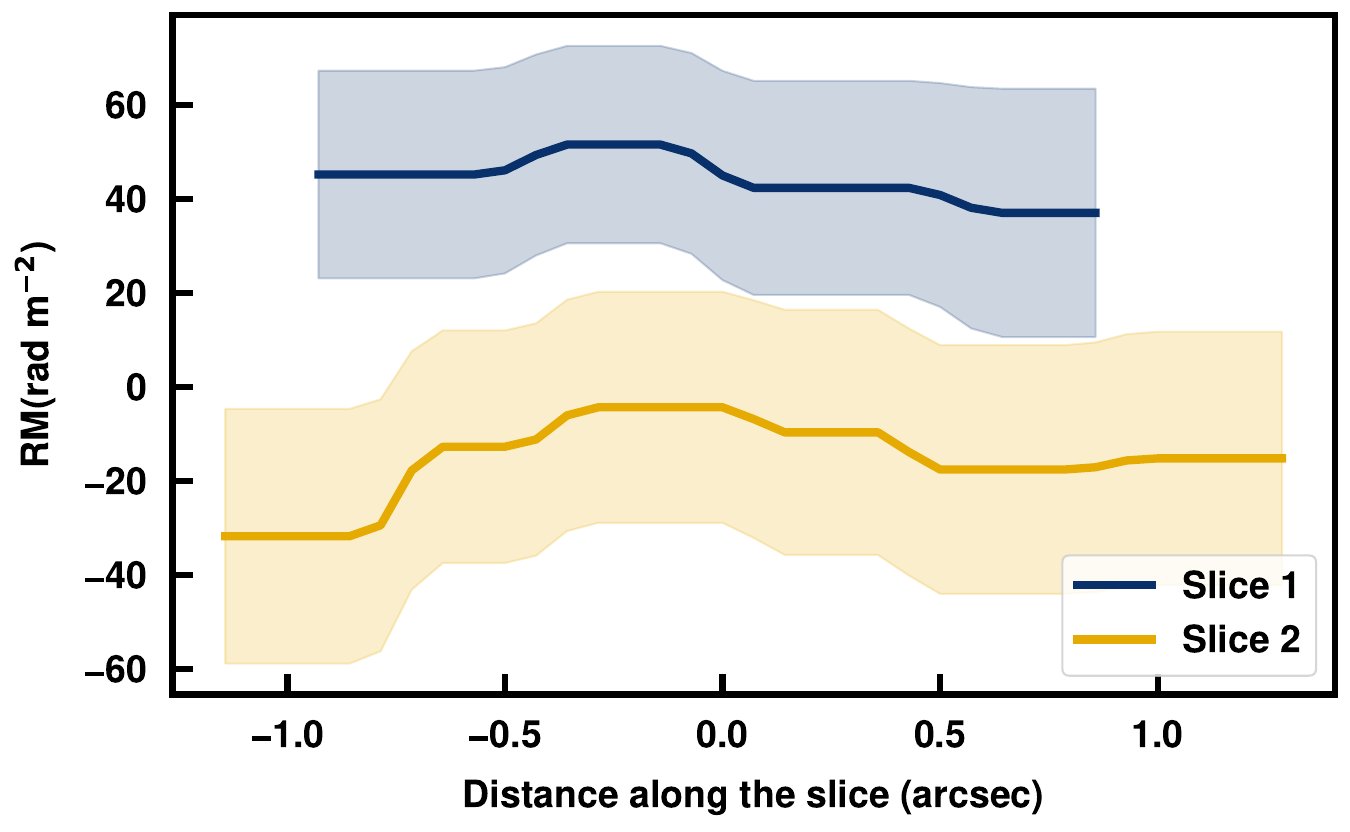}
\includegraphics[width=8cm]{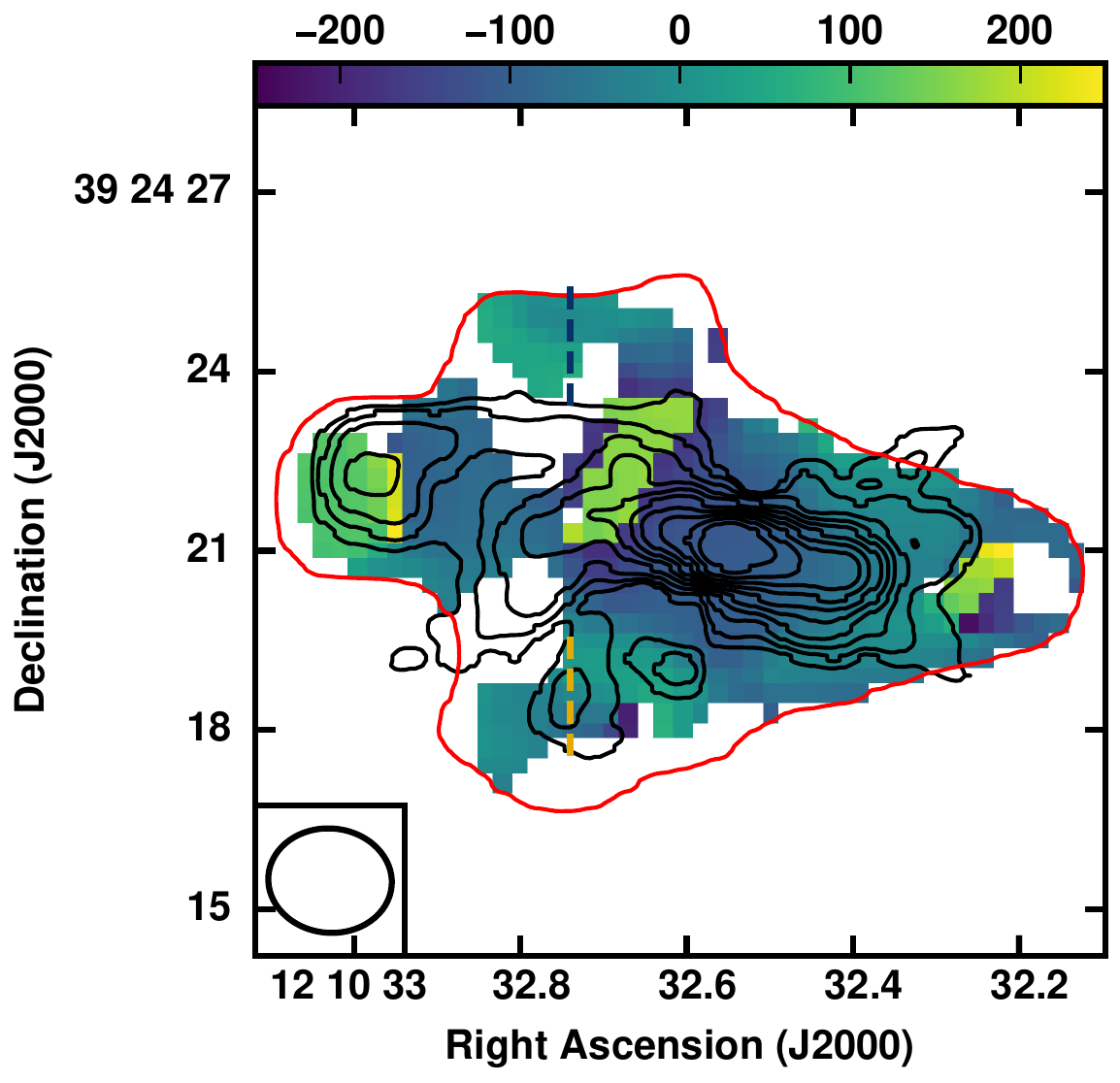}
\includegraphics[width=8cm]{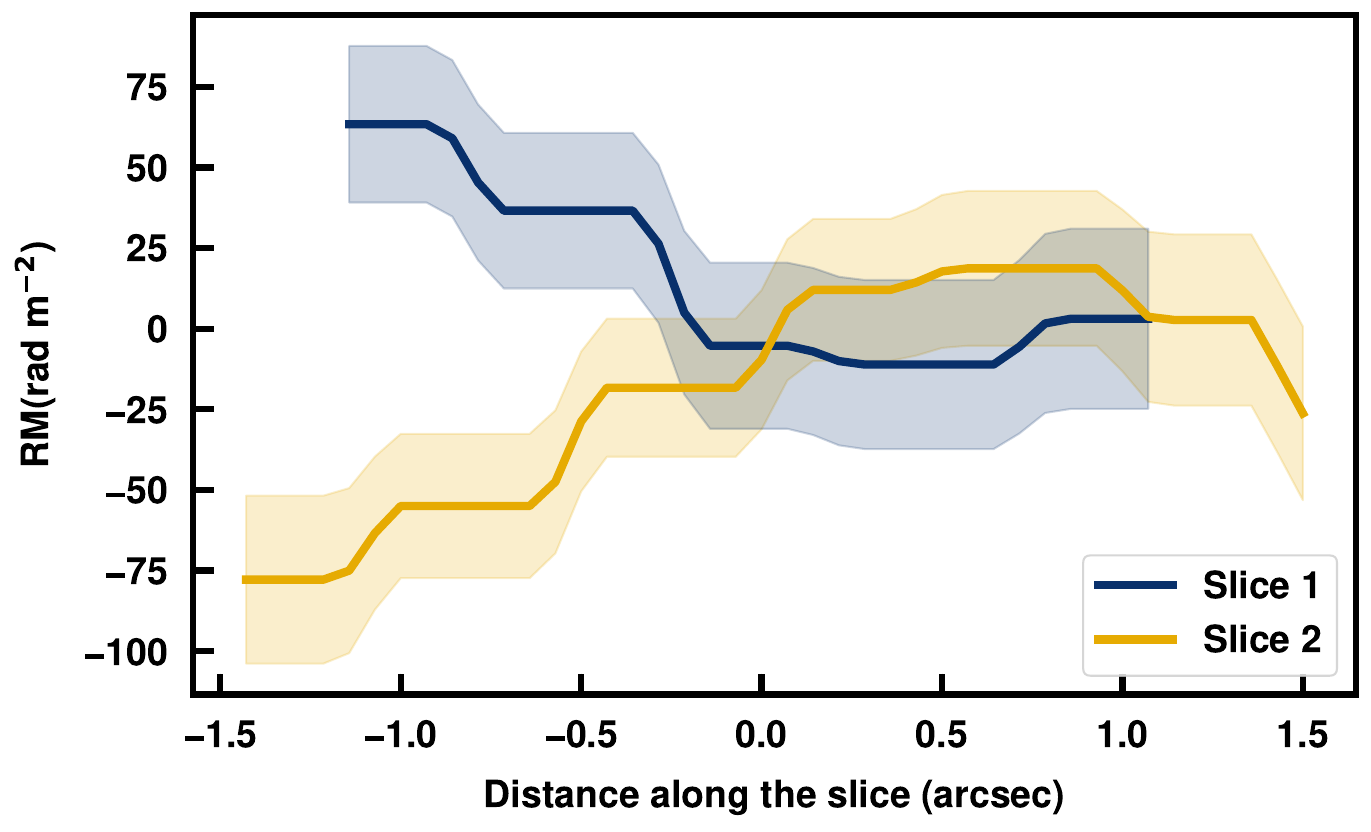}
\caption{\small Left panel: RM images showing RM slices taken across the different parts of the `wind' region. The northern and southern `wind' have been indicated with blue and yellow dashed lines, respectively. Right panel: The slices (considered from top to bottom) are shown as a function of distance along with the errors. The northern and southern `wind' show similar absolute value of RM but with a changed direction, indicating a reversal of the B-field direction along our line of sight between the two sides. Such an RM gradient may indicate the presence of helical fields along a conical envelope of synchrotron winds.}
\label{fig:helicalfield}
\end{figure*}

{Figure~\ref{fig:Ipolmap} presents the polarized intensity images at 3 and 10~GHz in the left and right panels, respectively. In both bands, the core exhibits the highest polarized intensity, which progressively decreases with distance from the core. The jet axis or the `spine' shows systematically higher polarized intensity than the surrounding `sheath', whereas the `sheath' displays a higher fractional polarization as in Figure \ref{fig:3GHzimage}.} We also present the transverse slices from the RM image showing gradients in the `wind' region in Figure \ref{fig:helicalfield}. These are consistent with the presence of 100-parsec-scale helical magnetic fields threading the `wind' component.

\begin{figure*}

\includegraphics[width=8.3cm,trim=150 45 50 120]{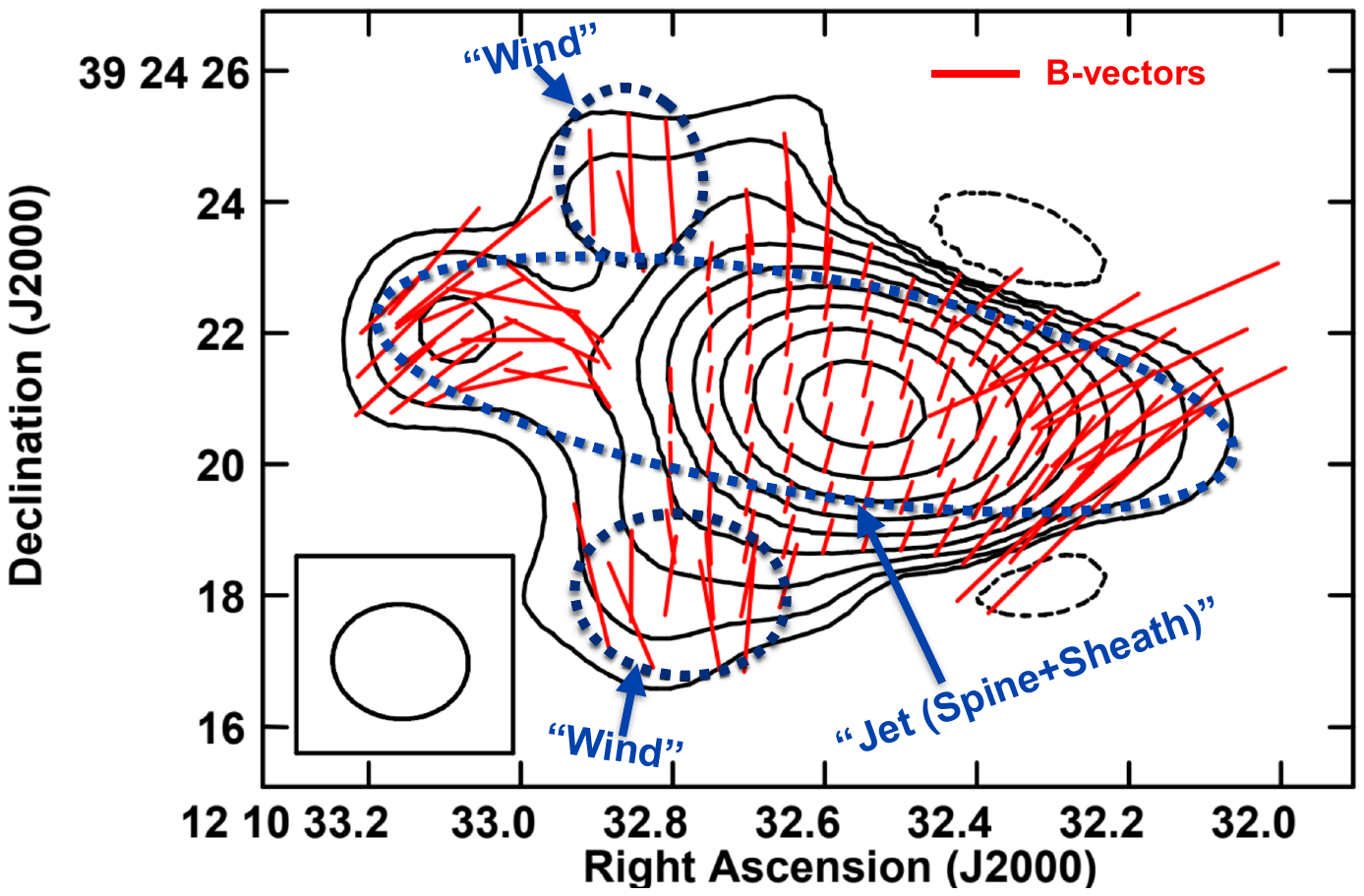}
\includegraphics[width=10.1cm,trim=30 60 20 120]{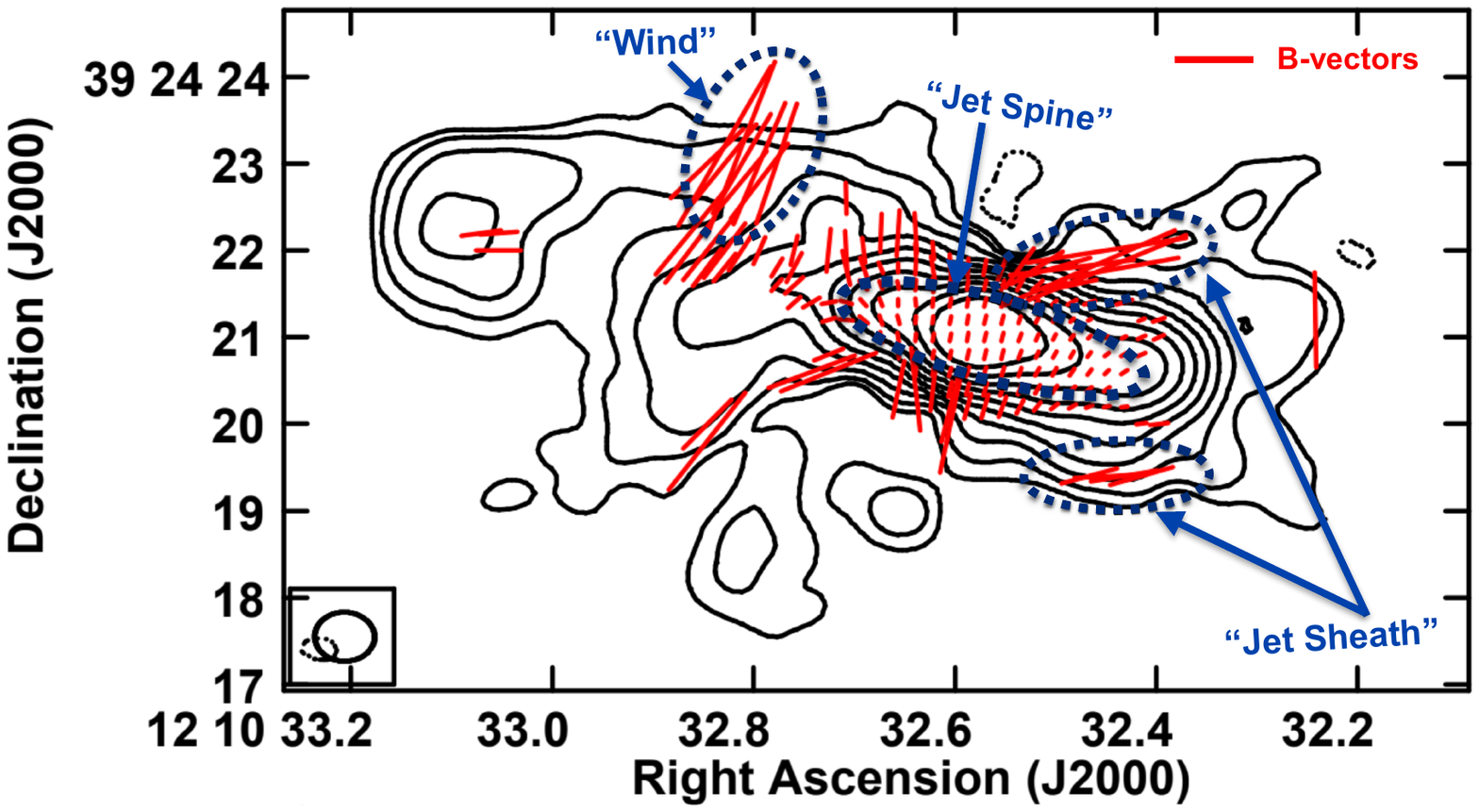}
\caption{\small 3 GHz (left) and 10 GHz (right) image showing B-field vectors corrected for Galactic plus source RM. The contours and the polarization vector lengths are the same as Figure~\ref{fig:3GHzimage}.}
\label{fig:RM-correctedB-vectors}
\end{figure*}

\begin{contribution}
The author order is consistent with their contributions to this work.
\end{contribution}

\facilities{VLA}
\software{CASA \citep{McMullin2007}}
\bibliography{ms}{}

@ARTICLE{Pudritz2006,
       author = {{Pudritz}, Ralph E. and {Rogers}, Conrad S. and {Ouyed}, Rachid},
        title = "{Controlling the collimation and rotation of hydromagnetic disc winds}",
      journal = {\mnras},
         year = 2006,
        month = feb,
       volume = {365},
       number = {4},
        pages = {1131-1148},
          doi = {10.1111/j.1365-2966.2005.09766.x},
archivePrefix = {arXiv},
       eprint = {astro-ph/0508295},
 primaryClass = {astro-ph},
       adsurl = {https://ui.adsabs.harvard.edu/abs/2006MNRAS.365.1131P}
}

@ARTICLE{Fendt2006,
       author = {{Fendt}, Christian},
        title = "{Collimation of Astrophysical Jets: The Role of the Accretion Disk Magnetic Field Distribution}",
      journal = {\apj},
         year = 2006,
        month = nov,
       volume = {651},
       number = {1},
        pages = {272-287},
          doi = {10.1086/507976},
archivePrefix = {arXiv},
       eprint = {astro-ph/0511611},
 primaryClass = {astro-ph},
       adsurl = {https://ui.adsabs.harvard.edu/abs/2006ApJ...651..272F}
}

@ARTICLE{Bentz2022,
       author = {{Bentz}, Misty C. and {Williams}, Peter R. and {Treu}, Tommaso},
        title = "{The Broad Line Region and Black Hole Mass of NGC 4151}",
      journal = {\apj},
         year = 2022,
        month = aug,
       volume = {934},
       number = {2},
          eid = {168},
        pages = {168},
          doi = {10.3847/1538-4357/ac7c0a},
archivePrefix = {arXiv},
       eprint = {2206.03513},
 primaryClass = {astro-ph.GA},
       adsurl = {https://ui.adsabs.harvard.edu/abs/2022ApJ...934..168B}
}

@ARTICLE{Begelman1989,
       author = {{Begelman}, Mitchell C. and {Cioffi}, Denis F.},
        title = "{Overpressured Cocoons in Extragalactic Radio Sources}",
      journal = {\apjl},
         year = 1989,
        month = oct,
       volume = {345},
        pages = {L21},
          doi = {10.1086/185542},
       adsurl = {https://ui.adsabs.harvard.edu/abs/1989ApJ...345L..21B}
}

@INCOLLECTION{Beck2013,
       author = {{Beck}, Rainer and {Wielebinski}, Richard},
        title = "{Magnetic Fields in Galaxies}",
    booktitle = {Planets, Stars and Stellar Systems. Volume 5: Galactic Structure and Stellar Populations},
         year = 2013,
       editor = {{Oswalt}, Terry D. and {Gilmore}, Gerard},
       volume = {5},
        pages = {641},
          doi = {10.1007/978-94-007-5612-0_13},
       adsurl = {https://ui.adsabs.harvard.edu/abs/2013pss5.book..641B}
}

@ARTICLE{Gallimore2024,
       author = {{Gallimore}, Jack F. and {Impellizzeri}, C.~M. Violette and {Aghelpasand}, Samaneh and {Gao}, Feng and {Hostetter}, Virginia and {Lankhaar}, Boy},
        title = "{The Discovery of Polarized Water Vapor Megamaser Emission in a Molecular Accretion Disk}",
      journal = {\apjl},
         year = 2024,
        month = nov,
       volume = {975},
       number = {1},
          eid = {L9},
        pages = {L9},
          doi = {10.3847/2041-8213/ad864f},
archivePrefix = {arXiv},
       eprint = {2410.10569},
 primaryClass = {astro-ph.GA},
       adsurl = {https://ui.adsabs.harvard.edu/abs/2024ApJ...975L...9G}
}

@INPROCEEDINGS{McMullin2007,
       author = {{McMullin}, J.~P. and {Waters}, B. and {Schiebel}, D. and {Young}, W. and {Golap}, K.},
        title = "{CASA Architecture and Applications}",
    booktitle = {Astronomical Data Analysis Software and Systems XVI},
         year = 2007,
       editor = {{Shaw}, R.~A. and {Hill}, F. and {Bell}, D.~J.},
       series = {Astronomical Society of the Pacific Conference Series},
       volume = {376},
        month = oct,
        pages = {127},
       adsurl = {https://ui.adsabs.harvard.edu/abs/2007ASPC..376..127M}
}

@ARTICLE{Farcy2025,
       author = {{Farcy}, Marion and {Hirschmann}, Michaela and {Somerville}, Rachel S. and {Choi}, Ena and {Koudmani}, Sophie and {Naab}, Thorsten and {Weinberger}, Rainer and {Bennett}, Jake S. and {Bhowmick}, Aklant K. and {Choi}, Hyunseop and {Hernquist}, Lars and {Hlavacek-Larrondo}, Julie and {Terrazas}, Bryan A. and {Valentino}, Francesco},
        title = "{MISTRAL: a model for AGN winds from radiatively efficient accretion in cosmological simulations}",
      journal = {\mnras},
         year = 2025,
        month = oct,
       volume = {543},
       number = {2},
        pages = {967-993},
          doi = {10.1093/mnras/staf1464},
archivePrefix = {arXiv},
       eprint = {2504.08041},
 primaryClass = {astro-ph.GA},
       adsurl = {https://ui.adsabs.harvard.edu/abs/2025MNRAS.543..967F}
}

@ARTICLE{Xiang2025,
       author = {{Xiang}, Xin and {Miller}, Jon M. and {Behar}, Ehud and {Boissay-Malaquin}, Rozenn and {Brenneman}, Laura and {Buhariwalla}, Margaret and {Byun}, Doyee and {Done}, Chris and {Gallo}, Luigi and {Gerolymatou}, Dimitra and {Hagen}, Scott and {Kaastra}, Jelle and {Paltani}, Stephane and {Porter}, Frederick S. and {Mushotzky}, Richard and {Noda}, Hirofumi and {Mehdipour}, Missagh and {Minezaki}, Takeo and {Tashiro}, Makoto and {Zoghbi}, Abderahmen},
        title = "{XRISM Spectroscopy of Accretion-driven Wind Feedback in NGC 4151}",
      journal = {\apjl},
         year = 2025,
        month = aug,
       volume = {988},
       number = {2},
          eid = {L54},
        pages = {L54},
          doi = {10.3847/2041-8213/adee9b},
archivePrefix = {arXiv},
       eprint = {2507.09210},
 primaryClass = {astro-ph.HE},
       adsurl = {https://ui.adsabs.harvard.edu/abs/2025ApJ...988L..54X}
}

@ARTICLE{Cioffi1992,
       author = {{Cioffi}, Denis F. and {Blondin}, John M.},
        title = "{The Evolution of Cocoons Surrounding Light, Extragalactic Jets}",
      journal = {\apj},
         year = 1992,
        month = jun,
       volume = {392},
        pages = {458},
          doi = {10.1086/171445},
       adsurl = {https://ui.adsabs.harvard.edu/abs/1992ApJ...392..458C}
}

@ARTICLE{Kraemer2018,
       author = {{Kraemer}, S.~B. and {Tombesi}, F. and {Bottorff}, M.~C.},
        title = "{Physical Conditions in Ultra-fast Outflows in AGN}",
      journal = {\apj},
         year = 2018,
        month = jan,
       volume = {852},
       number = {1},
          eid = {35},
        pages = {35},
          doi = {10.3847/1538-4357/aa9ce0},
archivePrefix = {arXiv},
       eprint = {1711.07965},
 primaryClass = {astro-ph.GA},
       adsurl = {https://ui.adsabs.harvard.edu/abs/2018ApJ...852...35K}
}

@ARTICLE{Kharb2021,
       author = {{Kharb}, P. and {Subramanian}, S. and {Das}, M. and {Vaddi}, S. and {Paragi}, Z.},
        title = "{The Nature of Jets in Double-peaked Emission-line AGN in the KISSR Sample}",
      journal = {\apj},
         year = 2021,
        month = oct,
       volume = {919},
       number = {2},
          eid = {108},
        pages = {108},
          doi = {10.3847/1538-4357/ac0c82},
archivePrefix = {arXiv},
       eprint = {2106.09304},
 primaryClass = {astro-ph.GA},
       adsurl = {https://ui.adsabs.harvard.edu/abs/2021ApJ...919..108K}
}

@ARTICLE{Emmering1992,
       author = {{Emmering}, Robert T. and {Blandford}, Roger D. and {Shlosman}, Isaac},
        title = "{Magnetic Acceleration of Broad Emission-Line Clouds in Active Galactic Nuclei}",
      journal = {\apj},
         year = 1992,
        month = feb,
       volume = {385},
        pages = {460},
          doi = {10.1086/170955},
       adsurl = {https://ui.adsabs.harvard.edu/abs/1992ApJ...385..460E}
}

@ARTICLE{Blandford1982,
       author = {{Blandford}, R.~D. and {Payne}, D.~G.},
        title = "{Hydromagnetic flows from accretion disks and the production of radio jets.}",
      journal = {\mnras},
         year = 1982,
        month = jun,
       volume = {199},
        pages = {883-903},
          doi = {10.1093/mnras/199.4.883},
       adsurl = {https://ui.adsabs.harvard.edu/abs/1982MNRAS.199..883B}
}

@ARTICLE{Ghosh2025IAU,
       author = {{Ghosh}, Salmoli and {Kharb}, Preeti and {Sebastian}, Biny and {Gallimore}, Jack and {Pasetto}, Alice and {O'Dea}, Christopher P. and {Heckman}, Timothy and {Baum}, Stefi A.},
        title = "{What drives kpc-scale outflows in Radio-Quiet AGN? Insights from a Polarimetric Study}",
      journal = {arXiv e-prints},
         year = 2025,
        month = sep,
          eid = {arXiv:2509.15355},
        pages = {arXiv:2509.15355},
          doi = {10.48550/arXiv.2509.15355},
archivePrefix = {arXiv},
       eprint = {2509.15355},
 primaryClass = {astro-ph.GA},
       adsurl = {https://ui.adsabs.harvard.edu/abs/2025arXiv250915355G}
}

@ARTICLE{Giroletti2009,
       author = {{Giroletti}, Marcello and {Panessa}, Francesca},
        title = "{The Faintest Seyfert Radio Cores Revealed by VLBI}",
      journal = {\apjl},
         year = 2009,
        month = dec,
       volume = {706},
       number = {2},
        pages = {L260-L264},
          doi = {10.1088/0004-637X/706/2/L260},
archivePrefix = {arXiv},
       eprint = {0910.5821},
 primaryClass = {astro-ph.CO},
       adsurl = {https://ui.adsabs.harvard.edu/abs/2009ApJ...706L.260G}
}

@ARTICLE{Sebastian2020,
       author = {{Sebastian}, Biny and {Kharb}, P. and {O'Dea}, C.~P. and {Gallimore}, J.~F. and {Baum}, S.~A.},
        title = "{A radio polarimetric study to disentangle AGN activity and star formation in Seyfert galaxies}",
      journal = {\mnras},
         year = 2020,
        month = nov,
       volume = {499},
       number = {1},
        pages = {334-354},
          doi = {10.1093/mnras/staa2473},
archivePrefix = {arXiv},
       eprint = {2008.06039},
 primaryClass = {astro-ph.GA},
       adsurl = {https://ui.adsabs.harvard.edu/abs/2020MNRAS.499..334S}
}

@ARTICLE{Silpa2021,
       author = {{Silpa}, S. and {Kharb}, P. and {O'Dea}, C.~P. and {Baum}, S.~A. and {Sebastian}, B. and {Mukherjee}, D. and {Harrison}, C.~M.},
        title = "{AGN jets and winds in polarized light: the case of Mrk 231}",
      journal = {\mnras},
         year = 2021,
        month = oct,
       volume = {507},
       number = {2},
        pages = {2550-2561},
          doi = {10.1093/mnras/stab2110},
archivePrefix = {arXiv},
       eprint = {2107.09466},
 primaryClass = {astro-ph.GA},
       adsurl = {https://ui.adsabs.harvard.edu/abs/2021MNRAS.507.2550S}
}

@ARTICLE{Couto2016,
       author = {{Couto}, J.~D. and {Kraemer}, S.~B. and {Turner}, T.~J. and {Crenshaw}, D.~M.},
        title = "{New Insights into the Spectral Variability and Physical Conditions of the X-Ray Absorbers in NGC 4151}",
      journal = {\apj},
         year = 2016,
        month = dec,
       volume = {833},
       number = {2},
          eid = {191},
        pages = {191},
          doi = {10.3847/1538-4357/833/2/191},
archivePrefix = {arXiv},
       eprint = {1610.05689},
 primaryClass = {astro-ph.GA},
       adsurl = {https://ui.adsabs.harvard.edu/abs/2016ApJ...833..191C}
}

@ARTICLE{Laing2006,
       author = {{Laing}, R.~A. and {Canvin}, J.~R. and {Cotton}, W.~D. and {Bridle}, A.~H.},
        title = "{Multifrequency observations of the jets in the radio galaxy NGC315}",
      journal = {\mnras},
         year = 2006,
        month = may,
       volume = {368},
       number = {1},
        pages = {48-64},
          doi = {10.1111/j.1365-2966.2006.10099.x},
archivePrefix = {arXiv},
       eprint = {astro-ph/0601660},
 primaryClass = {astro-ph},
       adsurl = {https://ui.adsabs.harvard.edu/abs/2006MNRAS.368...48L}
}

@ARTICLE{Kraemer2006,
       author = {{Kraemer}, S.~B. and {Crenshaw}, D.~M. and {Gabel}, J.~R. and {Kriss}, G.~A. and {Netzer}, H. and {Peterson}, B.~M. and {George}, I.~M. and {Gull}, T.~R. and {Hutchings}, J.~B. and {Mushotzky}, R.~F. and {Turner}, T.~J.},
        title = "{Simultaneous Ultraviolet and X-Ray Observations of the Seyfert Galaxy NGC 4151. II. Physical Conditions in the UV Absorbers}",
      journal = {\apjs},
         year = 2006,
        month = dec,
       volume = {167},
       number = {2},
        pages = {161-176},
          doi = {10.1086/508629},
archivePrefix = {arXiv},
       eprint = {astro-ph/0608383},
 primaryClass = {astro-ph},
       adsurl = {https://ui.adsabs.harvard.edu/abs/2006ApJS..167..161K}
}

@ARTICLE{Schurch2004,
       author = {{Schurch}, N.~J. and {Warwick}, R.~S. and {Griffiths}, R.~E. and {Kahn}, S.~M.},
        title = "{The complex soft X-ray spectrum of NGC 4151}",
      journal = {\mnras},
         year = 2004,
        month = may,
       volume = {350},
       number = {1},
        pages = {1-9},
          doi = {10.1111/j.1365-2966.2004.07632.x},
archivePrefix = {arXiv},
       eprint = {astro-ph/0401550},
 primaryClass = {astro-ph},
       adsurl = {https://ui.adsabs.harvard.edu/abs/2004MNRAS.350....1S}
}

@ARTICLE{Mehdipour2019,
       author = {{Mehdipour}, Missagh and {Costantini}, Elisa},
        title = "{Relation between winds and jets in radio-loud AGN}",
      journal = {\aap},
         year = 2019,
        month = may,
       volume = {625},
          eid = {A25},
        pages = {A25},
          doi = {10.1051/0004-6361/201935205},
archivePrefix = {arXiv},
       eprint = {1903.11605},
 primaryClass = {astro-ph.HE},
       adsurl = {https://ui.adsabs.harvard.edu/abs/2019A&A...625A..25M}
}

@ARTICLE{Ulvestad2005,
       author = {{Ulvestad}, James S. and {Wong}, Diane S. and {Taylor}, Gregory B. and {Gallimore}, Jack F. and {Mundell}, Carole G.},
        title = "{VLBA Identification of the Milliarcsecond Active Nucleus in the Seyfert Galaxy NGC 4151}",
      journal = {\aj},
         year = 2005,
        month = sep,
       volume = {130},
       number = {3},
        pages = {936-944},
          doi = {10.1086/432034},
archivePrefix = {arXiv},
       eprint = {astro-ph/0505141},
 primaryClass = {astro-ph},
       adsurl = {https://ui.adsabs.harvard.edu/abs/2005AJ....130..936U}
}

@ARTICLE{Pedlar1992,
       author = {{Pedlar}, A. and {Howley}, P. and {Axon}, D.~J. and {Unger}, S.~W.},
        title = "{A neutral hydrogen study of NGC 4151.}",
      journal = {\mnras},
         year = 1992,
        month = nov,
       volume = {259},
        pages = {369-380},
          doi = {10.1093/mnras/259.2.369},
       adsurl = {https://ui.adsabs.harvard.edu/abs/1992MNRAS.259..369P}
}

@ARTICLE{Laing1988,
       author = {{Laing}, R.~A.},
        title = "{The sidedness of jets and depolarization in powerful extragalactic radio sources}",
      journal = {\nat},
         year = 1988,
        month = jan,
       volume = {331},
       number = {6152},
        pages = {149-151},
          doi = {10.1038/331149a0},
       adsurl = {https://ui.adsabs.harvard.edu/abs/1988Natur.331..149L}
}

@ARTICLE{Garrington1988,
       author = {{Garrington}, S.~T. and {Leahy}, J.~P. and {Conway}, R.~G. and {Laing}, R.~A.},
        title = "{A systematic asymmetry in the polarization properties of double radio sources with one jet}",
      journal = {\nat},
         year = 1988,
        month = jan,
       volume = {331},
       number = {6152},
        pages = {147-149},
          doi = {10.1038/331147a0},
       adsurl = {https://ui.adsabs.harvard.edu/abs/1988Natur.331..147G}
}

@ARTICLE{Mundell1999b,
       author = {{Mundell}, C.~G. and {Pedlar}, A. and {Shone}, D.~L. and {Robinson}, A.},
        title = "{Gas dynamics in the barred Seyfert galaxy NGC 4151 - II. High-resolution HI study}",
      journal = {\mnras},
         year = 1999,
        month = apr,
       volume = {304},
       number = {3},
        pages = {481-494},
          doi = {10.1046/j.1365-8711.1999.02331.x},
archivePrefix = {arXiv},
       eprint = {astro-ph/9812183},
 primaryClass = {astro-ph},
       adsurl = {https://ui.adsabs.harvard.edu/abs/1999MNRAS.304..481M}
}

@ARTICLE{Peretti2025,
       author = {{Peretti}, E. and {Peron}, G. and {Tombesi}, F. and {Lamastra}, A. and {Saturni}, F.~G. and {Cerruti}, M. and {Ahlers}, M.},
        title = "{Gamma-ray emission from the Seyfert galaxy NGC 4151: multi-messenger implications for ultra-fast outflows}",
      journal = {\jcap},
         year = 2025,
        month = jul,
       volume = {2025},
       number = {7},
          eid = {013},
        pages = {013},
          doi = {10.1088/1475-7516/2025/07/013},
archivePrefix = {arXiv},
       eprint = {2303.03298},
 primaryClass = {astro-ph.HE},
       adsurl = {https://ui.adsabs.harvard.edu/abs/2025JCAP...07..013P}
}

@ARTICLE{Mundell2003,
       author = {{Mundell}, C.~G. and {Wrobel}, J.~M. and {Pedlar}, A. and {Gallimore}, J.~F.},
        title = "{The Nuclear Regions of the Seyfert Galaxy NGC 4151: Parsec-Scale H I Absorption and a Remarkable Radio Jet}",
      journal = {\apj},
         year = 2003,
        month = jan,
       volume = {583},
       number = {1},
        pages = {192-204},
          doi = {10.1086/345356},
archivePrefix = {arXiv},
       eprint = {astro-ph/0209540},
 primaryClass = {astro-ph},
       adsurl = {https://ui.adsabs.harvard.edu/abs/2003ApJ...583..192M}
}

@ARTICLE{Gaensler2008,
       author = {{Gaensler}, B.~M. and {Madsen}, G.~J. and {Chatterjee}, S. and {Mao}, S.~A.},
        title = "{The Vertical Structure of Warm Ionised Gas in the Milky Way}",
      journal = {\pasa},
         year = 2008,
        month = nov,
       volume = {25},
       number = {4},
        pages = {184-200},
          doi = {10.1071/AS08004},
archivePrefix = {arXiv},
       eprint = {0808.2550},
 primaryClass = {astro-ph},
       adsurl = {https://ui.adsabs.harvard.edu/abs/2008PASA...25..184G}
}

@ARTICLE{Lisakov2021,
       author = {{Lisakov}, M.~M. and {Kravchenko}, E.~V. and {Pushkarev}, A.~B. and {Kovalev}, Y.~Y. and {Savolainen}, T.~K. and {Lister}, M.~L.},
        title = "{An Oversized Magnetic Sheath Wrapping around the Parsec-scale Jet in 3C 273}",
      journal = {\apj},
         year = 2021,
        month = mar,
       volume = {910},
       number = {1},
          eid = {35},
        pages = {35},
          doi = {10.3847/1538-4357/abe1bd},
archivePrefix = {arXiv},
       eprint = {2102.04563},
 primaryClass = {astro-ph.HE},
       adsurl = {https://ui.adsabs.harvard.edu/abs/2021ApJ...910...35L}
}

@ARTICLE{Cordes2002,
       author = {{Cordes}, J.~M. and {Lazio}, T.~J.~W.},
        title = "{NE2001.I. A New Model for the Galactic Distribution of Free Electrons and its Fluctuations}",
      journal = {arXiv e-prints},
         year = 2002,
        month = jul,
          eid = {astro-ph/0207156},
        pages = {astro-ph/0207156},
          doi = {10.48550/arXiv.astro-ph/0207156},
archivePrefix = {arXiv},
       eprint = {astro-ph/0207156},
 primaryClass = {astro-ph},
       adsurl = {https://ui.adsabs.harvard.edu/abs/2002astro.ph..7156C}
}

@INPROCEEDINGS{Mundell2001,
       author = {{Mundell}, C.~G. and {Wrobel}, J.~M. and {Pedlar}, A. and {Gallimore}, J.~F.},
        title = "{Subparsec-scale HI in the nucleus of NGC 4151}",
    booktitle = {Galaxies and their Constituents at the Highest Angular Resolutions},
         year = 2001,
       editor = {{Schilizzi}, R.~T.},
       series = {IAU Symposium},
       volume = {205},
        month = jan,
        pages = {192},
          doi = {10.48550/arXiv.astro-ph/0012229},
archivePrefix = {arXiv},
       eprint = {astro-ph/0012229},
 primaryClass = {astro-ph},
       adsurl = {https://ui.adsabs.harvard.edu/abs/2001IAUS..205..192M}
}

@ARTICLE{Mundell1995,
       author = {{Mundell}, C.~G. and {Pedlar}, A. and {Baum}, S.~A. and {O'Dea}, C.~P. and {Gallimore}, J.~F. and {Brinks}, E.},
        title = "{MERLIN observations of neutral hydrogen absorption in the Seyfert nucleus of NGC 4151}",
      journal = {\mnras},
         year = 1995,
        month = jan,
       volume = {272},
       number = {2},
        pages = {355-362},
          doi = {10.1093/mnras/272.2.355},
       adsurl = {https://ui.adsabs.harvard.edu/abs/1995MNRAS.272..355M}
}

@ARTICLE{Feng2024,
       author = {{Feng}, Hai-Cheng and {Li}, Sha-Sha and {Bai}, J.~M. and {Liu}, H.~T. and {Lu}, Kai-Xing and {Pang}, Yu-Xuan and {Sun}, Mouyuan and {Wang}, Jian-Guo and {Zhang}, Yang-Wei and {Zhou}, Shuying},
        title = "{Velocity-resolved Reverberation Mapping of Changing-look Active Galactic Nucleus NGC 4151 during Outburst Stage. II. Results of Four Seasons of Observation}",
      journal = {\apj},
         year = 2024,
        month = dec,
       volume = {976},
       number = {2},
          eid = {176},
        pages = {176},
          doi = {10.3847/1538-4357/ad8568},
archivePrefix = {arXiv},
       eprint = {2409.01637},
 primaryClass = {astro-ph.GA},
       adsurl = {https://ui.adsabs.harvard.edu/abs/2024ApJ...976..176F}
}

@ARTICLE{Tombesi2011,
       author = {{Tombesi}, F. and {Cappi}, M. and {Reeves}, J.~N. and {Palumbo}, G.~G.~C. and {Braito}, V. and {Dadina}, M.},
        title = "{Evidence for Ultra-fast Outflows in Radio-quiet Active Galactic Nuclei. II. Detailed Photoionization Modeling of Fe K-shell Absorption Lines}",
      journal = {\apj},
         year = 2011,
        month = nov,
       volume = {742},
       number = {1},
          eid = {44},
        pages = {44},
          doi = {10.1088/0004-637X/742/1/44},
archivePrefix = {arXiv},
       eprint = {1109.2882},
 primaryClass = {astro-ph.HE},
       adsurl = {https://ui.adsabs.harvard.edu/abs/2011ApJ...742...44T}
}

@ARTICLE{Kochanek1990,
       author = {{Kochanek}, Christopher S. and {Hawley}, John F.},
        title = "{Hollow Conical Jet Models for SS 433: A Paradigm Lost?}",
      journal = {\apj},
         year = 1990,
        month = feb,
       volume = {350},
        pages = {561},
          doi = {10.1086/168411},
       adsurl = {https://ui.adsabs.harvard.edu/abs/1990ApJ...350..561K}
}

@ARTICLE{Blundell2005,
       author = {{Blundell}, Katherine M. and {Bowler}, Michael G.},
        title = "{Jet Velocity in SS 433: Its Anticorrelation with Precession-Cone Angle and Dependence on Orbital Phase}",
      journal = {\apjl},
         year = 2005,
        month = apr,
       volume = {622},
       number = {2},
        pages = {L129-L132},
          doi = {10.1086/429663},
archivePrefix = {arXiv},
       eprint = {astro-ph/0410457},
 primaryClass = {astro-ph},
       adsurl = {https://ui.adsabs.harvard.edu/abs/2005ApJ...622L.129B}
}

@ARTICLE{Romanova2009,
       author = {{Romanova}, M.~M. and {Ustyugova}, G.~V. and {Koldoba}, A.~V. and {Lovelace}, R.~V.~E.},
        title = "{Launching of conical winds and axial jets from the disc-magnetosphere boundary: axisymmetric and 3D simulations}",
      journal = {\mnras},
         year = 2009,
        month = nov,
       volume = {399},
       number = {4},
        pages = {1802-1828},
          doi = {10.1111/j.1365-2966.2009.15413.x},
archivePrefix = {arXiv},
       eprint = {0907.3394},
 primaryClass = {astro-ph.SR},
       adsurl = {https://ui.adsabs.harvard.edu/abs/2009MNRAS.399.1802R}
}

@ARTICLE{Laing1980,
       author = {{Laing}, R.~A.},
        title = "{A model for the magnetic-field structure in extended radio sources.}",
      journal = {\mnras},
         year = 1980,
        month = nov,
       volume = {193},
        pages = {439-449},
          doi = {10.1093/mnras/193.3.439},
       adsurl = {https://ui.adsabs.harvard.edu/abs/1980MNRAS.193..439L}
}

@ARTICLE{Zakamska2014,
       author = {{Zakamska}, Nadia L. and {Greene}, Jenny E.},
        title = "{Quasar feedback and the origin of radio emission in radio-quiet quasars}",
      journal = {\mnras},
         year = 2014,
        month = jul,
       volume = {442},
       number = {1},
        pages = {784-804},
          doi = {10.1093/mnras/stu842},
archivePrefix = {arXiv},
       eprint = {1402.6736},
 primaryClass = {astro-ph.GA},
       adsurl = {https://ui.adsabs.harvard.edu/abs/2014MNRAS.442..784Z}
}

@ARTICLE{Crenshaw2007,
       author = {{Crenshaw}, D.~M. and {Kraemer}, S.~B.},
        title = "{Mass Outflow from the Nucleus of the Seyfert 1 Galaxy NGC 4151}",
      journal = {\apj},
         year = 2007,
        month = apr,
       volume = {659},
       number = {1},
        pages = {250-256},
          doi = {10.1086/511970},
archivePrefix = {arXiv},
       eprint = {astro-ph/0612446},
 primaryClass = {astro-ph},
       adsurl = {https://ui.adsabs.harvard.edu/abs/2007ApJ...659..250C}
}

@ARTICLE{Williams2020,
       author = {{Williams}, D.~R.~A. and {Baldi}, R.~D. and {McHardy}, I.~M. and {Beswick}, R.~J. and {Panessa}, F. and {May}, D. and {Mold{\'o}n}, J. and {Argo}, M.~K. and {Bruni}, G. and {Dullo}, B.~T. and {Knapen}, J.~H. and {Brinks}, E. and {Fenech}, D.~M. and {Mundell}, C.~G. and {Muxlow}, T.~W.~B. and {Pahari}, M. and {Westcott}, J.},
        title = "{The curious activity in the nucleus of NGC 4151: jet interaction causing variability?}",
      journal = {\mnras},
         year = 2020,
        month = jul,
       volume = {495},
       number = {3},
        pages = {3079-3086},
          doi = {10.1093/mnras/staa1152},
archivePrefix = {arXiv},
       eprint = {2004.10552},
 primaryClass = {astro-ph.GA},
       adsurl = {https://ui.adsabs.harvard.edu/abs/2020MNRAS.495.3079W}
}

@INPROCEEDINGS{Rosa2006,
       author = {{La Rosa}, T.~N. and {Shore}, S.~N. and {Joseph}, T. and {Lazio}, W. and {Kassim}, Namir E.},
        title = "{The Strength and Structure of the Galactic Center Magnetic Field}",
    booktitle = {Journal of Physics Conference Series},
         year = 2006,
       editor = {{Sch{\"o}del}, Rainer and {Bower}, Geoffrey C. and {Muno}, Michael P. and {Nayakshin}, Sergei and {Ott}, Thomas},
       series = {Journal of Physics Conference Series},
       volume = {54},
        month = dec,
    publisher = {IOP},
        pages = {10-15},
          doi = {10.1088/1742-6596/54/1/002},
       adsurl = {https://ui.adsabs.harvard.edu/abs/2006JPhCS..54...10L}
}

@ARTICLE{Shapovalova2010,
       author = {{Shapovalova}, A.~I. and {Popovi{\'c}}, L. {\v{C}}. and {Burenkov}, A.~N. and {Chavushyan}, V.~H. and {Ili{\'c}}, D. and {Kova{\v{c}}evi{\'c}}, A. and {Bochkarev}, N.~G. and {Le{\'o}n-Tavares}, J.},
        title = "{Long-term variability of the optical spectra of NGC 4151. II. Evolution of the broad H{\ensuremath{\alpha}} and H{\ensuremath{\beta}} emission-line profiles}",
      journal = {\aap},
         year = 2010,
        month = jan,
       volume = {509},
          eid = {A106},
        pages = {A106},
          doi = {10.1051/0004-6361/200912311},
archivePrefix = {arXiv},
       eprint = {0910.2980},
 primaryClass = {astro-ph.GA},
       adsurl = {https://ui.adsabs.harvard.edu/abs/2010A&A...509A.106S}
}

@ARTICLE{Ruiz2003,
       author = {{Ruiz}, M. and {Young}, S. and {Packham}, C. and {Alexander}, D.~M. and {Hough}, J.~H.},
        title = "{Near-infrared imaging polarimetry and modelling of NGC 4151}",
      journal = {\mnras},
         year = 2003,
        month = apr,
       volume = {340},
       number = {3},
        pages = {733-738},
          doi = {10.1046/j.1365-8711.2003.06239.x},
       adsurl = {https://ui.adsabs.harvard.edu/abs/2003MNRAS.340..733R}
}

@ARTICLE{Feain2009,
       author = {{Feain}, I.~J. and {Ekers}, R.~D. and {Murphy}, T. and {Gaensler}, B.~M. and {Macquart}, J. -P. and {Norris}, R.~P. and {Cornwell}, T.~J. and {Johnston-Hollitt}, M. and {Ott}, J. and {Middelberg}, E.},
        title = "{Faraday Rotation Structure on Kiloparsec Scales in the Radio Lobes of Centaurus A}",
      journal = {\apj},
         year = 2009,
        month = dec,
       volume = {707},
       number = {1},
        pages = {114-125},
          doi = {10.1088/0004-637X/707/1/114},
archivePrefix = {arXiv},
       eprint = {0910.3458},
 primaryClass = {astro-ph.CO},
       adsurl = {https://ui.adsabs.harvard.edu/abs/2009ApJ...707..114F}
}

@INPROCEEDINGS{Proga2007,
       author = {{Proga}, D.},
        title = "{Theory of Winds in AGNs}",
    booktitle = {The Central Engine of Active Galactic Nuclei},
         year = 2007,
       editor = {{Ho}, L.~C. and {Wang}, J. -W.},
       series = {Astronomical Society of the Pacific Conference Series},
       volume = {373},
        month = oct,
        pages = {267},
          doi = {10.48550/arXiv.astro-ph/0701100},
archivePrefix = {arXiv},
       eprint = {astro-ph/0701100},
 primaryClass = {astro-ph},
       adsurl = {https://ui.adsabs.harvard.edu/abs/2007ASPC..373..267P}
}

@ARTICLE{Kharb2009,
       author = {{Kharb}, P. and {Gabuzda}, D.~C. and {O'Dea}, C.~P. and {Shastri}, P. and {Baum}, S.~A.},
        title = "{Rotation Measures Across Parsec-Scale Jets of Fanaroff-Riley Type I Radio Galaxies}",
      journal = {\apj},
         year = 2009,
        month = apr,
       volume = {694},
       number = {2},
        pages = {1485-1497},
          doi = {10.1088/0004-637X/694/2/1485},
archivePrefix = {arXiv},
       eprint = {0901.0913},
 primaryClass = {astro-ph.GA},
       adsurl = {https://ui.adsabs.harvard.edu/abs/2009ApJ...694.1485K}
}

@ARTICLE{Wang2011a,
       author = {{Wang}, Junfeng and {Fabbiano}, Giuseppina and {Elvis}, Martin and {Risaliti}, Guido and {Mundell}, Carole G. and {Karovska}, Margarita and {Zezas}, Andreas},
        title = "{A Deep Chandra ACIS Study of NGC 4151. II. The Innermost Emission Line Region and Strong Evidence for Radio Jet-NLR Cloud Collision}",
      journal = {\apj},
         year = 2011,
        month = jul,
       volume = {736},
       number = {1},
          eid = {62},
        pages = {62},
          doi = {10.1088/0004-637X/736/1/62},
archivePrefix = {arXiv},
       eprint = {1103.1912},
 primaryClass = {astro-ph.CO},
       adsurl = {https://ui.adsabs.harvard.edu/abs/2011ApJ...736...62W}
}

@ARTICLE{Wang2010,
       author = {{Wang}, Junfeng and {Fabbiano}, Giuseppina and {Risaliti}, Guido and {Elvis}, Martin and {Mundell}, Carole G. and {Dumas}, Gaelle and {Schinnerer}, Eva and {Zezas}, Andreas},
        title = "{Extended X-ray Emission in the H I Cavity of NGC 4151: Galaxy-scale Active Galactic Nucleus Feedback?}",
      journal = {\apjl},
         year = 2010,
        month = aug,
       volume = {719},
       number = {2},
        pages = {L208-L212},
          doi = {10.1088/2041-8205/719/2/L208},
archivePrefix = {arXiv},
       eprint = {1007.4472},
 primaryClass = {astro-ph.GA},
       adsurl = {https://ui.adsabs.harvard.edu/abs/2010ApJ...719L.208W}
}

@BOOK{Longair2011,
       author = {{Longair}, Malcolm S.},
        title = "{High Energy Astrophysics}",
         year = 2011,
       adsurl = {https://ui.adsabs.harvard.edu/abs/2011hea..book.....L}
}

@ARTICLE{Kraemer2020,
       author = {{Kraemer}, S.~B. and {Turner}, T.~J. and {Couto}, J.~D. and {Crenshaw}, D.~M. and {Schmitt}, H.~R. and {Revalski}, M. and {Fischer}, T.~C.},
        title = "{Mass outflow of the X-ray emission line gas in NGC 4151}",
      journal = {\mnras},
         year = 2020,
        month = apr,
       volume = {493},
       number = {3},
        pages = {3893-3910},
          doi = {10.1093/mnras/staa428},
archivePrefix = {arXiv},
       eprint = {2002.05806},
 primaryClass = {astro-ph.GA},
       adsurl = {https://ui.adsabs.harvard.edu/abs/2020MNRAS.493.3893K}
}

@ARTICLE{Uchida1985,
       author = {{Uchida}, Y. and {Shibata}, K.},
        title = "{Magnetodynamical acceleration of CO and optical bipolar flows from the region of star formation.}",
      journal = {\pasj},
         year = 1985,
        month = jan,
       volume = {37},
        pages = {515-535},
       adsurl = {https://ui.adsabs.harvard.edu/abs/1985PASJ...37..515U}
}

@ARTICLE{Pudritz1986,
       author = {{Pudritz}, R.~E. and {Norman}, C.~A.},
        title = "{Bipolar Hydromagnetic Winds from Disks around Protostellar Objects}",
      journal = {\apj},
         year = 1986,
        month = feb,
       volume = {301},
        pages = {571},
          doi = {10.1086/163924},
       adsurl = {https://ui.adsabs.harvard.edu/abs/1986ApJ...301..571P}
}

@ARTICLE{Contopoulos1995,
       author = {{Contopoulos}, J.},
        title = "{A Simple Type of Magnetically Driven Jets: an Astrophysical Plasma Gun}",
      journal = {\apj},
         year = 1995,
        month = sep,
       volume = {450},
        pages = {616},
          doi = {10.1086/176170},
       adsurl = {https://ui.adsabs.harvard.edu/abs/1995ApJ...450..616C}
}

@ARTICLE{Faucher2012,
       author = {{Faucher-Gigu{\`e}re}, Claude-Andr{\'e} and {Quataert}, Eliot},
        title = "{The physics of galactic winds driven by active galactic nuclei}",
      journal = {\mnras},
         year = 2012,
        month = sep,
       volume = {425},
       number = {1},
        pages = {605-622},
          doi = {10.1111/j.1365-2966.2012.21512.x},
archivePrefix = {arXiv},
       eprint = {1204.2547},
 primaryClass = {astro-ph.CO},
       adsurl = {https://ui.adsabs.harvard.edu/abs/2012MNRAS.425..605F}
}

@ARTICLE{Vleugels2025,
       author = {{Vleugels}, C. and {McClure}, M. and {Sturm}, A. and {Vlasblom}, M.},
        title = "{The H$_{2}$ jet and disk wind of the Class I protostar HOPS 315}",
      journal = {\aap},
         year = 2025,
        month = mar,
       volume = {695},
          eid = {A145},
        pages = {A145},
          doi = {10.1051/0004-6361/202452475},
       adsurl = {https://ui.adsabs.harvard.edu/abs/2025A&A...695A.145V}
}

@ARTICLE{Fiore2017,
       author = {{Fiore}, F. and {Feruglio}, C. and {Shankar}, F. and {Bischetti}, M. and {Bongiorno}, A. and {Brusa}, M. and {Carniani}, S. and {Cicone}, C. and {Duras}, F. and {Lamastra}, A. and {Mainieri}, V. and {Marconi}, A. and {Menci}, N. and {Maiolino}, R. and {Piconcelli}, E. and {Vietri}, G. and {Zappacosta}, L.},
        title = "{AGN wind scaling relations and the co-evolution of black holes and galaxies}",
      journal = {\aap},
         year = 2017,
        month = may,
       volume = {601},
          eid = {A143},
        pages = {A143},
          doi = {10.1051/0004-6361/201629478},
archivePrefix = {arXiv},
       eprint = {1702.04507},
 primaryClass = {astro-ph.GA},
       adsurl = {https://ui.adsabs.harvard.edu/abs/2017A&A...601A.143F}
}

@ARTICLE{Perucho2017,
       author = {{Perucho}, M. and {Bosch-Ramon}, V. and {Barkov}, M.~V.},
        title = "{Impact of red giant/AGB winds on active galactic nucleus jet propagation}",
      journal = {\aap},
         year = 2017,
        month = oct,
       volume = {606},
          eid = {A40},
        pages = {A40},
          doi = {10.1051/0004-6361/201630117},
archivePrefix = {arXiv},
       eprint = {1706.06301},
 primaryClass = {astro-ph.HE},
       adsurl = {https://ui.adsabs.harvard.edu/abs/2017A&A...606A..40P}
}

@ARTICLE{Gianolli2023,
       author = {{Gianolli}, V.~E. and {Kim}, D.~E. and {Bianchi}, S. and {Ag{\'\i}s-Gonz{\'a}lez}, B. and {Madejski}, G. and {Marin}, F. and {Marinucci}, A. and {Matt}, G. and {Middei}, R. and {Petrucci}, P. -O. and {Soffitta}, P. and {Tagliacozzo}, D. and {Tombesi}, F. and {Ursini}, F. and {Barnouin}, T. and {De Rosa}, A. and {Di Gesu}, L. and {Ingram}, A. and {Loktev}, V. and {Panagiotou}, C. and {Podgorny}, J. and {Poutanen}, J. and {Puccetti}, S. and {Ratheesh}, A. and {Veledina}, A. and {Zhang}, W. and {Agudo}, I. and {Antonelli}, L.~A. and {Bachetti}, M. and {Baldini}, L. and {Baumgartner}, W.~H. and {Bellazzini}, R. and {Bongiorno}, S.~D. and {Bonino}, R. and {Brez}, A. and {Bucciantini}, N. and {Capitanio}, F. and {Castellano}, S. and {Cavazzuti}, E. and {Chen}, C. -T. and {Ciprini}, S. and {Costa}, E. and {Del Monte}, E. and {Di Lalla}, N. and {Di Marco}, A. and {Donnarumma}, I. and {Doroshenko}, V. and {Dov{\v{c}}iak}, M. and {Ehlert}, S.~R. and {Enoto}, T. and {Evangelista}, Y. and {Fabiani}, S. and {Ferrazzoli}, R. and {Garc{\'\i}a}, J.~A. and {Gunji}, S. and {Heyl}, J. and {Iwakiri}, W. and {Jorstad}, S.~G. and {Kaaret}, P. and {Karas}, V. and {Kislat}, F. and {Kitaguchi}, T. and {Kolodziejczak}, J.~J. and {Krawczynski}, H. and {La Monaca}, F. and {Latronico}, L. and {Liodakis}, I. and {Maldera}, S. and {Manfreda}, A. and {Marscher}, A.~P. and {Marshall}, H.~L. and {Massaro}, F. and {Mitsuishi}, I. and {Mizuno}, T. and {Muleri}, F. and {Negro}, M. and {Ng}, C. -Y. and {O'Dell}, S.~L. and {Omodei}, N. and {Oppedisano}, C. and {Papitto}, A. and {Pavlov}, G.~G. and {Peirson}, A.~L. and {Perri}, M. and {Pesce-Rollins}, M. and {Pilia}, M. and {Possenti}, A. and {Ramsey}, B.~D. and {Rankin}, J. and {Roberts}, O.~J. and {Romani}, R.~W. and {Sgr{\`o}}, C. and {Slane}, P. and {Spandre}, G. and {Swartz}, D.~A. and {Tamagawa}, T. and {Tavecchio}, F. and {Taverna}, R. and {Tawara}, Y. and {Tennant}, A.~F. and {Thomas}, N.~E. and {Trois}, A. and {Tsygankov}, S.~S. and {Turolla}, R. and {Vink}, J. and {Weisskopf}, M.~C. and {Wu}, K. and {Xie}, F. and {Zane}, S.},
        title = "{Uncovering the geometry of the hot X-ray corona in the Seyfert galaxy NGC 4151 with IXPE}",
      journal = {\mnras},
         year = 2023,
        month = aug,
       volume = {523},
       number = {3},
        pages = {4468-4476},
          doi = {10.1093/mnras/stad1697},
archivePrefix = {arXiv},
       eprint = {2303.12541},
 primaryClass = {astro-ph.GA},
       adsurl = {https://ui.adsabs.harvard.edu/abs/2023MNRAS.523.4468G}
}

@ARTICLE{Bhatnagar2013,
       author = {{Bhatnagar}, S. and {Rau}, U. and {Golap}, K.},
        title = "{Wide-field wide-band Interferometric Imaging: The WB A-Projection and Hybrid Algorithms}",
      journal = {\apj},
         year = 2013,
        month = jun,
       volume = {770},
       number = {2},
          eid = {91},
        pages = {91},
          doi = {10.1088/0004-637X/770/2/91},
archivePrefix = {arXiv},
       eprint = {1304.4987},
 primaryClass = {astro-ph.IM},
       adsurl = {https://ui.adsabs.harvard.edu/abs/2013ApJ...770...91B}
}

@ARTICLE{Johnston1982,
       author = {{Johnston}, K.~J. and {Elvis}, M. and {Kjer}, D. and {Shen}, B.~S.~P.},
        title = "{Radio jets in NGC 4151.}",
      journal = {\apj},
         year = 1982,
        month = nov,
       volume = {262},
        pages = {61-65},
          doi = {10.1086/160396},
       adsurl = {https://ui.adsabs.harvard.edu/abs/1982ApJ...262...61J}
}

@ARTICLE{Ghosh2025b,
       author = {{Ghosh}, Salmoli and {Kharb}, Preeti and {Sajjanhar}, Esha and {Pasetto}, Alice and {Sebastian}, Biny},
        title = "{Magnetic Field in the Lobes of the Seyfert Galaxy NGC 3516: Suggestions of a Helical Field}",
      journal = {\apj},
         year = 2025,
        month = aug,
       volume = {989},
       number = {1},
          eid = {40},
        pages = {40},
          doi = {10.3847/1538-4357/ade98d},
archivePrefix = {arXiv},
       eprint = {2506.11854},
 primaryClass = {astro-ph.GA},
       adsurl = {https://ui.adsabs.harvard.edu/abs/2025ApJ...989...40G}
}

@ARTICLE{Williams2017,
       author = {{Williams}, D.~R.~A. and {McHardy}, I.~M. and {Baldi}, R.~D. and {Beswick}, R.~J. and {Argo}, M.~K. and {Dullo}, B.~T. and {Knapen}, J.~H. and {Brinks}, E. and {Fenech}, D.~M. and {Mundell}, C.~G. and {Muxlow}, T.~W.~B. and {Panessa}, F. and {Rampadarath}, H. and {Westcott}, J.},
        title = "{Radio jets in NGC 4151: where eMERLIN meets HST}",
      journal = {\mnras},
         year = 2017,
        month = dec,
       volume = {472},
       number = {4},
        pages = {3842-3853},
          doi = {10.1093/mnras/stx2205},
archivePrefix = {arXiv},
       eprint = {1708.07011},
 primaryClass = {astro-ph.GA},
       adsurl = {https://ui.adsabs.harvard.edu/abs/2017MNRAS.472.3842W}
}

@ARTICLE{Booler1982,
       author = {{Booler}, R.~V. and {Pedlar}, A. and {Davies}, R.~D.},
        title = "{High-resolution 1666-MHz observqations of the nucleus of NGC 4151.}",
      journal = {\mnras},
         year = 1982,
        month = apr,
       volume = {199},
        pages = {229-237},
          doi = {10.1093/mnras/199.2.229},
       adsurl = {https://ui.adsabs.harvard.edu/abs/1982MNRAS.199..229B}
}

@ARTICLE{Hovatta2012,
       author = {{Hovatta}, Talvikki and {Lister}, Matthew L. and {Aller}, Margo F. and {Aller}, Hugh D. and {Homan}, Daniel C. and {Kovalev}, Yuri Y. and {Pushkarev}, Alexander B. and {Savolainen}, Tuomas},
        title = "{MOJAVE: Monitoring of Jets in Active Galactic Nuclei with VLBA Experiments. VIII. Faraday Rotation in Parsec-scale AGN Jets}",
      journal = {\aj},
         year = 2012,
        month = oct,
       volume = {144},
       number = {4},
          eid = {105},
        pages = {105},
          doi = {10.1088/0004-6256/144/4/105},
archivePrefix = {arXiv},
       eprint = {1205.6746},
 primaryClass = {astro-ph.CO},
       adsurl = {https://ui.adsabs.harvard.edu/abs/2012AJ....144..105H}
}

@ARTICLE{Pasetto2021,
       author = {{Pasetto}, Alice and {Carrasco-Gonz{\'a}lez}, Carlos and {G{\'o}mez}, Jos{\'e} L. and {Mart{\'\i}}, Jos{\'e}-Maria and {Perucho}, Manel and {O'Sullivan}, Shane P. and {Anderson}, Craig and {D{\'\i}az-Gonz{\'a}lez}, Daniel Jacobo and {Fuentes}, Antonio and {Wardle}, John},
        title = "{Reading M87's DNA: A Double Helix Revealing a Large-scale Helical Magnetic Field}",
      journal = {\apjl},
         year = 2021,
        month = dec,
       volume = {923},
       number = {1},
          eid = {L5},
        pages = {L5},
          doi = {10.3847/2041-8213/ac3a88},
archivePrefix = {arXiv},
       eprint = {2112.06971},
 primaryClass = {astro-ph.GA},
       adsurl = {https://ui.adsabs.harvard.edu/abs/2021ApJ...923L...5P}
}

@ARTICLE{Asada2002,
       author = {{Asada}, Keiichi and {Inoue}, Makoto and {Uchida}, Yutaka and {Kameno}, Seiji and {Fujisawa}, Kenta and {Iguchi}, Satoru and {Mutoh}, Mutsumi},
        title = "{A Helical Magnetic Field in the Jet of 3C 273}",
      journal = {\pasj},
         year = 2002,
        month = jun,
       volume = {54},
        pages = {L39-L43},
          doi = {10.1093/pasj/54.3.L39},
archivePrefix = {arXiv},
       eprint = {astro-ph/0205497},
 primaryClass = {astro-ph},
       adsurl = {https://ui.adsabs.harvard.edu/abs/2002PASJ...54L..39A}
}

@BOOK{Pacholczyk1970,
       author = {{Pacholczyk}, A.~G.},
        title = "{Radio astrophysics. Nonthermal processes in galactic and extragalactic sources}",
         year = 1970,
       adsurl = {https://ui.adsabs.harvard.edu/abs/1970ranp.book.....P}
}

@ARTICLE{Sullivan2013a,
       author = {{O'Sullivan}, S.~P. and {Feain}, I.~J. and {McClure-Griffiths}, N.~M. and {Ekers}, R.~D. and {Carretti}, E. and {Robishaw}, T. and {Mao}, S.~A. and {Gaensler}, B.~M. and {Bland-Hawthorn}, J. and {Stawarz}, {\L}.},
        title = "{Thermal Plasma in the Giant Lobes of the Radio Galaxy Centaurus A}",
      journal = {\apj},
         year = 2013,
        month = feb,
       volume = {764},
       number = {2},
          eid = {162},
        pages = {162},
          doi = {10.1088/0004-637X/764/2/162},
       adsurl = {https://ui.adsabs.harvard.edu/abs/2013ApJ...764..162O}
}

@ARTICLE{Burn1966,
       author = {{Burn}, B.~J.},
        title = "{On the depolarization of discrete radio sources by Faraday dispersion}",
      journal = {\mnras},
         year = 1966,
        month = jan,
       volume = {133},
        pages = {67},
          doi = {10.1093/mnras/133.1.67},
       adsurl = {https://ui.adsabs.harvard.edu/abs/1966MNRAS.133...67B}
}

@ARTICLE{Baum1993,
       author = {{Baum}, S.~A. and {O'Dea}, C.~P. and {Dallacassa}, D. and {de Bruyn}, A.~G. and {Pedlar}, A.},
        title = "{Kiloparsec-Scale Radio Emission in Seyfert Galaxies: Evidence for Starburst-driven Superwinds?}",
      journal = {\apj},
         year = 1993,
        month = dec,
       volume = {419},
        pages = {553},
          doi = {10.1086/173508},
       adsurl = {https://ui.adsabs.harvard.edu/abs/1993ApJ...419..553B}
}

@ARTICLE{Ulvestad1998,
   author = {{Ulvestad}, J.~S. and {Roy}, A.~L. and {Colbert}, E.~J.~M. and 
	{Wilson}, A.~S.},
    title = "{A Subparsec Radio Jet or Disk in NGC 4151}",
  journal = {\apj},
     year = 1998,
    month = mar,
   volume = 496,
    pages = {196},
      doi = {10.1086/305382},
   adsurl = {http://adsabs.harvard.edu/cgi-bin/nph-bib_query?bibcode=1998ApJ...496..196U&db_key=AST}
}

@ARTICLE{UrryPadovani1995,
   author = {{Urry}, C.~M. and {Padovani}, P.},
    title = "{Unified Schemes for Radio-Loud Active Galactic Nuclei}",
  journal = {\pasp},
     year = 1995,
    month = sep,
   volume = 107,
    pages = {803},
   adsurl = {http://adsabs.harvard.edu/cgi-bin/nph-bib_query?bibcode=1995PASP..107..803U&db_key=AST}
}

@ARTICLE{vanBreugel1984,
   author = {{van Breugel}, W. and {Fomalont}, E.~B.},
    title = "{Is 3C 310 blowing bubbles?}",
  journal = {\apjl},
     year = 1984,
    month = jul,
   volume = 282,
    pages = {L55-L58},
      doi = {10.1086/184304},
   adsurl = {http://adsabs.harvard.edu/cgi-bin/nph-bib_query?bibcode=1984ApJ...282L..55V&db_key=AST}
}

@ARTICLE{Pacholczyk1976,
   author = {{Pacholczyk}, A.~G. and {Scott}, J.~S.},
    title = "{In situ particle acceleration and physical conditions in radio tail galaxies}",
  journal = {\apj},
     year = 1976,
    month = jan,
   volume = 203,
    pages = {313-315},
   adsurl = {http://adsabs.harvard.edu/cgi-bin/nph-bib_query?bibcode=1976ApJ...203..313P&db_key=AST}
}

@ARTICLE{OdeaOwen1987,
   author = {{O'Dea}, C.~P. and {Owen}, F.~N.},
    title = "{Astrophysical implications of the multifrequency VLA observations of NGC 1265}",
  journal = {\apj},
     year = 1987,
    month = may,
   volume = 316,
    pages = {95-112},
      doi = {10.1086/165182},
   adsurl = {http://adsabs.harvard.edu/cgi-bin/nph-bib_query?bibcode=1987ApJ...316...95O&db_key=AST}
}

@ARTICLE{Hjellming1981,
   author = {{Hjellming}, R.~M. and {Johnston}, K.~J.},
    title = "{An analysis of the proper motions of SS 433 radio jets}",
  journal = {\apjl},
     year = 1981,
    month = jun,
   volume = 246,
    pages = {L141-L145},
      doi = {10.1086/183571},
   adsurl = {http://adsabs.harvard.edu/cgi-bin/nph-bib_query?bibcode=1981ApJ...246L.141H&db_key=AST}
}

@ARTICLE{vanderlaan1969,
   author = {{van der Laan}, H. and {Perola}, G.~C.},
    title = "{Aspects of Radio Galaxy Evolution}",
  journal = {\aap},
     year = 1969,
    month = dec,
   volume = 3,
    pages = {468},
   adsurl = {http://adsabs.harvard.edu/abs/1969A%26A.....3..468V}
}

@ARTICLE{Blustin2009,
   author = {{Blustin}, A.~J. and {Fabian}, A.~C.},
    title = "{Radio constraints on the volume filling factors of AGN winds}",
  journal = {\mnras},
archivePrefix = "arXiv",
   eprint = {0904.0209},
 primaryClass = "astro-ph.CO",
     year = 2009,
    month = jul,
   volume = 396,
    pages = {1732-1736},
      doi = {10.1111/j.1365-2966.2009.14856.x},
   adsurl = {http://adsabs.harvard.edu/abs/2009MNRAS.396.1732B}
}

@ARTICLE{Ho2001,
   author = {{Ho}, L.~C. and {Peng}, C.~Y.},
    title = "{Nuclear Luminosities and Radio Loudness of Seyfert Nuclei}",
  journal = {\apj},
   eprint = {astro-ph/0102502},
     year = 2001,
    month = jul,
   volume = 555,
    pages = {650-662},
      doi = {10.1086/321524},
   adsurl = {http://adsabs.harvard.edu/abs/2001ApJ...555..650H}
}

@ARTICLE{Laing2008,
       author = {{Laing}, R.~A. and {Bridle}, A.~H. and {Parma}, P. and {Murgia}, M.},
        title = "{Structures of the magnetoionic media around the Fanaroff-Riley Class I radio galaxies 3C31 and Hydra A}",
      journal = {\mnras},
         year = 2008,
        month = dec,
       volume = {391},
       number = {2},
        pages = {521-549},
          doi = {10.1111/j.1365-2966.2008.13895.x},
archivePrefix = {arXiv},
       eprint = {0809.2411},
 primaryClass = {astro-ph},
       adsurl = {https://ui.adsabs.harvard.edu/abs/2008MNRAS.391..521L}
}

@ARTICLE{Laing2014,
   author = {{Laing}, R.~A. and {Bridle}, A.~H.},
    title = "{Systematic properties of decelerating relativistic jets in low-luminosity radio galaxies}",
  journal = {\mnras},
archivePrefix = "arXiv",
   eprint = {1311.1015},
     year = 2014,
    month = feb,
   volume = 437,
    pages = {3405-3441},
      doi = {10.1093/mnras/stt2138},
   adsurl = {http://adsabs.harvard.edu/abs/2014MNRAS.437.3405L}
}

@ARTICLE{Laing2002,
   author = {{Laing}, R.~A. and {Bridle}, A.~H.},
    title = "{Relativistic models and the jet velocity field in the radio galaxy 3C 31}",
  journal = {\mnras},
   eprint = {astro-ph/0206215},
     year = 2002,
    month = oct,
   volume = 336,
    pages = {328-352},
      doi = {10.1046/j.1365-8711.2002.05756.x},
   adsurl = {http://adsabs.harvard.edu/abs/2002MNRAS.336..328L}
}

@ARTICLE{Gabuzda2004,
   author = {{Gabuzda}, D.~C. and {Murray}, {\'E}. and {Cronin}, P.},
    title = "{Helical magnetic fields associated with the relativistic jets of four BL Lac objects}",
  journal = {\mnras},
   eprint = {astro-ph/0405394},
     year = 2004,
    month = jul,
   volume = 351,
    pages = {L89-L93},
      doi = {10.1111/j.1365-2966.2004.08037.x},
   adsurl = {http://adsabs.harvard.edu/abs/2004MNRAS.351L..89G}
}

@ARTICLE{Merloni2007,
   author = {{Merloni}, A. and {Heinz}, S.},
    title = "{Measuring the kinetic power of active galactic nuclei in the radio mode}",
  journal = {\mnras},
archivePrefix = "arXiv",
   eprint = {0707.3356},
     year = 2007,
    month = oct,
   volume = 381,
    pages = {589-601},
      doi = {10.1111/j.1365-2966.2007.12253.x},
   adsurl = {http://adsabs.harvard.edu/abs/2007MNRAS.381..589M}
}

@ARTICLE{Kellermann1989,
       author = {{Kellermann}, K.~I. and {Sramek}, R. and {Schmidt}, M. and {Shaffer}, D.~B. and {Green}, R.},
        title = "{VLA Observations of Objects in the Palomar Bright Quasar Survey}",
      journal = {\aj},
         year = 1989,
        month = oct,
       volume = {98},
        pages = {1195},
          doi = {10.1086/115207},
       adsurl = {https://ui.adsabs.harvard.edu/abs/1989AJ.....98.1195K}
}

@ARTICLE{Cecil2001,
       author = {{Cecil}, Gerald and {Bland-Hawthorn}, Joss and {Veilleux}, Sylvain and {Filippenko}, Alexei V.},
        title = "{Jet- and Wind-driven Ionized Outflows in the Superbubble and Star-forming Disk of NGC 3079}",
      journal = {\apj},
         year = 2001,
        month = jul,
       volume = {555},
       number = {1},
        pages = {338-355},
          doi = {10.1086/321481},
archivePrefix = {arXiv},
       eprint = {astro-ph/0101010},
 primaryClass = {astro-ph},
       adsurl = {https://ui.adsabs.harvard.edu/abs/2001ApJ...555..338C}
}

@ARTICLE{Hopkins2010,
       author = {{Hopkins}, Philip F. and {Elvis}, Martin},
        title = "{Quasar feedback: more bang for your buck}",
      journal = {\mnras},
         year = 2010,
        month = jan,
       volume = {401},
       number = {1},
        pages = {7-14},
          doi = {10.1111/j.1365-2966.2009.15643.x},
archivePrefix = {arXiv},
       eprint = {0904.0649},
 primaryClass = {astro-ph.CO},
       adsurl = {https://ui.adsabs.harvard.edu/abs/2010MNRAS.401....7H}
}

@ARTICLE{CASA2022,
       author = {{THE CASA TEAM} and {Bean}, Ben and {Bhatnagar}, Sanjay and {Castro}, Sandra and {Donovan Meyer}, Jennifer and {Emonts}, Bjorn and {Garcia}, Enrique and {Garwood}, Robert and {Golap}, Kumar and {Gonzalez Villalba}, Justo and {Harris}, Pamela and {Hayashi}, Yohei and {Hoskins}, Josh and {Hsieh}, Mingyu and {Jagannathan}, Preshanth and {Kawasaki}, Wataru and {Keimpema}, Aard and {Kettenis}, Mark and {Lopez}, Jorge and {Marvil}, Joshua and {Masters}, Joseph and {McNichols}, Andrew and {Mehringer}, David and {Miel}, Renaud and {Moellenbrock}, George and {Montesino}, Federico and {Nakazato}, Takeshi and {Ott}, Juergen and {Petry}, Dirk and {Pokorny}, Martin and {Raba}, Ryan and {Rau}, Urvashi and {Schiebel}, Darrell and {Schweighart}, Neal and {Sekhar}, Srikrishna and {Shimada}, Kazuhiko and {Small}, Des and {Steeb}, Jan-Willem and {Sugimoto}, Kanako and {Suoranta}, Ville and {Tsutsumi}, Takahiro and {van Bemmel}, Ilse M. and {Verkouter}, Marjolein and {Wells}, Akeem and {Xiong}, Wei and {Szomoru}, Arpad and {Griffith}, Morgan and {Glendenning}, Brian and {Kern}, Jeff},
        title = "{CASA, the Common Astronomy Software Applications for Radio Astronomy}",
      journal = {arXiv e-prints},
         year = 2022,
        month = oct,
          eid = {arXiv:2210.02276},
        pages = {arXiv:2210.02276},
archivePrefix = {arXiv},
       eprint = {2210.02276},
 primaryClass = {astro-ph.IM},
       adsurl = {https://ui.adsabs.harvard.edu/abs/2022arXiv221002276T}
}

@ARTICLE{Colbert1996a,
       author = {{Colbert}, Edward J.~M. and {Baum}, Stefi A. and {Gallimore}, Jack F. and {O'Dea}, Christopher P. and {Lehnert}, Matthew D. and {Tsvetanov}, Zlatan I. and {Mulchaey}, John S. and {Caganoff}, Saul},
        title = "{Large-Scale Outflows in Edge-on Seyfert Galaxies. I. Optical Emission-Line Imaging and Optical Spectroscopy}",
      journal = {\apjs},
         year = 1996,
        month = jul,
       volume = {105},
        pages = {75},
          doi = {10.1086/192307},
archivePrefix = {arXiv},
       eprint = {astro-ph/9512169},
 primaryClass = {astro-ph},
       adsurl = {https://ui.adsabs.harvard.edu/abs/1996ApJS..105...75C}
}

@ARTICLE{Colbert1996b,
       author = {{Colbert}, Edward J.~M. and {Baum}, Stefi A. and {Gallimore}, Jack F. and {O'Dea}, Christopher P. and {Christensen}, Jennifer A.},
        title = "{Large-Scale Outflows in Edge-on Seyfert Galaxies. II. Kiloparsec-Scale Radio Continuum Emission}",
      journal = {\apj},
         year = 1996,
        month = aug,
       volume = {467},
        pages = {551},
          doi = {10.1086/177633},
archivePrefix = {arXiv},
       eprint = {astro-ph/9604022},
 primaryClass = {astro-ph},
       adsurl = {https://ui.adsabs.harvard.edu/abs/1996ApJ...467..551C}
}

@ARTICLE{Reuter1994,
       author = {{Reuter}, H. -P. and {Klein}, U. and {Lesch}, H. and {Wielebinski}, R. and {Kronberg}, P.~P.},
        title = "{The magnetic field in the halo of M 82. Polarized radio emission at mbda lambda 6.2 and 3.6 cm.}",
      journal = {\aap},
         year = 1994,
        month = feb,
       volume = {282},
        pages = {724-730},
       adsurl = {https://ui.adsabs.harvard.edu/abs/1994A&A...282..724R}
}

@ARTICLE{Ishibashi2011,
       author = {{Ishibashi}, W. and {Courvoisier}, T.~J. -L.},
        title = "{Synchrotron radio emission in radio-quiet AGNs}",
      journal = {\aap},
         year = 2011,
        month = jan,
       volume = {525},
          eid = {A118},
        pages = {A118},
          doi = {10.1051/0004-6361/201014987},
archivePrefix = {arXiv},
       eprint = {1010.5591},
 primaryClass = {astro-ph.HE},
       adsurl = {https://ui.adsabs.harvard.edu/abs/2011A&A...525A.118I}
}

@ARTICLE{Crenshaw2003,
       author = {{Crenshaw}, D. Michael and {Kraemer}, Steven B. and {George}, Ian M.},
        title = "{Mass Loss from the Nuclei of Active Galaxies}",
      journal = {\araa},
         year = 2003,
        month = jan,
       volume = {41},
        pages = {117-167},
          doi = {10.1146/annurev.astro.41.082801.100328},
       adsurl = {https://ui.adsabs.harvard.edu/abs/2003ARA&A..41..117C}
}

@ARTICLE{Radcliffe2021,
       author = {{Radcliffe}, J.~F. and {Barthel}, P.~D. and {Garrett}, M.~A. and {Beswick}, R.~J. and {Thomson}, A.~P. and {Muxlow}, T.~W.~B.},
        title = "{The radio emission from active galactic nuclei}",
      journal = {\aap},
         year = 2021,
        month = may,
       volume = {649},
          eid = {L9},
        pages = {L9},
          doi = {10.1051/0004-6361/202140791},
archivePrefix = {arXiv},
       eprint = {2104.04519},
 primaryClass = {astro-ph.GA},
       adsurl = {https://ui.adsabs.harvard.edu/abs/2021A&A...649L...9R}
}

@ARTICLE{Panessa2019,
       author = {{Panessa}, Francesca and {Baldi}, Ranieri Diego and {Laor}, Ari and {Padovani}, Paolo and {Behar}, Ehud and {McHardy}, Ian},
        title = "{The origin of radio emission from radio-quiet active galactic nuclei}",
      journal = {Nature Astronomy},
         year = 2019,
        month = apr,
       volume = {3},
        pages = {387-396},
          doi = {10.1038/s41550-019-0765-4},
archivePrefix = {arXiv},
       eprint = {1902.05917},
 primaryClass = {astro-ph.GA},
       adsurl = {https://ui.adsabs.harvard.edu/abs/2019NatAs...3..387P}
}

@ARTICLE{Carral1990,
       author = {{Carral}, Patricia and {Turner}, Jean L. and {Ho}, Paul T.~P.},
        title = "{15 GHz Compact Structure in Galactic Nuclei}",
      journal = {\apj},
         year = 1990,
        month = oct,
       volume = {362},
        pages = {434},
          doi = {10.1086/169280},
       adsurl = {https://ui.adsabs.harvard.edu/abs/1990ApJ...362..434C}
}

@ARTICLE{Veilleux2005,
       author = {{Veilleux}, Sylvain and {Cecil}, Gerald and {Bland-Hawthorn}, Joss},
        title = "{Galactic Winds}",
      journal = {\araa},
         year = 2005,
        month = sep,
       volume = {43},
       number = {1},
        pages = {769-826},
          doi = {10.1146/annurev.astro.43.072103.150610},
archivePrefix = {arXiv},
       eprint = {astro-ph/0504435},
 primaryClass = {astro-ph},
       adsurl = {https://ui.adsabs.harvard.edu/abs/2005ARA&A..43..769V}
}

@ARTICLE{Yongkang2025,
       author = {{Yongkang}, Yan and {Peng}, Zhang and {Zhou}, Lu and {Tong}, Bao and {Dejian}, Liu and {Gaochao}, Liu and {Qingzhong}, Liu and {Jingzhi}, Yan and {Xiangyun}, Zeng},
        title = "{Detection and characterization of quasiperiodic oscillations in the Seyfert galaxy NGC 4151}",
      journal = {\prd},
         year = 2025,
        month = feb,
       volume = {111},
       number = {4},
          eid = {043027},
        pages = {043027},
          doi = {10.1103/PhysRevD.111.043027},
archivePrefix = {arXiv},
       eprint = {2504.17436},
 primaryClass = {astro-ph.HE},
       adsurl = {https://ui.adsabs.harvard.edu/abs/2025PhRvD.111d3027Y}
}

@ARTICLE{Beck1996,
       author = {{Beck}, Rainer and {Brandenburg}, Axel and {Moss}, David and {Shukurov}, Anvar and {Sokoloff}, Dmitry},
        title = "{Galactic Magnetism: Recent Developments and Perspectives}",
      journal = {\araa},
         year = 1996,
        month = jan,
       volume = {34},
        pages = {155-206},
          doi = {10.1146/annurev.astro.34.1.155},
       adsurl = {https://ui.adsabs.harvard.edu/abs/1996ARA&A..34..155B}
}

@ARTICLE{May2017,
       author = {{May}, D. and {Steiner}, J.~E.},
        title = "{A two-stage outflow in NGC 1068}",
      journal = {\mnras},
         year = 2017,
        month = jul,
       volume = {469},
       number = {1},
        pages = {994-1025},
          doi = {10.1093/mnras/stx886},
archivePrefix = {arXiv},
       eprint = {2007.07932},
 primaryClass = {astro-ph.GA},
       adsurl = {https://ui.adsabs.harvard.edu/abs/2017MNRAS.469..994M}
}

@ARTICLE{Strickland2000,
       author = {{Strickland}, David K. and {Stevens}, Ian R.},
        title = "{Starburst-driven galactic winds - I. Energetics and intrinsic X-ray emission}",
      journal = {\mnras},
         year = 2000,
        month = may,
       volume = {314},
       number = {3},
        pages = {511-545},
          doi = {10.1046/j.1365-8711.2000.03391.x},
archivePrefix = {arXiv},
       eprint = {astro-ph/0001395},
 primaryClass = {astro-ph},
       adsurl = {https://ui.adsabs.harvard.edu/abs/2000MNRAS.314..511S}
}

@ARTICLE{Crenshaw2012,
       author = {{Crenshaw}, D.~M. and {Kraemer}, S.~B.},
        title = "{Feedback from Mass Outflows in Nearby Active Galactic Nuclei. I. Ultraviolet and X-Ray Absorbers}",
      journal = {\apj},
         year = 2012,
        month = jul,
       volume = {753},
       number = {1},
          eid = {75},
        pages = {75},
          doi = {10.1088/0004-637X/753/1/75},
archivePrefix = {arXiv},
       eprint = {1204.6694},
 primaryClass = {astro-ph.CO},
       adsurl = {https://ui.adsabs.harvard.edu/abs/2012ApJ...753...75C}
}

@INPROCEEDINGS{Kraemer2021,
       author = {{Kraemer}, S.~B. and {Turner}, T.~J. and {Crenshaw}, D.~M. and {Schmitt}, H.~R. and {Revalski}, M. and {Fischer}, T.~C.},
        title = "{Mass outflow of the X-ray emission line gas in NGC 4151}",
    booktitle = {Galaxy Evolution and Feedback across Different Environments},
         year = 2021,
       editor = {{Storchi Bergmann}, Thaisa and {Forman}, William and {Overzier}, Roderik and {Riffel}, Rog{\'e}rio},
       series = {IAU Symposium},
       volume = {359},
        month = jan,
        pages = {131-135},
          doi = {10.1017/S1743921320001660},
       adsurl = {https://ui.adsabs.harvard.edu/abs/2021IAUS..359..131K}
}

@ARTICLE{Xu2014,
       author = {{Xu}, Jun and {Han}, Jin-Lin},
        title = "{A compiled catalog of rotation measures of radio point sources}",
      journal = {Research in Astronomy and Astrophysics},
         year = 2014,
        month = aug,
       volume = {14},
       number = {8},
          eid = {942-958},
        pages = {942-958},
          doi = {10.1088/1674-4527/14/8/005},
archivePrefix = {arXiv},
       eprint = {1405.1920},
 primaryClass = {astro-ph.GA},
       adsurl = {https://ui.adsabs.harvard.edu/abs/2014RAA....14..942X}
}
\bibliographystyle{aasjournalv7}
\end{document}